\documentclass[preprint]{aastex}
\usepackage{txfonts}

\newcommand{\dmf}{$\Delta m_{15}(B)$}

\newcommand{\ubvri}{\hbox{$U\!BV\!RI$ }}          
\newcommand{\ubv}{\hbox{$U\!BV$ }}
\newcommand{\bvri}{\hbox{$BV\!RI$ }}
\newcommand{\bvi}{\hbox{$BV\!I$ }}
\newcommand{\bvr}{\hbox{$BV\!R$ }}

\newcommand{\vi}{\hbox{$V\!-\!I$}}               

\newcommand{\iub}{\hbox{$u\!-\!b$}}               
\newcommand{\ibv}{\hbox{$b\!-\!v$}}               
\newcommand{\ivr}{\hbox{$v\!-\!r$}}               
\newcommand{\ivi}{\hbox{$v\!-\!i$}}               

\newcommand{\vV}{\hbox{$v\!-\!V$}}                

\slugcomment{to appear in the \emph{Astronomical Journal}}
\shorttitle{UBVRI SN Ia light curves}
\shortauthors{S. Jha et al.}

\received{2004 December 6}
\accepted{2005 September 3}
\begin{document}

\title{UBVRI Light Curves of 44 Type Ia Supernovae}

\author{Saurabh~Jha\altaffilmark{1},
Robert~P.~Kirshner,
Peter~Challis,
Peter~M.~Garnavich,
Thomas~Matheson,
Alicia~M.~Soderberg, 
Genevieve~J.~M.~Graves,
Malcolm~Hicken, 
Jo\~ao~F.~Alves, 
H\'ector~G.~Arce, 
Zoltan~Balog, 
Pauline~Barmby, 
Elizabeth~J.~Barton, 
Perry~Berlind, 
Ann~E.~Bragg, 
C\'esar~Brice\~no, 
Warren~R.~Brown, 
James~H.~Buckley, 
Nelson~Caldwell, 
Michael~L.~Calkins, 
Barbara~J.~Carter, 
Kristi~Dendy~Concannon, 
R.~Hank~Donnelly, 
Kristoffer~A.~Eriksen, 
Daniel~G.~Fabricant, 
Emilio~E.~Falco, 
Fabrizio~Fiore, 
Michael~R.~Garcia, 
Mercedes~G\'omez, 
Norman~A.~Grogin, 
Ted~Groner, 
Paul~J.~Groot, 
Karl~E.~Haisch,~Jr., 
Lee~Hartmann, 
Carl~W.~Hergenrother, 
Matthew~J.~Holman, 
John~P.~Huchra, 
Ray~Jayawardhana, 
Diab~Jerius, 
Sheila~J.~Kannappan, 
Dong-Woo~Kim, 
Jan~T.~Kleyna, 
Christopher~S.~Kochanek, 
Daniel~M.~Koranyi, 
Martin~Krockenberger, 
Charles~J.~Lada, 
Kevin~L.~Luhman, 
Jane~X.~Luu, 
Lucas~M.~Macri, 
Jeff~A.~Mader, 
Andisheh~Mahdavi, 
Massimo~Marengo, 
Brian~G.~Marsden, 
Brian~A.~McLeod, 
Brian~R.~McNamara, 
S.~Thomas~Megeath, 
Dan~Moraru, 
Amy~E.~Mossman, 
August~A.~Muench, 
Jose~A.~Mu\~noz, 
James~Muzerolle, 
Orlando~Naranjo, 
Kristin~Nelson-Patel, 
Michael~A.~Pahre, 
Brian~M.~Patten, 
James~Peters,
Wayne~Peters, 
John~C.~Raymond, 
Kenneth~Rines, 
Rudolph~E.~Schild, 
Gregory~J.~Sobczak, 
Timothy~B.~Spahr, 
John~R.~Stauffer, 
Robert~P.~Stefanik, 
Andrew~H.~Szentgyorgyi, 
Eric~V.~Tollestrup, 
Petri~V\"ais\"anen, 
Alexey~Vikhlinin, 
Zhong~Wang, 
S.~P.~Willner, 
Scott~J.~Wolk, 
Joseph~M.~Zajac, 
Ping~Zhao, 
and Krzysztof~Z.~Stanek
}
\affil{Harvard-Smithsonian Center for Astrophysics, 60 Garden Street,
  Cambridge, MA 02138}
\email{saurabh@astron.berkeley.edu}
\altaffiltext{1}{present address: Department of Astronomy and Miller
  Institute for Basic Research, 601 Campbell Hall, University of
  California, Berkeley, CA 94720-3411}

\begin{abstract}
We present \ubvri photometry of 44 type-Ia supernovae (SN Ia) observed
from 1997 to 2001 as part of a continuing monitoring campaign at the
Fred Lawrence Whipple Observatory of the Harvard-Smithsonian Center
for Astrophysics. The data set comprises 2190 observations and is the
largest homogeneously observed and reduced sample of SN Ia to date,
nearly doubling the number of well-observed, nearby SN Ia with
published multicolor CCD light curves. The large sample of $U$-band
photometry is a unique addition, with important connections to SN Ia
observed at high redshift. The decline rate of SN Ia $U$-band light
curves correlates well with the decline rate in other bands, as does
the \ub\ color at maximum light. However, the $U$-band peak magnitudes
show an increased dispersion relative to other bands even after
accounting for extinction and decline rate, amounting to an additional
$\sim$40\% intrinsic scatter compared to $B$-band. 
\end{abstract}


\section{Introduction}

Over the last decade, type Ia supernovae (SN Ia) have become
increasingly sharp tools for precision cosmology, with applications of
these exquisite distance indicators ranging from our galactic
neighbors to establish the Hubble constant, to halfway across the
observable Universe to uncover cosmic deceleration and acceleration
(Riess et al.~2004; Barris et al.~2004; Knop et al.~2003 and
references therein). These cosmological applications of SN Ia
rely on accurate, high-precision, and unscheduled measurements of
their light curves in multiple passbands over a period of weeks,
presenting a challenge to would-be observers.

The project of collecting a large sample of nearby SN Ia with
high-quality, multicolor CCD photometry to be used in cosmological
studies began in earnest in 1990, with the Cal\'an/Tololo survey
(Hamuy et al.~1993) that combined a photographic search for SN in
the southern sky with a program of CCD followup photometry obtained
with the help of visiting astronomers. Hamuy et al.~(1996) present
Johnson/Cousins \bvi photometry of 29 SN Ia from this project (27 of
which were discovered as part of the survey itself) out to redshifts
$z \simeq 0.1$.

In 1993, astronomers at the Harvard-Smithsonian Center for
Astrophysics (CfA) began a campaign of CCD photometric and
spectroscopic monitoring of newly-discovered supernovae at the Fred
L. Whipple Observatory (FLWO) on Mt. Hopkins in southern Arizona, and
this program has been ongoing ever since. We employ a similar
cooperative observing strategy for the follow-up photometry, whereby
the SN monitoring program is allocated a small amount of time each
night ($\sim$20 minutes), with the observations being carried out by
the scheduled observer. Our SN program is also allocated approximately
one dedicated night per month that is used for photometry of the
fainter objects, photometric calibration of the SN fields, and
template observations after the SN have faded.

Our cooperative observing strategy has been very successful so
far. FLWO \bvri observations of 22 type Ia supernovae discovered
between 1993 and 1996 have been published by Riess et al.~(1999) and
we have also undertaken \ubvri photometry and in-depth analysis of a
number of individual SN Ia observed as part of this program: SN
1998bu (Jha et al.~1999), SN 1999by (Garnavich et al.~2004), SN 1998aq
(Riess et al.~2005) and SN 2001V (Mandel et al.~2005,
in preparation).

Here we report our \ubvri photometry for 44 SN Ia discovered between
1997 and 2000. The full data set presented here consists of 2190
observations on 338 nights, and is the largest set of homogeneously
observed and reduced SN Ia data published to date.

\section{Data and Reduction}

\subsection{Discovery}

Our program of supernova photometry consists solely of follow-up; we
search only our email, not the sky, to find new supernovae. A number
of observers, both amateur and professional, are engaged in searching
for supernovae. We rely on these searches, as well as prompt
notification of candidates, coordinated by Dan Green and Brian Marsden
of the IAU's Central Bureau for Astronomical Telegrams (CBAT), with
confirmed SN reported in the IAU Circulars. In some cases the SN
discoverers provide spectroscopic classification of the new objects,
but generally spectroscopy is obtained by others, and reported
separately in the IAU Circulars. With our spectroscopic SN follow-up
program at the F. L. Whipple Observatory 1.5m telescope and FAST
spectrograph (Fabricant et al.~1998), we have classified a large
fraction of the new, nearby supernovae reported over the last several
years and compiled a large spectroscopic database (Matheson et
al.~2005, in preparation).

Given a newly discovered and classified supernova, several factors
help determine whether or not we include it in our monitoring
program. Because of their importance, SN Ia are often given higher
priority over other types, but factors such as ease of observability
(southern targets and those discovered far to the west are less
appealing), supernova phase (objects whose spectra indicate they are
after maximum light are given lower priority), redshift (more nearby
objects are favored), as well as the number of objects we are already
monitoring are significant. Our final sample of well-observed SN Ia is
not obtained from a single well-defined set of criteria, and selection
effects in both the searches and follow-up may make this sample
unsuitable for some applications (such as determining the intrinsic
luminosity function of SN Ia, for example). A thorough discussion of
the selection biases in the Cal\'an/Tololo supernova search and
follow-up campaign can be found in Hamuy \& Pinto (1999).

The discovery data for the sample of SN Ia presented here are given in
Table \ref{ch3-tab-disc}. All of the SN Ia listed were discovered with
CCD images, except for SN 1997bp which was discovered visually, and SN
1999ef and SN 1999gh which were discovered photographically. New,
systematic CCD supernova searches have provided the great majority of
our sample: the Beijing Astronomical Observatory Supernova Survey (Li
et al.~1996; designated as BAO in Table \ref{ch3-tab-disc}), the UK
Nova/Supernova Patrol (Armstrong \& Hurst~1996; UK), the Puckett
Observatory Supernova Search (Puckett~1998; POSS), the Tenagra
Observatories supernova patrol (Schwartz~1997; TO), and the Lick
Observatory Supernova Search (Treffers et al.~1997; LOSS). In
addition, we note in Table \ref{ch3-tab-disc} supernovae whose
classification as SN Ia is from our spectroscopic monitoring program
described above (designated as ``CfA'').

\subsection{Observations}

All the photometry presented here was obtained with the F. L. Whipple
Observatory (FLWO) 1.2m telescope, with either the ``AndyCam'' CCD
camera or the ``4Shooter'' 2x2 CCD mosaic (Szentgyorgyi et al.~2005,
in preparation). Both instruments use thinned, backside illuminated,
anti-reflective coated Loral 2048$^2$ CCD detectors, situated at the
f/8 Cassegrain focus. The pixel scale is approximately $0\farcs33$ per
pixel, yielding a field of view of over 11\arcmin\ on a side for each
chip. All the data were taken in a 2x2 binned mode, resulting in a
sampling of $0\farcs66$ per pixel that is well matched to the typical
image quality ($1\farcs5$ to 2\arcsec\ FWHM).  We have ensured that
all data used are within the linear regime of the detectors.
Observations using the 4Shooter taken before October 1998 were made
with the ``chip 1'' CCD detector, while those taken afterwards were
made on ``chip 3'', which has slightly improved quantum efficiency but
slightly inferior cosmetic characteristics.

Both instruments have good near-ultraviolet and near-infrared
response, and our observations have been in the Johnson \ubv and
Kron-Cousins $RI$ bandpasses. The data were taken with two \ubvri
filtersets, the ``SAO'' set and the newer ``Harris''
set. Observations before December 1998 were taken with the SAO
filterset (the same described by Riess et al.~1999 and Jha et
al.~1999), while those taken after May 1999 were taken with the Harris
set. Between December 1998 and May 1999 only the Harris \ubvr\ filters
were available, and the $I$ filter used was from the SAO
filterset. Because of the importance of knowing precisely the
bandpasses used for a given observation (particularly for supernova
photometry), we discuss these in greater detail in \S \ref{ch3-sec-calib}.

Our observing approach, combining nightly requests for one or two
objects with monthly dedicated nights, allows us to sample the light
curves with the appropriate cadence. Generally observations are more
frequent when the SN Ia are near maximum light, and less frequent
(but deeper) as each SN Ia fades. During the period of these
observations, the FLWO 1.2m was equipped with the 4Shooter or AndyCam
usually only during dark time, with an infrared imager on the
telescope when the moon was near full. This unfortunately leads to
$\sim$1 to 2 week gaps in our light curves, but in most cases the
light curves are still well-defined and suitable for distance
analyses.

\subsection{Differential Photometry}

To measure the brightness of the supernova in any image, we perform
the photometry differentially with respect to stars in the field of
view, allowing for useful measurements even in non-photometric
conditions. In general we use as many of these comparison stars (or
``field standards'') as feasible, choosing stars that are bright
enough to be precisely measured but faint enough to not saturate the
detector in the late-time, deeper images. In addition, we try to
choose comparison stars that cover a range of color comparable to
those exhibited by SN Ia over their evolution, though it is often not
possible to find stars in the field that are as blue as SN Ia at or
before maximum light. Figure \ref{ch3-fig-finder} shows $R$-band
finder charts for all of the supernovae and their associated
comparison stars.

All of the CCD observations were reduced uniformly, with bad pixel
masking, bias subtraction and flat-field correction using the NOAO
Image Reduction and Analysis Facility (IRAF) CCDPROC
package\footnote{IRAF is distributed by the National Optical Astronomy
Observatories, which are operated by the Association of Universities
for Research in Astronomy, Inc., under cooperative agreement with the
National Science Foundation.}. In addition, we remove, to the extent
possible, the small but non-negligible, amount of fringing for
observations in the $I$-band via a fringe frame created from combined
night-sky exposures of sparse fields.

A major complication in supernova photometry arises in separating
light from the SN itself from light from the underlying galaxy at the
SN position. Poor subtraction of the background light can have
significant effects on the supernova light curve shapes and colors
(cf. discussion in Riess et al.~1999; Boisseau
\& Wheeler 1991). For this reason, we take observations of the
supernova fields the following year, after the SN has faded, to use as
templates that are subtracted from all of the previous images.  We
have used galaxy subtraction to perform the differential photometry of
all the SN Ia except for SN 2000cx, which was located very far from
the nucleus of its (elliptical) host galaxy, where the galaxy
background was negligible and template subtraction only added
undesirable correlated noise. For SN 2000cx, we performed point-spread
function (PSF) fitting photometry on the SN and comparison stars,
using the DAOPHOT/ALLSTAR (Stetson 1987, 1994) package in IRAF.

For the other 43 objects, we employed template subtraction as follows.
Generally a number of late-time images were taken in each passband
with exposure times comparable to or slightly longer than the deepest
images with the SN present, and we chose the set of images with the
best seeing to serve as the templates. For each passband, all of the
images were registered to the template, and the image subtraction was
performed using the ISIS subtraction package (Alard \& Lupton~1998) as
modified by B. Schmidt (personal communication), to allow for more
robust selection of regions in the two images suitable for
determination of the convolution kernel (avoiding saturated stars,
cosmic rays, and cosmetic defects).  We subtracted the template from
each SN image, and replaced a small region around the SN with the
template-subtracted version. In the typical case, where the template
image quality was better than the SN image, we convolved the template
to the SN image, subtracted, and replaced the SN neighborhood from the
subtracted image back into the original SN image. In the rare case,
where the SN image quality was better than the template, we degraded
the SN image to match the PSF of the template image, subtracted, and
replaced the subtracted SN neighborhood back into the convolved
(degraded) SN image. This procedure ensures that the PSF of the SN
matches the PSF of the comparison stars. We also added artificial
stars of known brightness into the SN images, mimicking the SN
subtraction procedure on these stars. Finally, we performed aperture
photometry as well as DoPHOT PSF-fitting photometry (Schechter, Mateo,
\& Saha 1993) on the SN, comparison stars and artificial stars in the
galaxy-subtracted images. We have checked that the recovered
magnitudes of the added stars match their input magnitudes, and that
the aperture and PSF photometry gave consistent results, generally to
better than 0.01 mag. We have also verified that this photometry
derived via galaxy subtraction is consistent with direct PSF
photometry for SN where the galaxy background is exceptionally
smooth. For our final differential photometry, we have chosen to use
the aperture photometry of the SN and comparison stars, with an
aperture radius given by 0.75 times the FWHM of the PSF.

This general strategy is identical to that used by the High-Z
Supernova Search Team (Schmidt et al.~1998) in analysis of
high-redshift SN Ia; while the actual software is in a state of
constant evolution, we have used one incarnation for all the light
curves presented here. The result of this process is homogeneous and
reliable differential \ubvri photometry of each supernova and its
associated comparison stars on the natural system of the observations
(i.e., instrumental magnitudes).

\subsection{Calibration \label{ch3-sec-calib}}

We calibrate each of the supernova fields following the precepts of
Harris, Fitzgerald \& Reed\ (1981), using the all-sky \ubvri standard
stars of Landolt~(1992). On photometric nights, we typically observe
on the order of 10 to 15 Landolt fields over a wide range of airmass
(generally from 1.1 to $\sim$2). We perform aperture photometry on the
reduced Landolt fields using the APPHOT package in IRAF, using a 6
pixel aperture radius ($\sim$4\arcsec) that is then corrected to a 15
pixel radius ($\sim$10\arcsec) via a curve of growth defined by a few
isolated, bright stars in each image. We then determine the zeropoints
and transformation coefficients linear in airmass and color from the
instrumental magnitudes $ubvri$ to the standard Landolt \ubvri
magnitudes and \ub, \bv, \vr, and \vi\ colors. For nights when many
standard stars were observed, we check the linear solution by also
fitting a quadratic term in color as well as a color times airmass
term; in all cases the coefficients for the higher order terms are
negligible, and so we use only the linear solutions. Because of the
different detector and filterset combinations we have used, we take
care to keep track of the transformation coefficients separately. As
expected, for a given detector/filterset combination, the variations
in the zeropoints and airmass terms are small but significant, while
the color terms are always consistent within the fit uncertainties.

Once we have the standard solution for a photometric night, we apply
this solution to the instrumental aperture magnitudes of the
comparison stars in each SN field, measured in exactly the same way as
the Landolt standard stars. This yields the standard \ubvri magnitudes
of the comparison stars in each SN field. For most of the fields, we
have several calibrations, enabling us to average the results and
identify and eliminate outliers. For a handful of SN, however, we have
only one night of photometric calibration, a somewhat perilous
situation. Nevertheless, for every one of these objects we have
checked that other SN fields taken on the same night have photometry
that is consistent on other nights, bolstering our confidence that the
photometry of objects with only one night of calibration is not
significantly in error. In Table \ref{ch3-tab-compstar}, we present
the final comparison star $V$ magnitudes and colors with their
uncertainties (in the mean), as well as the number of photometric
nights averaged to yield the results. We also give positions of the
supernovae and comparison stars, referenced to the USNO-A2.0 catalogue
(Monet et al.~1998), with a typical root-mean-square (RMS) uncertainty
of $\pm 0\farcs3$. The location of the supernovae and comparison stars
are shown in Figure \ref{ch3-fig-finder}.

We present the average color terms for each detector/filterset
combination in Table \ref{ch3-tab-colterm}, along with the internal
uncertainties in the mean.  We do not have data on any photometric
nights when the AndyCam and the Harris filters were on the telescope,
and thus we could not use observations of standard stars to determine
the color terms for this detector/filterset combination. Instead we
used the color terms based on the calibrated comparison stars
themselves (allowing for a variable zeropoint for each frame, given
the non-photometric conditions). For the other detector/filterset
combinations, we successfully used this method to check the color
terms for consistency.

Armed with the comparison star standard magnitudes and the color terms
for each detector/filterset combination, we determined the zeropoint
for each SN image by transforming the comparison star standard
magnitudes to instrumental magnitudes (using the appropriate color
term) and comparing to them to the observed comparison star
magnitudes. Because the SN is observed at the same time (and thus,
airmass) as the comparison stars, the airmass term is consumed into
the zeropoint, which is robustly determined from the flux-weighted
average of the comparison stars. We then use this zeropoint to
determine calibrated instrumental magnitudes for the SN, and use the
linear color term transformation to arrive at the final Landolt
standard magnitudes for the supernova. We keep track of and propagate
the uncertainties throughout this procedure, including photon noise in
the instrumental magnitudes, dispersion in the photometric solution,
uncertainties in the transformation coefficients, and internal
uncertainty in the zeropoint for each image. The final standard system
\ubvri magnitudes of the supernovae, along with the uncertainties and
the detector/filterset combination are given in Tables
\ref{ch3-tab-sn97E} to \ref{ch3-tab-sn00fa}. The \ubvri light curves
of the 44 SN Ia are shown in Figure \ref{ch3-fig-ltcurves} relative to
maximum light (defined in the $B$-band) and corrected for
time-dilation to the SN rest frame (cf. Table \ref{ch3-tab-sndata}, \S
\ref{ch3-sec-sndata}).

We have used linear color transformations between the supernova
instrumental magnitudes and standard magnitudes as has been
conventional when presenting SN Ia light curves, but these may be
inappropriate due to the strong, broad features present in SN spectra,
as compared to the stars from which the color terms are
derived. Fortunately, our primary concern is accurate photometry of SN
Ia near and soon after maximum light, when the SN flux is still
dominated by the continuum in this ``photospheric'' phase, where the
linear transformations derived from stars would be most
appropriate. Furthermore, for most of the detector/filter
combinations, the color terms do not strongly suggest the effective
wavelengths are far from the standard bandpasses.  The ultimate test,
though, is in the light curves, which also give no evidence for
systematic differences between observations taken with different
detector/filterset combinations. For instance, the smoothness of the
light curve of SN 1998es (Table \ref{ch3-tab-sn98es}), observed with
both instruments with multiple filtersets, is evidence of the internal
consistency and homogeneity of the photometry. This is particularly
important in the $U$-band, for which this sample represents the first
large collection of SN Ia photometry, but which is also notoriously
difficult to transform to a standard system (see, e.g., Suntzeff et
al.~1999; Jha et al.~1999).

Though we have strived to ensure that the transformations to the
standard system result in consistent, homogeneous photometry, the
future uses of these data might nonetheless be limited by the accuracy
of these transformations.  It may be more convenient and useful to
have the data as measured on the natural system.  Given the color
terms in Table \ref{ch3-tab-colterm}, it is straightforward to
transform the data back to the natural system (and the natural system
magnitudes are available on request). This is only useful, however, in
conjunction with the natural system passbands. We have synthesized
these passbands by combining the primary and secondary mirror
reflectivities (taken simply as two reflections off an aluminum
surface), the measured filter transmissions, and the measured detector
quantum efficiencies (QE).\footnote{We reiterate the footnote of
Suntzeff et al.~(1999), that the Bessell~(1990) passband convention
that we adopt also includes a term in the passband that is a linearly
increasing function of wavelength. In this convention, then, the
magnitude measured with a photon counting detector is $m = -2.5 \log
\left\{ \int F_\lambda(\lambda)\,R(\lambda)\,d\lambda \right\} + $
const, where $F_\lambda(\lambda)$ is the source flux density and
$R(\lambda)$ is the bandpass response.} We have assumed that the shape
of the QE curves for the two 4Shooter chips is identical. The
synthesized passbands are shown in Figure \ref{ch3-fig-allpb}, along
with the standard $U\!X$ and \bvri passbands of
Bessell~(1990). Because the $U$ passband is defined by the atmospheric
cutoff in the blue, we follow the Bessell convention of realizing this
passband at airmass 1.0 (using the IRAF Kitt Peak atmospheric
extinction curve, adjusted to match the average observed extinction
coefficients), whereas the $BVRI$ passbands are extra-atmospheric
(i.e., airmass 0). As shown in Figure \ref{ch3-fig-allpb}, the
correspondence between the natural system passbands and the Bessell
standard response curves is quite good, save for the $I$-band in the
SAO filterset. The synthesized passbands are also tabulated in Table
\ref{ch3-tab-passbands}.

Through synthetic photometry, we have verified that the natural system
passbands yield color terms consistent with those directly measured,
cf. Table \ref{ch3-tab-colterm}. We have also tried to constrain the
natural system passbands directly, though observations of
spectrophotometric standard stars on the photometric night of 2001
October 24 UT with the FLWO 1.2m telescope using chip 3 of the
4Shooter and the Harris filterset. We took multiple \ubvri
observations of the following eight tertiary spectrophotometric
standard stars (Massey et al~1988; Hamuy et al.~1992), over a wide
airmass range throughout the night: BD $+$28\arcdeg4211, Feige 34,
Feige 110, G191B2B, Hiltner 600, LTT 9239, LTT 9491, and Wolf 1346.
All of these stars also have published spectrophotometry in the red to
1\micron\ (Massey \& Gronwall 1990; Hamuy et al.~1994), allowing us to
measure synthetic \bvri magnitudes. The ground-based spectrophotometry
does not extend far enough to the blue with enough precision to
synthesize $U$ magnitudes (the Bessell UX passband extends down to 300
nm), and so for the $U$-band we have used the results of Bohlin,
Dickinson, \& Calzetti~(2001), who give HST/STIS fluxes for five of
the standards (BD $+$28\arcdeg4211, Feige 34, Feige 110, G191B2B, and
LTT 9491) extending below the atmospheric limit.

For each passband, we model the response curve as a cubic spline
through a number of spline points spaced equally over the wavelength
region where we expect a nonzero response. For each observation in the
passband ($\sim$20 each in \bvri and 13 in $U$), we correct the
standard star spectrum for atmospheric extinction (as above, to zero
airmass for \bvri and 1.0 airmass for $U$), and synthesize photometry
using the model passband. We find the best-fit model passband by
minimizing the residuals between the synthetic and observed
magnitudes, using a downhill-simplex (amoeba) method (Press et
al.~1992). Our model is specified by the amplitudes (restricted to
between zero and one) at the fixed spline points, with the
normalization adjusted to yield a fixed zeropoint. The number of
spline points in our model is somewhat arbitrary, limited by the
number of individual measurements ($\sim$20 in \bvri and 13 in
$U$). We have found that, in general, having fewer spline points is
generally advantageous, avoiding pathological cases and overfitting
the measurements at the expense of detailed information about the
shape of the response curve. We have also imposed constraints that the
model passband is ``reasonable''; it is forced to zero at the ends and
not allowed to be wildly oscillatory.

Given these constraints, the best-fit model passbands from the
spectrophotometric data are shown in Figure \ref{ch3-fig-specphot},
along with the 4Shooter/Harris passbands synthesized from the CCD QE
curves, filter transmissions, etc. from Figure \ref{ch3-fig-allpb},
and the Bessell~(1990) passbands. Because of the somewhat arbitrary
nature of the model, as well as uncertainties in the photometry, these
best-fit response curves should be viewed as ``typical'' realizations
of the true response, rather than exact representations. There is a
range of models that fit the data reasonably well (with a dispersion
of $\sim$0.02 mag in \bvri and $\sim$0.04 mag in $U$, similar to the
scatter typically exhibited by the Landolt standards), and this range
overlaps well with the calculated passbands.  A few of the
discrepancies between the solid and dashed curves seem to be robust;
in particular, the spectrophotometric data favor a $B$ response which
is narrower than the filter transmission would predict. To test this
definitively, we would need a larger data set, with more
spectrophotometric standards. 

Though we have only tried this exercise with one detector/filterset
combination, the results suggest that the match between the best-fit
model passband and the calculated passbands is generally good, with
the calculated passband yielding photometry always within 2$\sigma$ of
the best-fit. Furthermore, the constancy of the color terms for a
particular detector/filterset indicates that effects such variable
detector response or mirror reflectivity (due to cleanliness, for
instance) do not significantly affect the natural system
bandpasses. We thus conclude that the response curves shown in Figure
\ref{ch3-fig-allpb} and Table \ref{ch3-tab-passbands} are good
representations of the natural system.

\section{Results}

\subsection{Comparison with Published Photometry \label{ch3-sec-compub}}

A number of the supernovae presented here have published photometry
from other groups. Because of the difficulties in supernova photometry
(correcting for galaxy contamination, transformation to the standard
system, etc.), systematic differences between SN photometry from
different telescopes are common. These differences are generally small,
at the level of a few hundredths of a magnitude (see, e.g.,
Suntzeff et al.~1999; Jha et al.~1999; Riess et al.~1999), though
larger differences can occur with worse filter mismatches. In this
paper, we strive to present photometry that is internally as
homogeneous as possible, but it is still useful to compare these data
with independent observations. When the systematic differences are
small, combining these independent data sets is highly desirable,
producing dramatic improvements in the light curve sampling.

\subsubsection{SN 1997bp}

Altavilla et al.~(2004) present photometry of 18 SN Ia from ESO (La
Silla) and Asiago Observatories including four objects also presented
here. For SN 1997bp in NGC 4680, the two data sets are quite
complementary in supernova phase, with the Altavilla et al.~photometry
filling in a gap in our light curve just after maximum light. Based on
the few contemporaneous points, the photometry shows good agreement in
\bvri, with offsets $\lesssim$0.05 mag. However, the $U$-band
photometry is more discordant; the Altavilla et al.~measurements of SN
1997bp are $\sim$0.15 mag fainter in $U$ than the photometry
presented here.

\subsubsection{SN 1997br}

Li et al.~(1999) present extensive \bvri photometry of SN 1997br in
ESO 576-40 from observations at the Beijing Astronomical Observatory
0.6m and the Lick Observatory 0.76m Katzman Automatic Imaging
Telescope (KAIT). There is good agreement in the $V$ and $I$-band
photometry presented by Li et al.~and that presented here
(r.m.s. offsets $\lesssim 0.05$ mag), but there are larger, systematic
differences in $B$ (an r.m.s. offset of 0.08 mag, with the Li et
al.~photometry fainter before maximum light but brighter at later
times $\gtrsim 30$ days after maximum light). The most significant
discrepancy is in the $R$ photometry, where the Li et al.~photometry
is fainter than the FLWO photometry by $\sim$0.18 mag on average, and
approaching $\sim$0.25 mag even near maximum light.  The field
comparison stars we have in common show good agreement\footnote{The
finder chart presented by Li et al.~(1999) seems to indicate their
star E corresponds to our comparison star 6, but the photometry in
their Table 1 matches our photometry of comparison star 5, which is
somewhat fainter and much redder than star 6. Because of its
faintness, Li et al.~do not assign much weight to this star, so it is
unlikely to explain the discrepant $R$ magnitudes.}. However, the
color terms presented by Li et al.~are relatively large in R, e.g.,
$(\ivr)/(\vr) = 1.20$ for the KAIT observations, and the photometry
differences correlate well with the SN color, implying that the
transformation to the standard system is the likely culprit.

Altavilla et al.~(2004) report 3 epochs of \bvri photometry of SN
1997br and these show good agreement ($\lesssim 0.05$ mag) with
the FLWO photometry presented here (also showing a similar offset when
compared to the Li et al.~$R$-band data). Altavilla et al.~also
present two $U$-band points, in fairly good accord ($\lesssim 0.1$ mag)
with the FLWO photometry.

\subsubsection{SN 1997cn}

Turatto et al.~(1998) present \ubvri photometry of SN 1997cn in NGC
5490 from a number of telescopes at ESO, La Silla. Our photometry
agrees well with theirs in $B$ and $V$; in $U$ our photometry is generally
brighter by $\sim$0.15 mag, but is consistent within the photometric
uncertainties for this faint object. Our $R$ and $I$-band photometry is also
brighter, by $\sim$0.08 mag. We have one comparison star in common with
Turatto et al.~(their star 2 is our star 9) and our photometry for
this star agree within the reported uncertainties in all bands.

\subsubsection{SN 1998de}

Extensive \bvri observations of SN 1998de in NGC 252 are presented by
Modjaz et al.~(2001). The data presented there have been K-corrected
to the SN rest frame, and to facilitate direct comparison with our
observations, M. Modjaz has kindly supplied us with their standard
magnitudes before K-correction. Our data set is relatively sparse
compared to that presented by Modjaz et al.\footnote{This is due to
the fact that the SN peaked at the end of July, just as FLWO undergoes
a month-long shutdown because of the southern Arizona monsoons.} but
the agreement is very good before maximum light ($\lesssim 0.05$
mag). Our $I$-band data taken about 45 days past maximum light show a
large discrepancy ($\sim$0.4 mag), likely a result of the
transformation to the standard system at a phase when the SN spectrum
is highly non-stellar. Comparison star C of Modjaz et al.~is the same
as our star 8, and our calibration is consistent.

\subsubsection{SN 1999aa}

Krisciunas et al.~(2000) present \bvri observations of SN 1999aa in
NGC 2595 that very nicely complement the data presented here. In
addition the photometric agreement is superb, with r.m.s. offsets
$\lesssim$ 0.03 mag near maximum light and $\lesssim$ 0.06 mag at late
times. Combining the data sets yields an excellent light curve for
this object.

Altavilla et al.~(2004) present 3 epochs of \ubvri photometry of SN
1999aa, with good accord in \bvr at the level $\sim$0.04 mag, with
larger discrepancies in $I$ ($\sim$0.1 mag at 30 days past maximum
light and $\sim$0.2 mag at 60 days past maximum light). The $U$-band
agreement is also good: $\sim$0.05 mag at +30 days and $\sim$0.1 mag
at +60 days.

\subsubsection{SN 1999cl}

Krisciunas et al.~(2000) also present \bvri observations of the nearby
SN 1999cl in NGC 4501 (M88). The data are not as extensive as for SN
1999aa, nor is the photometric agreement as good. The two sets agree
relatively well in all bands at maximum light ($\sim$0.03 mag), but
the photometry of Krisciunas et al.~at about a month past maximum is
brighter than our (single) late time point at that epoch by $0.1$ to
$0.3$ mag in the different bands. Moreover, the discrepancy is larger
in the red. This is a good indication of contamination from the host
galaxy; indeed, Krisciunas et al.~note that SN 1999cl might be an
object where galaxy subtraction would improve their aperture
photometry performed without a template. Our late-time images after
the SN had faded show that the host galaxy makes a non-negligible
contribution to the flux at the position of the supernova. Based on
this discrepancy, Krisciunas et al.~have reanalyzed their data for SN
1999cl with subtraction of host-galaxy template images, and the new
results bring the photometry into much better agreement
(K. Krisciunas, personal communication).

\subsubsection{SN 1999ek}

Extensive \bvri photometry of SN 1999ek in UGC 3329 is provided by
Krisciunas et al.~(2004), supplemented by the handful of data points
presented here. Comparing the one epoch common to both data sets shows
good agreement ($\sim$0.05 mag) in $B$ and $I$, as well as excellent
agreement ($\sim$0.01 mag) in $V$ and $R$. In addition, Krisciunas et
al.~list \bvri magnitudes for two of the field comparison stars we
have used, with excellent agreement ($\sim$0.01 mag) in all bands.

\subsubsection{SN 1999gp and SN 2000ce}

Krisciunas et al.~(2001) present \bvri photometry of five SN Ia,
including SN 1999gp in UGC 1993 (with galaxy subtraction) and SN
2000ce in UGC 4195. For SN 1999gp, the two sets of photometry match
extremely well ($\lesssim 0.03$ mag), with only a small ($\sim$0.05
mag) consistent difference in the $R$ band photometry. This
discrepancy can be traced directly to the comparison stars, as the
ones in common show an identical offset. Our comparison star
photometry for the SN 1999gp field comes from 5 photometric nights,
with consistent $R$ photometry on all epochs. We thus recommend that
the Krisciunas et al.~SN 1999gp $R$ photometry be adjusted 0.05 mag
brighter to be consistent with the data presented here. As in the case
of SN 1999aa, the data sets are nicely complementary.

The light curve of SN 2000ce also benefits from the combined data
sets. In fact, the overlap is very slight (we have two epochs in
common, and only one for all the bands simultaneously). Nonetheless,
the agreement of the photometry at these epochs is good
($\lesssim$0.04 mag).

\subsubsection{SN 2000cx}

Li et al.~(2001) and Candia et al.~(2003) present an immense data set
in \ubvri for the unique SN 2000cx in NGC 524, with an additional two
epochs of \ubvri reported in Altavilla et al.~(2004). The photometry
presented here is also quite extensive, except for the fact the SN was
discovered in mid July, just prior to the aforementioned August
shutdown of FLWO. Thus, our data set consists of only set of points near
maximum light, before a large number of observations beginning a month
later. The data taken together comprise the most optical photometry of
any SN Ia, and generally show good photometric agreement, at the level
of $\sim$0.05 mag, as far as 100 days past maximum light (see Figure
3 of Candia et al.). At even later times, the agreement is still
generally good, though there are some larger discrepancies,
worst in $I$-band where the FLWO data and the KAIT data of Li
et al.~differ by $\sim$0.4 mag. Candia et al.~provide
more detailed comparisons of subsets of this large data set.

Though we have described photometric agreement from different
telescopes at the level of $\lesssim$ 0.05 mag as ``good'', it
nonetheless remains the case that these differences are systematic and
often exceed the nominal published uncertainties. The problem is
almost certainly caused by variations in the photometric passbands at
different sites that cannot be corrected by a simple linear
transformation based on a broad-band color. Some of these
discrepancies can be overcome by corrections derived from direct
application of instrumental passbands to supernova spectrophotometry
(e.g., Jha et al.~1999). Stritzinger et al.~(2002) have formalized
this idea through ``S-corrections'' determined in analogy to
K-corrections. However, the calculated S-corrections have not always
proved effective in reconciling discordant photometry. In addition,
accurate S-corrections require accurate knowledge of both instrumental
bandpasses and SN spectrophotometry, neither of which are always
available. These issues in combining photometry from different sites
are compounded in cosmological applications of SN Ia over a wide range
of redshifts, and will be an important source of systematic
uncertainty that must be controlled in the era of precision
cosmology.

\subsection{SN and Host Galaxy Properties \label{ch3-sec-sndata}}

In Table \ref{ch3-tab-sndata} we list basic data about each SN Ia.
The host-galaxy heliocentric redshifts listed are taken from the
Updated Zwicky Catalog (Falco et al.~1999) if possible, and from the
NASA/IPAC Extragalactic Database\footnote{The NASA/IPAC Extragalactic
Database (NED) is operated by the Jet Propulsion Laboratory,
California Institute of Technology, under contract with the National
Aeronautics and Space Administration.} (NED) otherwise, where we favor
optical redshifts over \ion{H}{1} redshifts if there is a
discrepancy. For three objects, host galaxy redshifts were not
available, and we report them here based on spectroscopy with the FLWO
1.5m telescope plus FAST spectrograph (Fabricant et al.~1998) and
cross-correlation with galaxy templates: the host of SN 1997dg,
$cz_{\rm helio} = 9238 \pm 14 \; {\rm km \; s^{-1}}$; the host of SN
1998dx (UGC 11149), $cz_{\rm helio} = 16197 \pm 32 \; {\rm km \;
s^{-1}}$; and the host of SN 2000cf (MCG $+$11$-$19$-$25), $cz_{\rm
helio} = 10920 \pm 20 \; {\rm km \; s^{-1}}$.

The supernovae in the sample range from heliocentric redshifts of 1968
to 16197 ${\rm km \; s^{-1}}$, with median and mean redshifts of 4888
and 5274 ${\rm km \; s^{-1}}$, respectively. The mean redshift is
significantly less than both the original CfA sample of Riess et
al.~(1999; $\overline{cz} \simeq 7500 \; {\rm km \; s^{-1}}$) and the
Cal\'an/Tololo sample of Hamuy et al.~(1996a; $\overline{cz} \simeq
13500 \; {\rm km \; s^{-1}}$).  Nonetheless, most of the objects are
in the Hubble flow; 39 of the 44 SN Ia have $cz \geq 2500 \; {\rm km
\; s^{-1}}$ in the CMB rest frame, a slightly larger fraction than the
original CfA sample (17 out of 22).

The host-galaxy morphology information shown in Table
\ref{ch3-tab-sndata} is taken from NED, and the supernova offset from
the nucleus is taken from the IAU CBAT list of
supernovae\footnote{\texttt{http://cfa-www.harvard.edu/iau/lists/Supernovae.html}}.
Gallagher et al.~(2005) present an analysis of
correlations between these properties and SN luminosity. In Table
\ref{ch3-tab-sndata} we also list the Galactic reddening towards each
supernova, derived from the dust maps of Schlegel, Finkbeiner \&
Davis~(1998).

\subsection{Light Curve Properties}

In Table \ref{ch3-tab-dm15} we list the times of maximum light in $B$
for each supernova, as determined from either a direct polynomial fit
to the $B$ light curve, or from MLCS2k2 fits (Jha, Riess, \& Kirshner
2005, in preparation). We also present the epoch of the first
observation in our data set (measured in the SN rest frame). Over half
the objects (25 out of 44) have observations before maximum light, and
seventy percent (31 out of 44) have observations earlier than 5 days
past maximum light.

We have also fit the \bvi light curves of our supernova sample to
determine maximum light magnitudes and the parameter \dmf, that has
been shown to correlate with the supernova intrinsic luminosity
(Phillips~1993). Though originally defined as the measured decline
rate of the supernova in $B$ from maximum to 15 days past maximum
light, we follow Hamuy et al.~(1995, 1996a) where \dmf\ is a parameter
in a multi-dimensional fit to template light curves (each with a
predefined \dmf). We have followed the recipe of Hamuy et al.~(1996a)
in our fits, using a parabolic fit through the minimum reduced
$\chi^2$ in a fit of the \bvi light curves to each of a set of
templates (``de''-K-corrected and time-dilated to the observer's frame,
for each SN). We have used the six \bvi templates presented by Hamuy
et al.~(1996b), and augmented this sample with templates based on an
additional four well-observed SN Ia in order to produce more robust
measurements of \dmf: SN 1995al (Riess et al.~1999; \dmf\ = 0.83), SN
1998aq (Riess et al.~2005; \dmf\ = 1.13), SN 1998bu
(Suntzeff et al.~1999; Jha et al.~1999; \dmf\ = 1.01) and SN 1999by
(Garnavich et al.~2004; \dmf\ = 1.90). We were able to get reliable
\dmf\ measurements for all but four of the SN Ia\footnote{The four
objects include SN 1998D and SN 1999cw, for which the first
observation was well after maximum light; SN 1998co, for which the
data are quite sparse; and SN 2000cx, whose light curve is unique
among all SN Ia (Li et al.~2001).}; these values (not corrected for
host-galaxy reddening) and their uncertainties (estimated from the
curvature of the best-fit parabola) are listed in Table
\ref{ch3-tab-dm15}. We also present the \bvi magnitudes at maximum
light (in $B$) for each SN determined from the best-fit template.

To further explore the light curve properties of this sample, and in
particular, to study the $U$-band light curves, we have also fit the
light curves to templates, based on the timescale stretch
parameterization developed by the Supernova Cosmology Project
(Perlmutter et al.~1997, 1999; Goldhaber et al.~2001). The stretch
template presented by Goldhaber et al.~(2001) is only for the
$B$-band; we would like to fit the \ubv light curves, for which the
simple stretching of the time axis does a good job of fitting the
observed data. To construct $U$ and $V$-band templates, one
possibility is to use composite light curves, combining a large number
of supernovae to produce an average template. However, because some
objects are better sampled in different bands, the average templates
produced this way might not consistently represent a supernova of
``average'' light curve shape and/or luminosity. For this reason, we
have constructed
\ubv templates based on photometry of a single supernova, the
well-observed SN 1998aq (Riess et al.~2005). To retain
consistency with the Goldhaber et al.~(2001) normalization, we have
corrected our SN 1998aq \ubv stretch templates to $s = 1$, by fitting
the $B$ template to the SCP1997 template presented in that paper.

In fitting our stretch templates to the data, we generally follow the
methodology of Goldhaber et al.~(2001) as applied in their analysis of
the Cal\'an/Tololo sample (Hamuy et al.~1996a). We restrict the light
curves to between $-$10 and $+$40 days in the SN rest frame, and we
only include objects with photometry commencing earlier than 5 days
after maximum light. Because we are interested in understanding the
general light-curve properties of these SN Ia, we allow the fits to be
as unrestrictive as possible: we fit for the stretch individually in
each of the three bands, and allow the times of maxima to vary in each
band (plus or minus a few days), as well as individually fitting for
the \ubv peak magnitudes\footnote{We fit the data in magnitude space,
rather than flux space, out of convenience. Because we are only
fitting the light curves between $-$10 and $+$40 days, the difference
between the two approaches is negligible. Determining rise-time
information at very early epochs clearly benefits from fitting in flux
space, where negative and zero fluxes are common.}. We also impose an
error floor on the photometry equal to 0.007 times the peak flux, as
did Goldhaber et al.~(2001, see their Table 7); while this is
negligible near maximum, it becomes the dominant uncertainty in the
photometry at late times (for instance, corresponding to $\pm 0.2$ mag
in the $U$-band at $+$40 days). As in the \dmf\ fits above, we fit the
data in the observer's frame (de-K-correcting and time-dilating the
templates).

The limits on the epoch of first observation, and the requirement
that we need $\gtrsim$ 5 points between $-$10 and $+$40 days in each
of the three bands for a meaningful fit limits the application of this
method to 22 of the 44 SN Ia presented here. The results are
presented in Table \ref{ch3-tab-stretch}, listing the \ubv peak magnitudes
and timescale stretch factors, along with the differences in
the time of maximum light in $U$ and $V$ relative to $t_{\rm Bmax}$,
all with error estimates given by the formal uncertainties in the fit.

\section{Discussion: $U$-band Light Curves \label{ch3-sec-uband}}

The $U$-band photometry presented here, while just a fraction of the
whole dataset, is the first large sample of homogeneously observed and
reduced $U$ photometry of SN Ia. The \bvri properties of SN Ia are
well-studied, and while our data provide a much expanded sample of
\bvri light curves, here we focus on the new element, the $U$-band
data. Though a number of other SN Ia individually also have published
$U$-band photoelectric or CCD photometry, the difficulties of
transforming this photometry (with the variety of instruments,
filters, sensitivities, etc.; see, e.g., Schaefer~1995; Suntzeff et
al.~1999) to a standard system leads us first to examine the $U$-band
properties of SN Ia from FLWO observations alone, as we have taken
care to ensure internal consistency.

Figure \ref{ch3-fig-ucomp} shows the composite $U$-band light curve of
the 44 SN Ia presented in this paper, along with six other SN Ia
with $U$ data from the FLWO 1.2m: SN 1995al and SN 1996X (for which
\bvri light curves were presented by Riess et al.~1999), SN 1998aq
(Riess et al.~2005), SN 1998bu (Jha et al.~1999), SN
1999by (Garnavich et al.~2004), and SN 2001V (Mandel et al.~2005, in
preparation). Of the \ubvri passbands, the SN Ia light curve declines
fastest in $U$, with an average SN Ia dropping $\sim$1.5 mag in $U$
over the first 15 days after $B_{\rm max}$, as compared to only a
$\sim$1.1 mag drop in $B$ and $\sim$0.5 mag drop in $V$ over that
time period. Over the first 30 days after $B_{\rm max}$, the declines
in $U$,$B$, and $V$ are $\sim$3.2, $\sim$2.6, and $\sim$1.4 mag,
respectively. At late time, $t \gtrsim 35$ days after $B$ maximum
light, the U-band light curves follow the typical exponential decline,
decaying at $0.020 \pm 0.001 \; {\rm mag \; day^{-1}}$.

In Figure \ref{ch3-fig-tu} we plot the distribution of the epoch of
$U$-band maximum light relative to $B$-band maximum light, using the
stretch templates results for the 22 SN Ia listed in Table
\ref{ch3-tab-stretch}, along with the 6 additional SN Ia listed
above. As can also be seen in Figure \ref{ch3-fig-ucomp}, the SN Ia
clearly peak earlier in the $U$-band than in $B$, with an average time
offset of $-$2.3 days and a dispersion of only 0.4 days. The earlier
peak in $U$ also implies the decline rate in $U$ relative to maximum
light in $U$ is not so different from the decline rate in $B$ relative
to maximum light in $B$.  A typical SN Ia that drops $\sim$1.1 mag in
$B$ over the first 15 days after maximum light (as above), declines by
$\sim$1.2 mag in $U$ over the first 15 days after $U$ maximum. We note
that our precise photometry confirms the result of Leibundgut et
al.~(1991), who found that maximum in light in $U$ occurs $\sim$2.8
days before maximum light in $B$, based on a compilation of
heterogeneous photoelectric
\ubv photometry.

The decline rate in $U$ is well correlated with the decline rate in
$B$, as shown in Figure \ref{ch3-fig-subv}, which plots the timescale
stretch factors for the 28 SN Ia described above.  However, as the
figure also illustrates, there is a significant scatter. The
relationship between the stretch factor in $V$ and the stretch factor
in $B$ is considerably tighter. Nonetheless, these correlations imply
that $U$ light curves can provide leverage in determining the
intrinsic luminosities of SN Ia. The best-fit linear relations between
$s_U$, $s_B$, and $s_V$ are given in the figure. Given the scatter,
the relations are consistent with a ``universal'' stretch, $s = s_U =
s_B = s_V$, though the data for a number of objects individually favor
slightly different stretch factors in each band. The slope of the
luminosity/stretch relation is $\sim$1.7 (Nugent, Kim, \& Perlmutter
2002), meaning that the dispersion in the $s_U$-$s_B$ relation
($\sigma \simeq 0.08$) translates into an uncertainty of $\sigma
\simeq 0.14$ mag in luminosity, comparable to the typical dispersion
in measuring SN Ia distances (e.g., in the stretch/luminosity relation
itself). Similarly, the dispersion in the $s_V$-$s_B$ relation
corresponds to $\sigma \simeq 0.09$ mag.

We can also examine the correlation between the timescale stretch
factors and \dmf\ for these 28 SN Ia (cf. Table \ref{ch3-tab-dm15});
the results are shown in Figure \ref{ch3-fig-sdm15}. The correlation
between \dmf\ and $s$ is clear, with $s_V$ and $s_B$ producing a
tighter relationship. It also appears that much of the dispersion
comes at the low \dmf\ (large $s$) end of the diagram, implying that
there may be larger intrinsic variation in the light curves of the
most luminous SN Ia. The dispersions in \dmf\ are 0.17, 0.12, 0.10
for the relations with $s_U$, $s_B$, and $s_V$, respectively. Using
the luminosity-\dmf\ relationship presented by Phillips et al.~(1999),
the luminosity scatter corresponding to these dispersions are 0.14,
0.10 and 0.08 mag, similar to the results above directly comparing
stretch to luminosity. We note that the relations between \dmf\ and
$s$ presented in Figure \ref{ch3-fig-sdm15} match well the results of
Garnavich et al.~(2004; see their Figure 6).

In addition to the $U$-band light curve shapes, we can explore the
\ub\ color with this data set. We display 27 SN Ia\footnote{We show 27
SN Ia rather than 28, because we exclude the highly-reddened SN
1999cl, for which there is strong evidence from near-infrared
photometry that the extinction law varies significantly from the
canonical $R_V = 3.1$ law (Krisciunas et al.~2000; Jha, Riess, \&
Kirshner 2005, in preparation).} in the color-color diagram shown in
the top panel of Figure \ref{ch3-fig-ubbv}. We note that the
stretch-template fits to the peak magnitudes include the effects of
K-correction, which can be significant, particularly in the $U$-band
($K_{UU} \simeq 0.12$ mag for $z = 0.03$ at maximum light; Jha, Riess,
\& Kirshner 2005, in preparation). We have also corrected the colors
for (the generally small) Galactic reddening (cf. Table
\ref{ch3-tab-sndata}), assuming the $R_V = 3.1$ extinction law of
Cardelli, Clayton, \& Mathis~(1989). For 23 of the 27 SN Ia, we were
also able to correct for the host-galaxy extinction, via measurement
of the tail \bv\ evolution and the method of Lira~(1995) and Phillips
et al.~(1999), as described in detail in Jha, Riess, \& Kirshner
(2005, in preparation). The colors corrected for host-galaxy reddening
are shown in the bottom panel of Figure \ref{ch3-fig-ubbv}.  These
results sharpen those of Schaefer~(1995) and Branch, Nugent, \&
Fisher~(1997), who display relations between the \ub\ and \bv\ maximum
light colors of SN Ia based on a handful of objects with heterogeneous
photometry from diverse sources.

The lower panel figure shows a tight relation between the intrinsic
\bv\ and \ub\ color at maximum light. In this plot, normal SN Ia have
\bv\ $\simeq -0.1$ (e.g., Phillips et al.~1999)\footnote{Phillips et
al.~(1999) find the ``pseudo''-color $B_{\rm Bmax} - V_{\rm Vmax}
\simeq -0.07$ for normal SN Ia. Because $V_{\rm Vmax} \simeq V_{\rm
Bmax} - 0.02$, their result implies $(\bv)_{\rm Bmax} \simeq -0.09$
for normal SN Ia.}, and there is a strong clustering of objects at
this value. Note, however, the wide span of \ub\ colors (from about
$-$0.2 to $-$0.8) for these normal SN Ia.  This is not an artifact of
the reddening correction, nor can it be explained by variation in the
extinction law in these external galaxies. If there were strong
variations in the extinction law, because of the patchiness of
interstellar dust, we would expect the top panel of Figure
\ref{ch3-fig-ubbv} to show a swarm of points at the lower left
(corresponding to an unreddened locus) with the remainder of the
points fanning out toward the upper right (corresponding to different
amounts of extinction and reddening), which is clearly not what we
see. We conclude that the intrinsic variation in \ub\ color at maximum
light is significantly greater than the variation seen in \bv.

Do these color variations correlate with light-curve shape or
luminosity? There is strong evidence that objects with intrinsically
red \bv\ colors at maximum are the fast-declining, low-luminosity SN
Ia (see, e.g., Garnavich et al.~2004 and references therein). The
bottom panel of Figure \ref{ch3-fig-ubbv} shows that the red objects
in \bv\ are also red in \ub. A direct check on the relation between
color and light-curve shape is shown in Figure \ref{ch3-fig-ubbvsv},
which plots the intrinsic \ub\ and \bv\ maximum light colors against
the measured timescale stretch factor (in $V$). The relationship
between \bv\ and $s_V$ shown in the lower panel is in good accord with
the results presented by Phillips et al.~(1999) and Garnavich et
al.~(2004). The \ub\ results in the top panel show that the \ub\ color
is well-correlated with stretch (and therefore, luminosity) over the
whole range of luminosity in the sample. However, the scatter is also
greater in \ub, implying that there is a significant intrinsic
dispersion $U$-band peak brightness even after accounting for
variations in light-curve shape. A simple linear fit to the data in
the top panel of Figure \ref{ch3-fig-ubbvsv} implies that this
intrinsic dispersion is $\sigma_U \simeq 0.12$ mag. It would be
interesting to check whether this increased dispersion is related to
other factors, such as progenitor metallicity, as some theoretical
studies have indicated that the these factors may have more
significant effects in $U$ than in \bvri (e.g., H\"oflich, Wheeler, \&
Thielemann~1998).

It is clear that the analysis of these $U$-band light curves and their
relation to light curves in \bvri and ultimately, precise distances,
is intimately tied to the luminosity and extinction of each SN. To
further explore these relations, a profitable strategy would be to
incorporate the $U$-band light curves into the general framework of
the Multicolor Light Curve Shape analysis presented by Riess, Press,
\& Kirshner~(1996). We present the methods and results of this
incorporation in Jha, Riess, \& Kirshner (2005, in preparation).

\acknowledgements

We thank the avid supernova searchers who scan the sky and allow us to
be successful in finding supernovae in our inboxes. We are also
grateful for the efforts of Dan Green at the IAU CBAT for enabling our
follow-up observations. We thank Paul Green, Scott Kenyon, Jeff
McClintock, and Kenny Wood for assistance with the observations, Brian
Schmidt for robust software, and Adam Riess, Nick Suntzeff, Dimitar
Sasselov, and Alyssa Goodman for enlightening discussions and
comments. We appreciate the helpful suggestions of the referee, Mario
Hamuy, in improving the paper. This work was supported in part by an
NSF Graduate Research Fellowship and the Miller Institute for Basic
Research in Science.


\begin{figure}
\begin{center}
\includegraphics[height=7.5in]{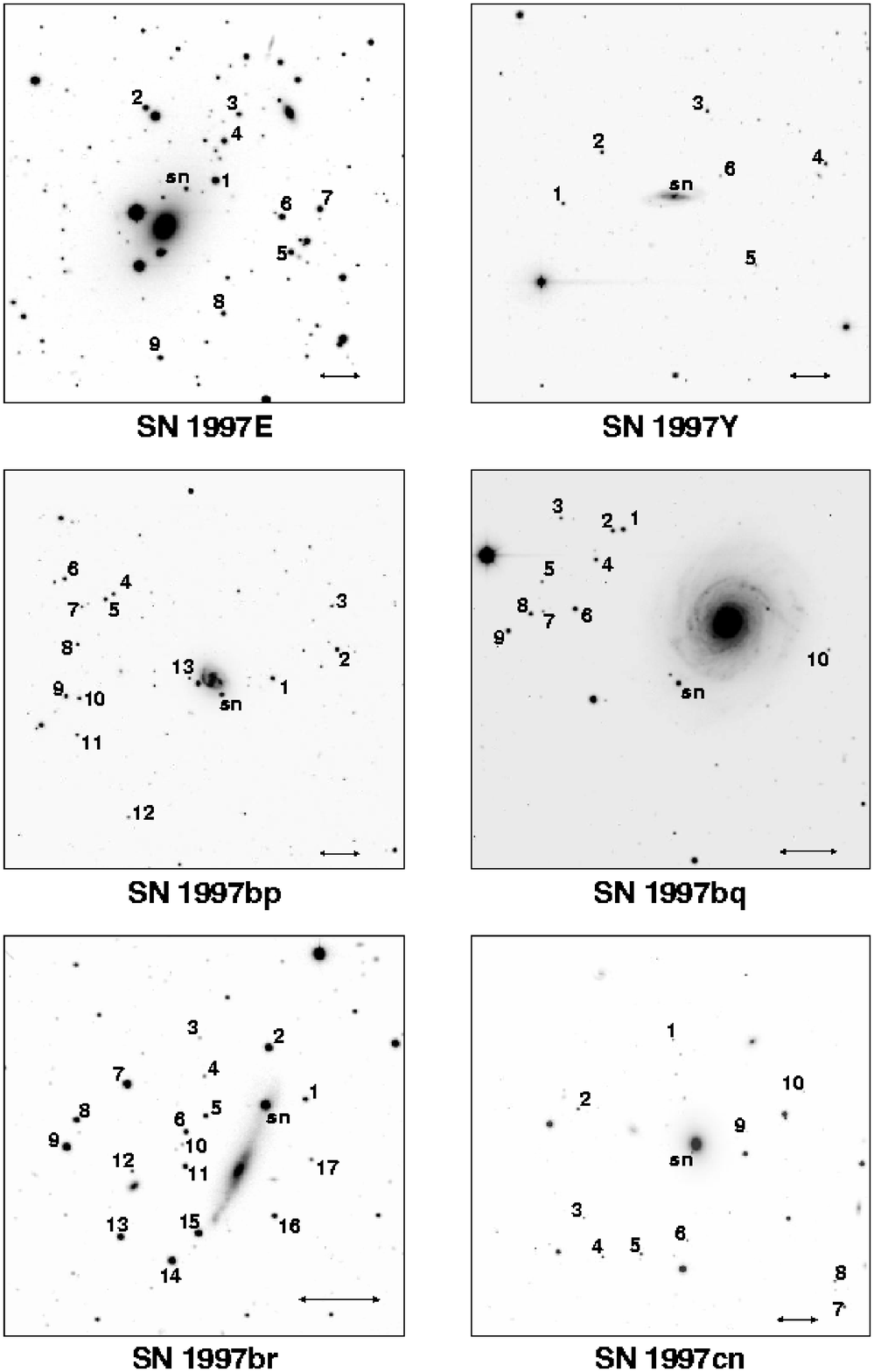}
\end{center}
\caption[Finder charts]{\singlespace Finder charts for the 44 SN Ia
presented here and associated comparison stars. The images are a
combination of all the $R$-band SN images. North is up and east is to
the left. The horizontal double-arrow in the lower right delineates
1\arcmin. \label{ch3-fig-finder}}
\end{figure}

\addtocounter{figure}{-1}
\begin{figure}
\begin{center}
\includegraphics[height=7.5in]{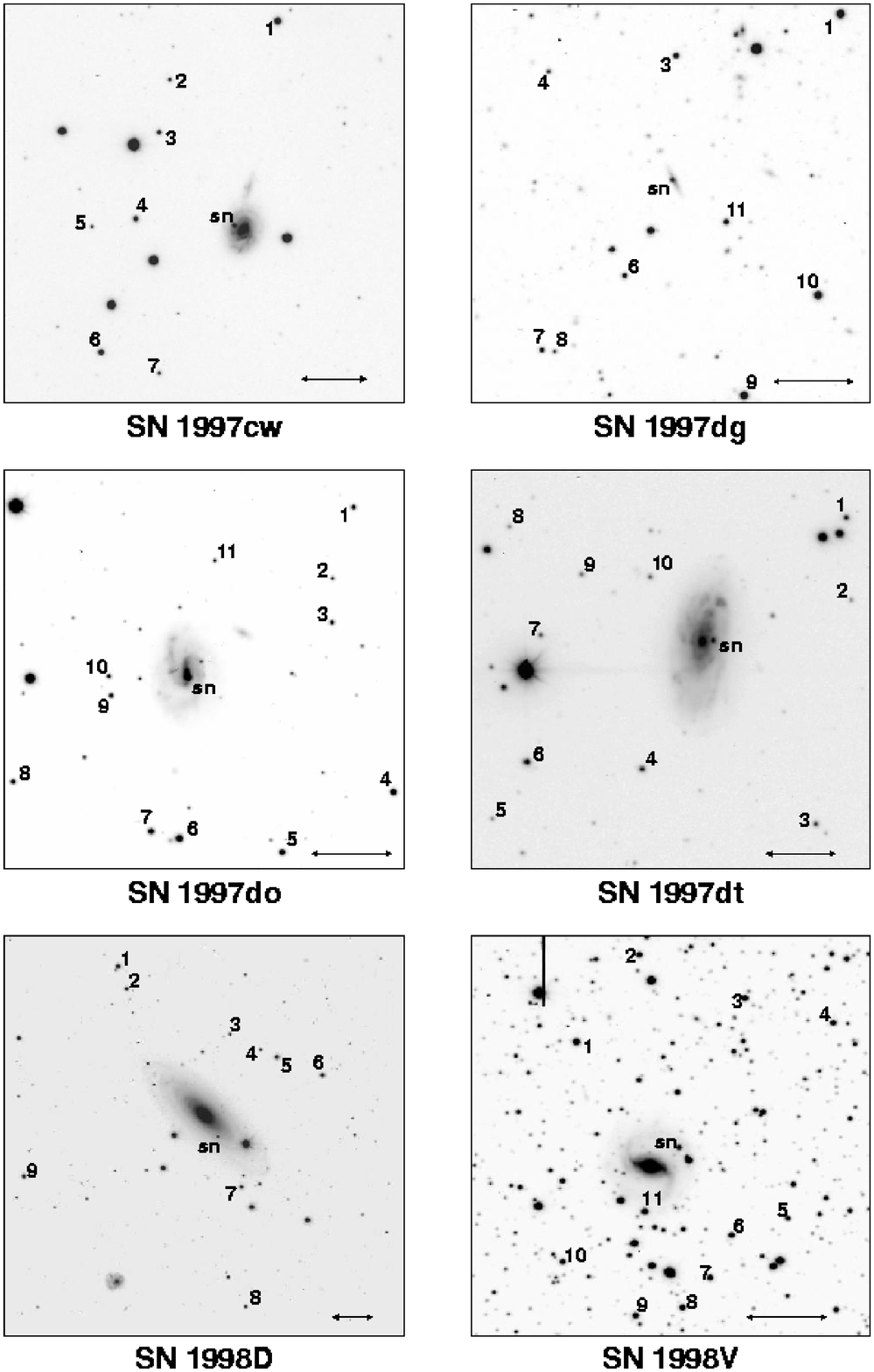}
\end{center}
\caption[Finder charts continued]{Continued}
\end{figure}

\addtocounter{figure}{-1}
\begin{figure}
\begin{center}
\includegraphics[height=7.5in]{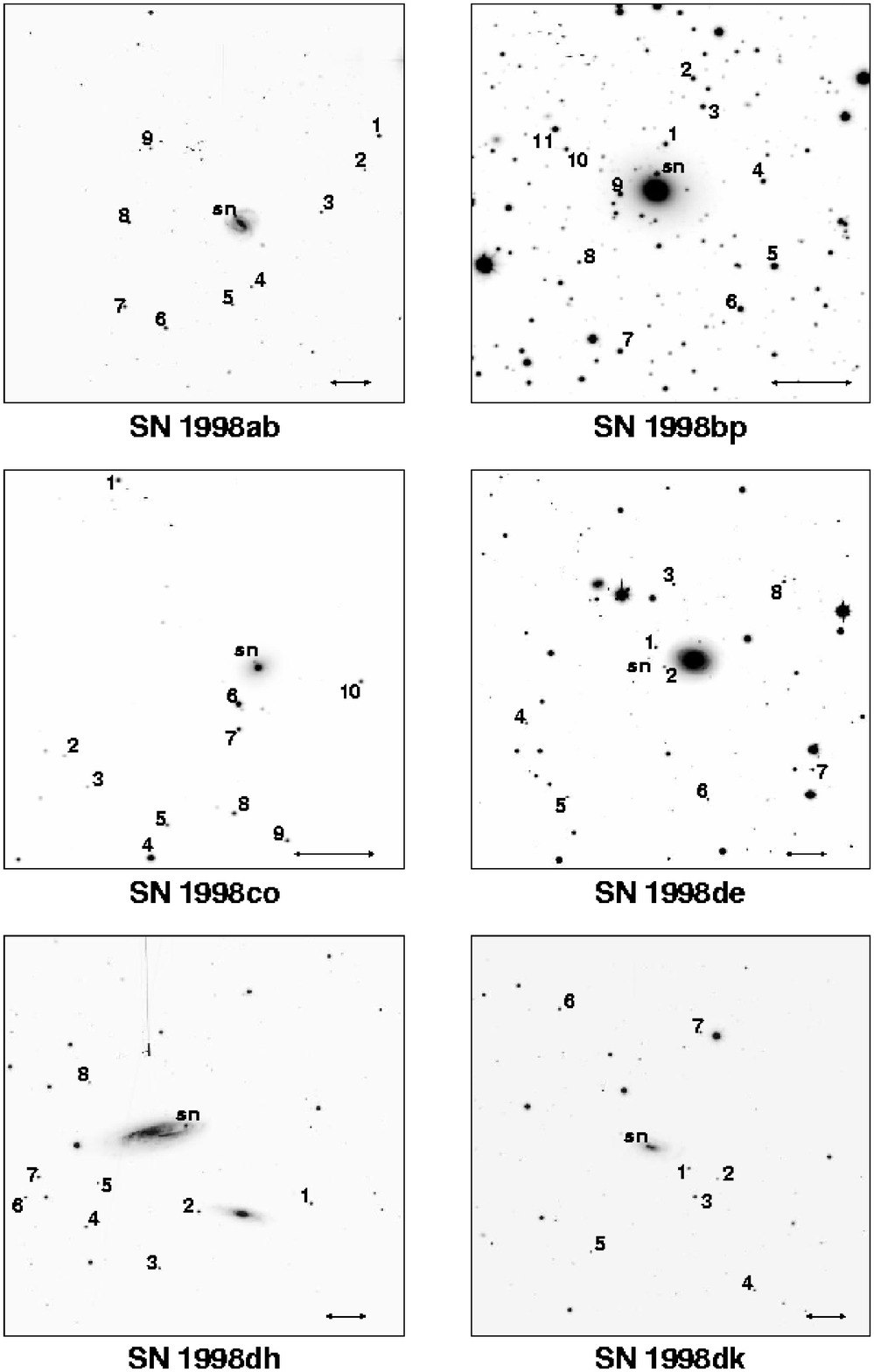}
\end{center}
\caption[Finder charts continued]{Continued}
\end{figure}

\addtocounter{figure}{-1}
\begin{figure}
\begin{center}
\includegraphics[height=7.5in]{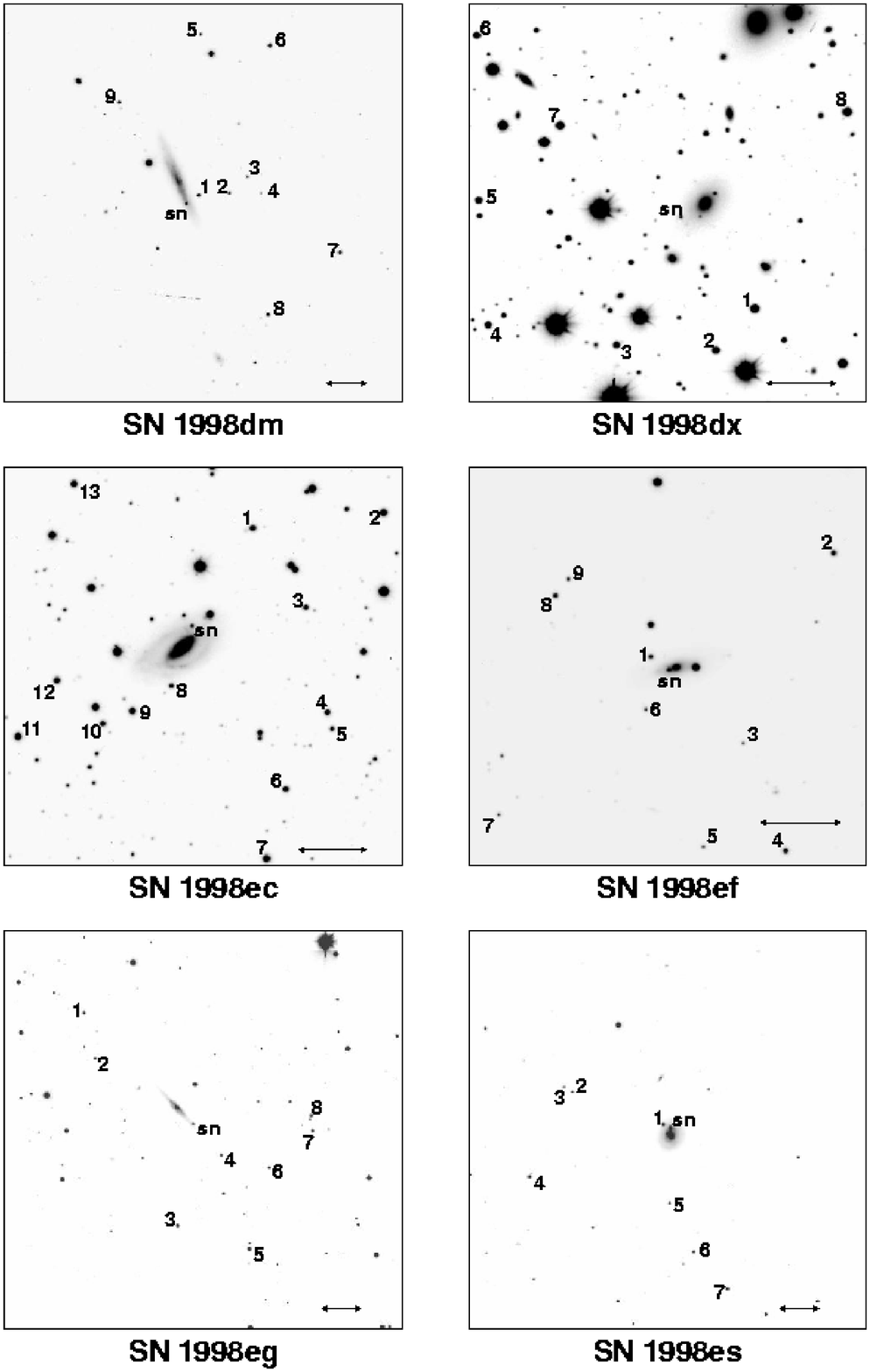}
\end{center}
\caption[Finder charts continued]{Continued}
\end{figure}

\addtocounter{figure}{-1}
\begin{figure}
\begin{center}
\includegraphics[height=7.5in]{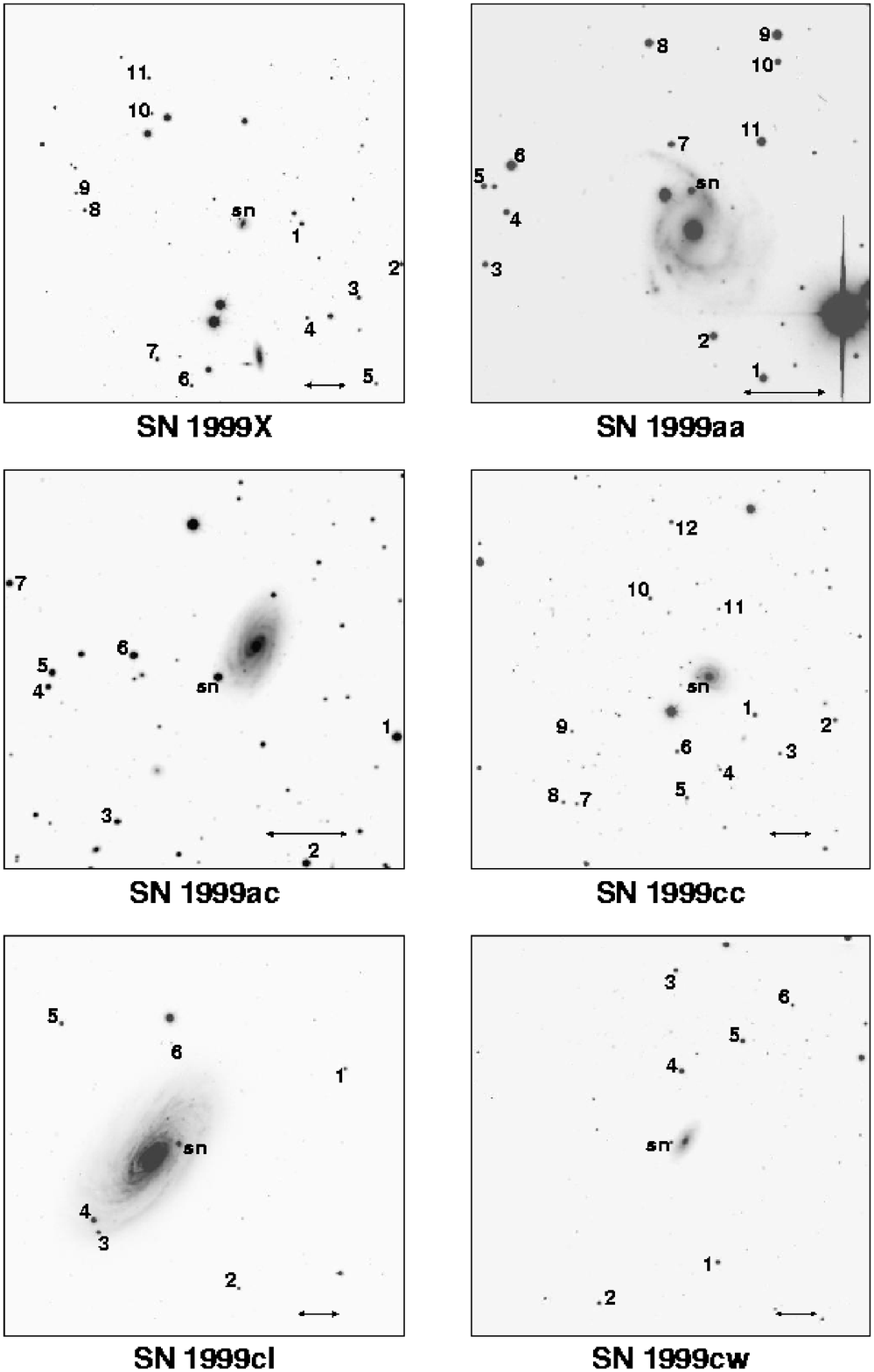}
\end{center}
\caption[Finder charts continued]{Continued}
\end{figure}

\addtocounter{figure}{-1}
\begin{figure}
\begin{center}
\includegraphics[height=7.5in]{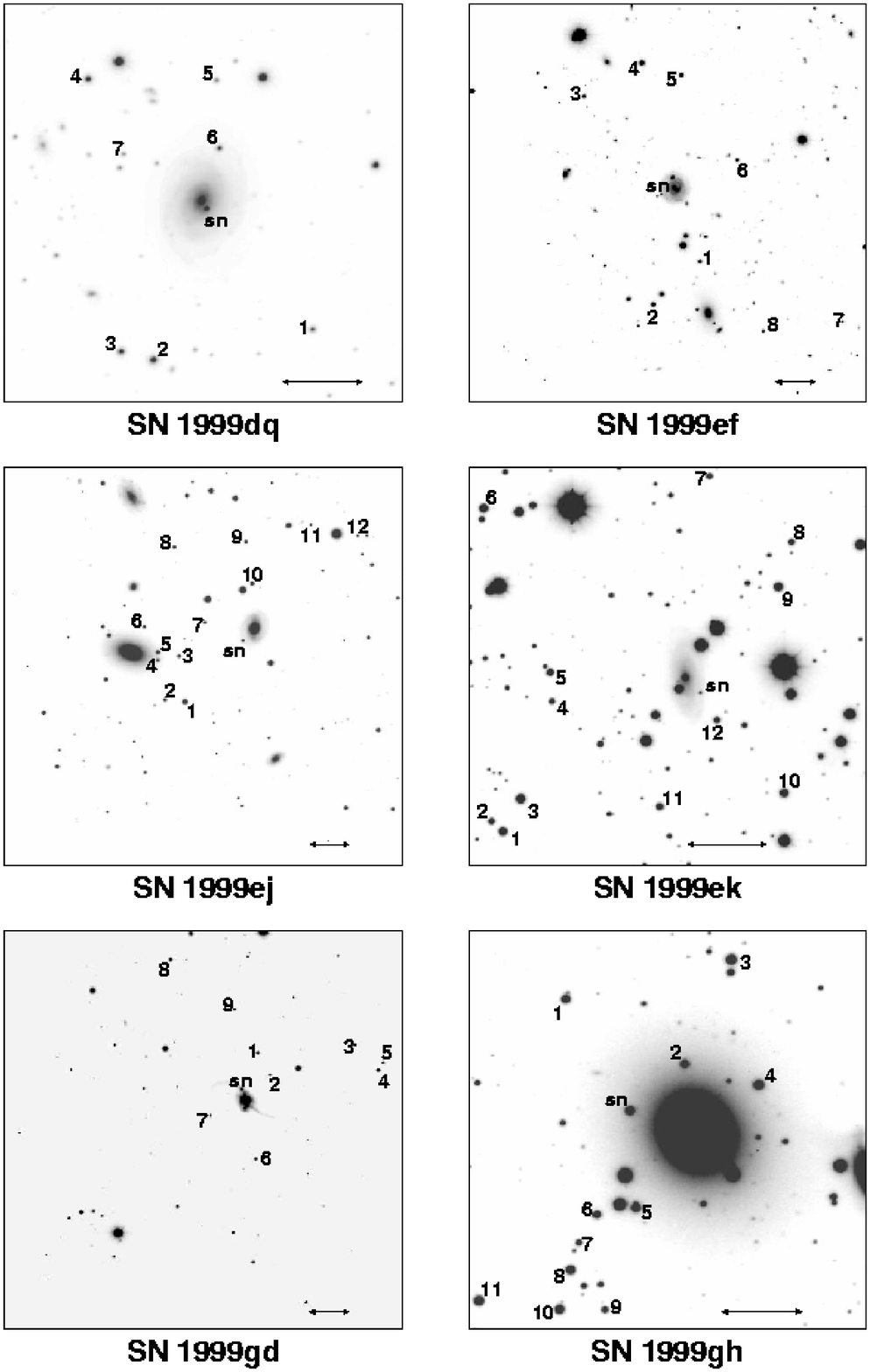}
\end{center}
\caption[Finder charts continued]{Continued}
\end{figure}

\addtocounter{figure}{-1}
\begin{figure}
\begin{center}
\includegraphics[height=7.5in]{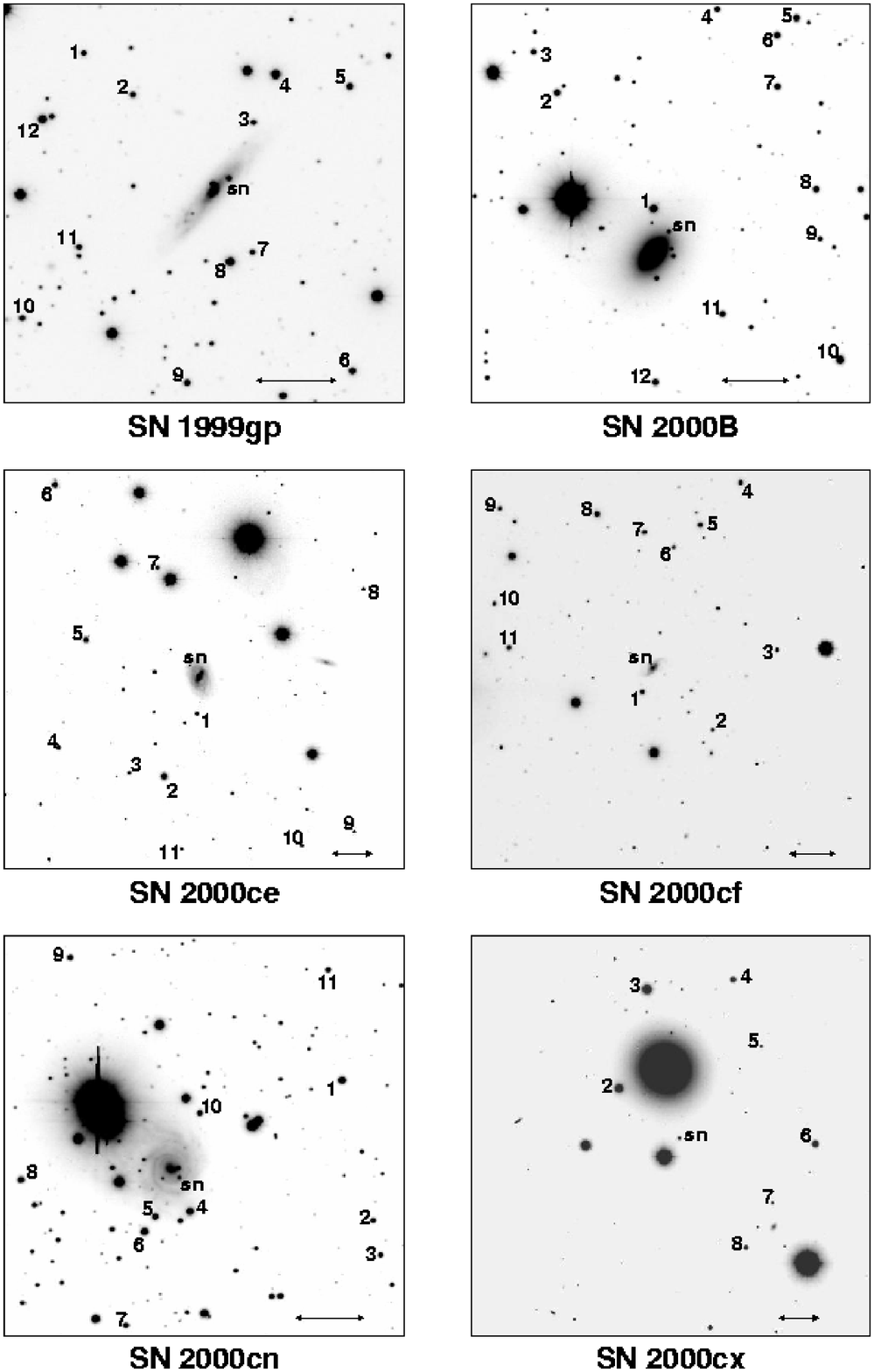}
\end{center}
\caption[Finder charts continued]{Continued}
\end{figure}

\addtocounter{figure}{-1}
\begin{figure}
\begin{center}
\includegraphics[height=2.6in]{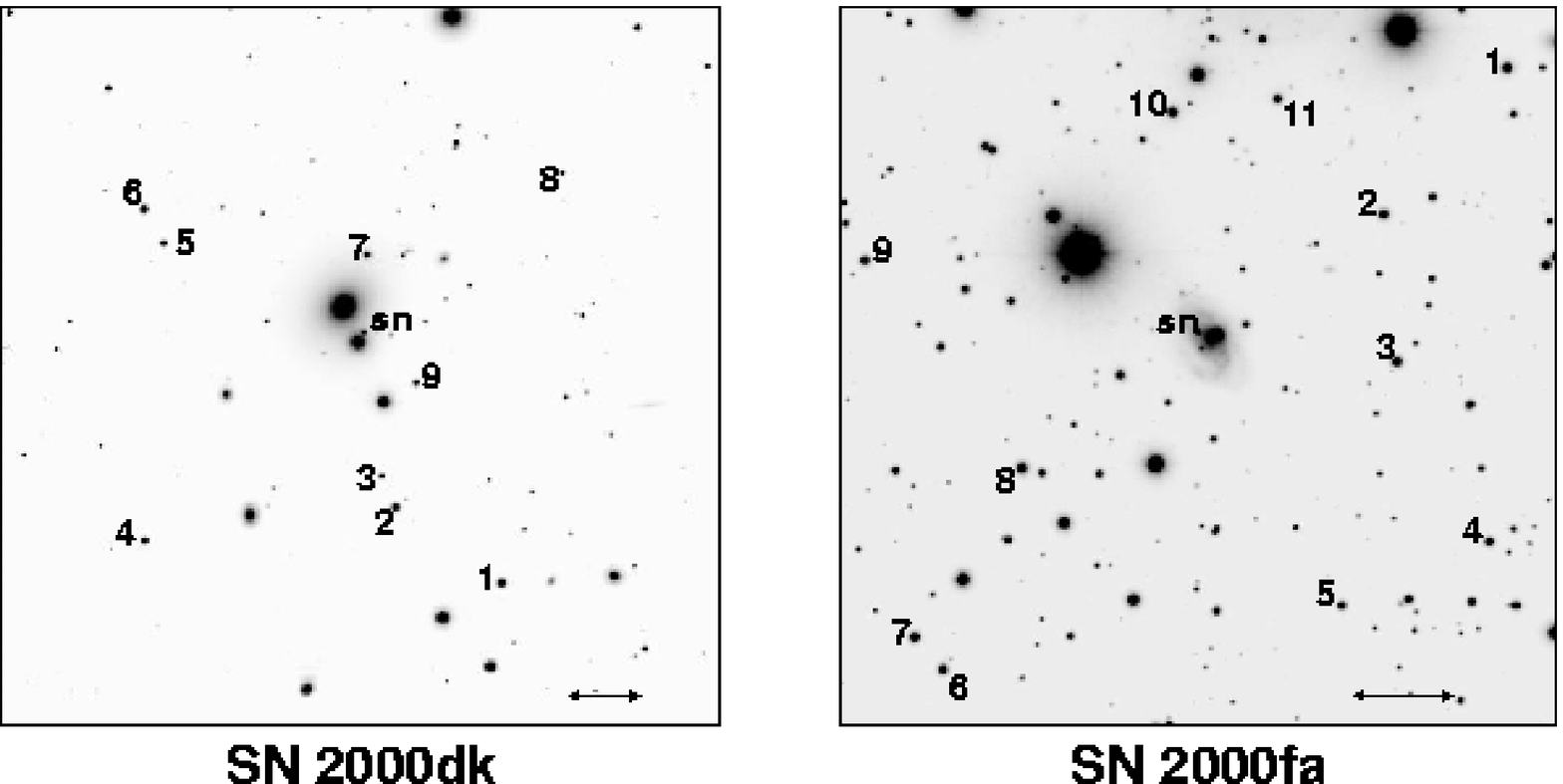}
\end{center}
\caption[Finder charts continued]{Continued}
\end{figure}

\clearpage

\begin{figure}
\begin{center}
\includegraphics[height=7.5in]{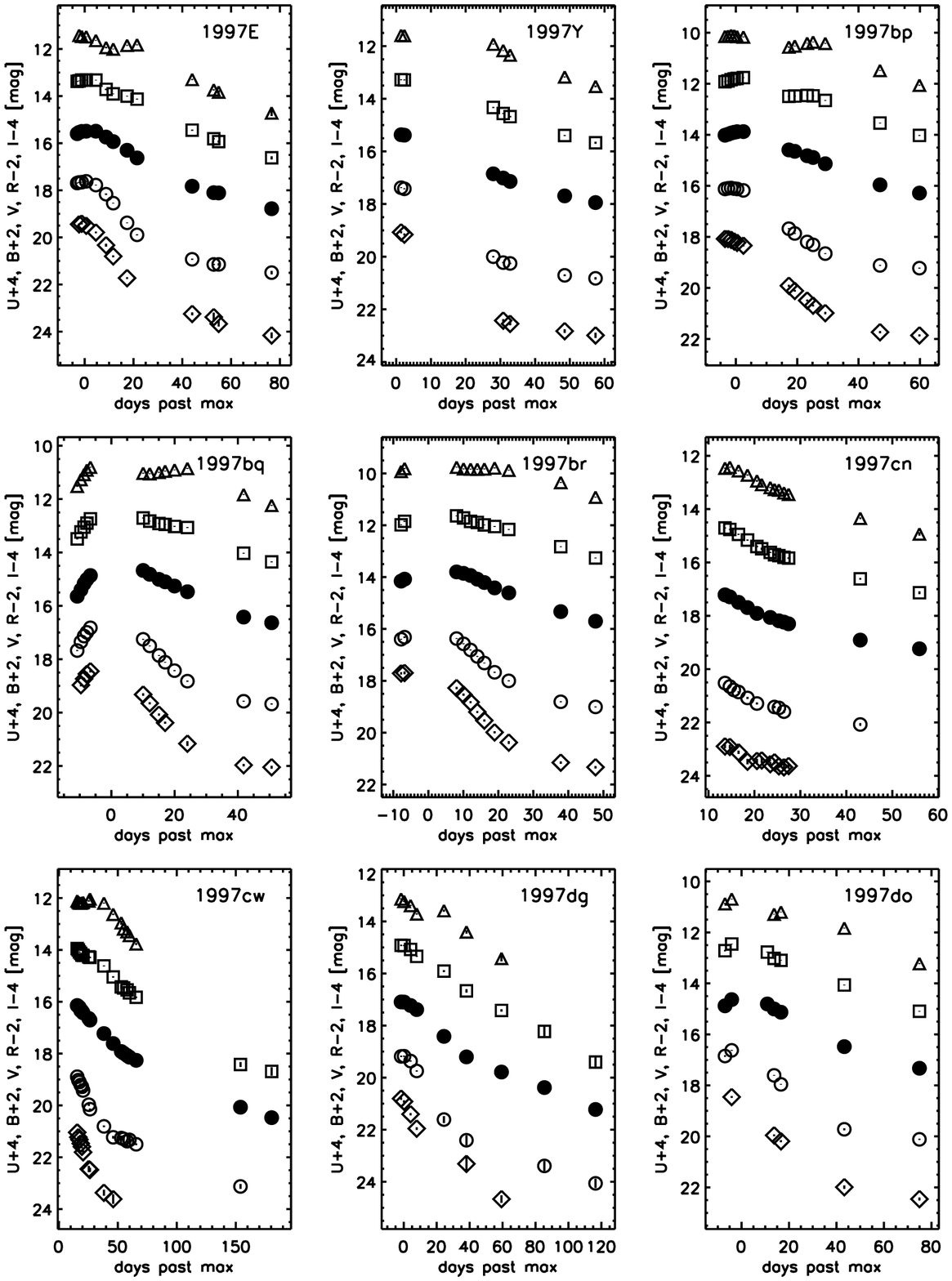}
\end{center}
\caption[Light curves]{\singlespace \ubvri photometry of 44 SN Ia. The
$U$ ({\emph{diamonds}}), $B$ ({\emph{open circles}}), $V$
({\emph{filled circles}}), $R$ ({\emph{squares}}), and $I$
({\emph{triangles}}) light curves are shown shown relative to $B$
maximum and have been corrected for time-dilation to the SN rest
frame. \label{ch3-fig-ltcurves}}
\end{figure}

\addtocounter{figure}{-1}
\begin{figure}
\begin{center}
\includegraphics[height=7.5in]{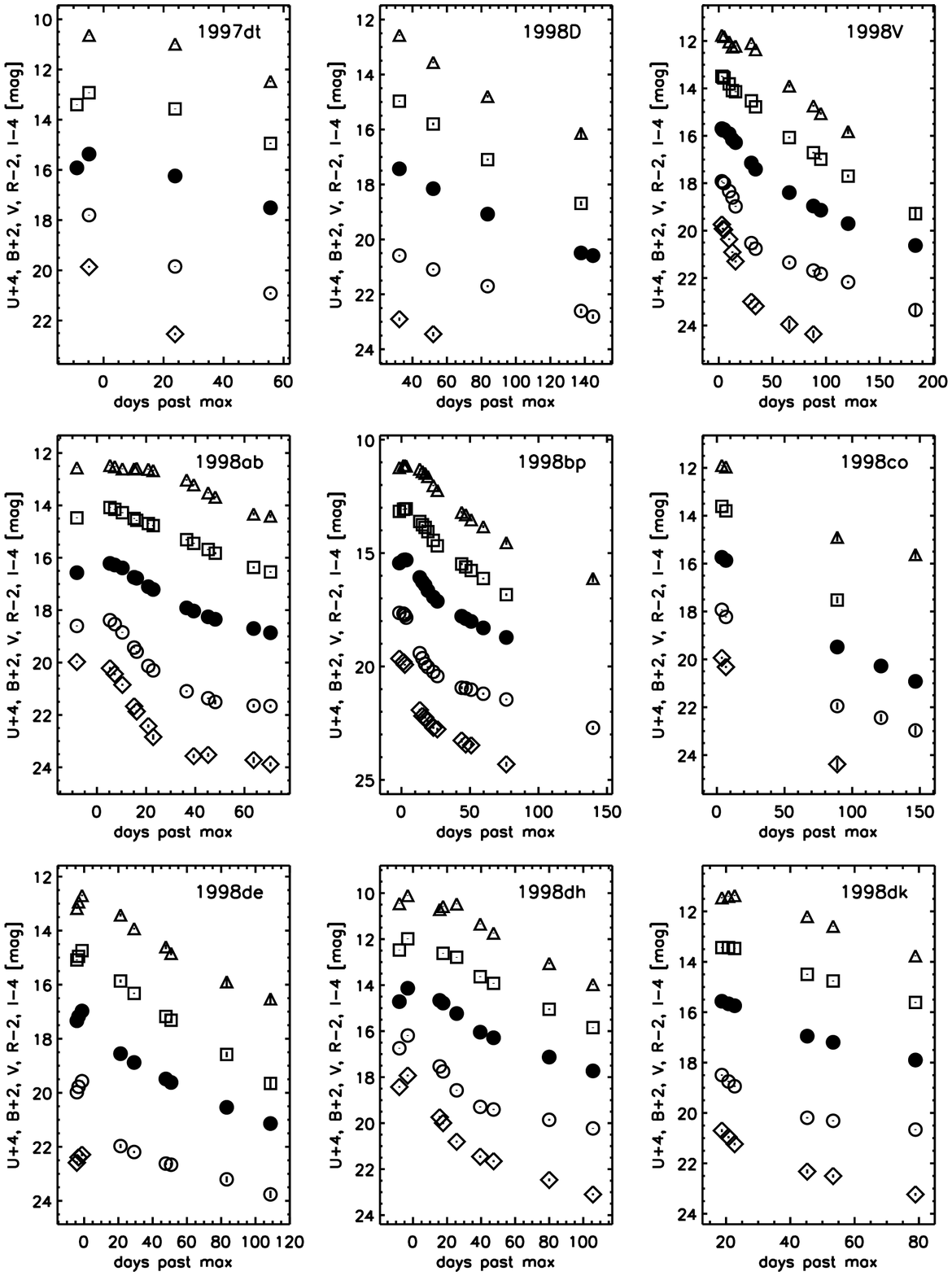}
\end{center}
\caption[Light curves continued]{Continued}
\end{figure}

\addtocounter{figure}{-1}
\begin{figure}
\begin{center}
\includegraphics[height=7.5in]{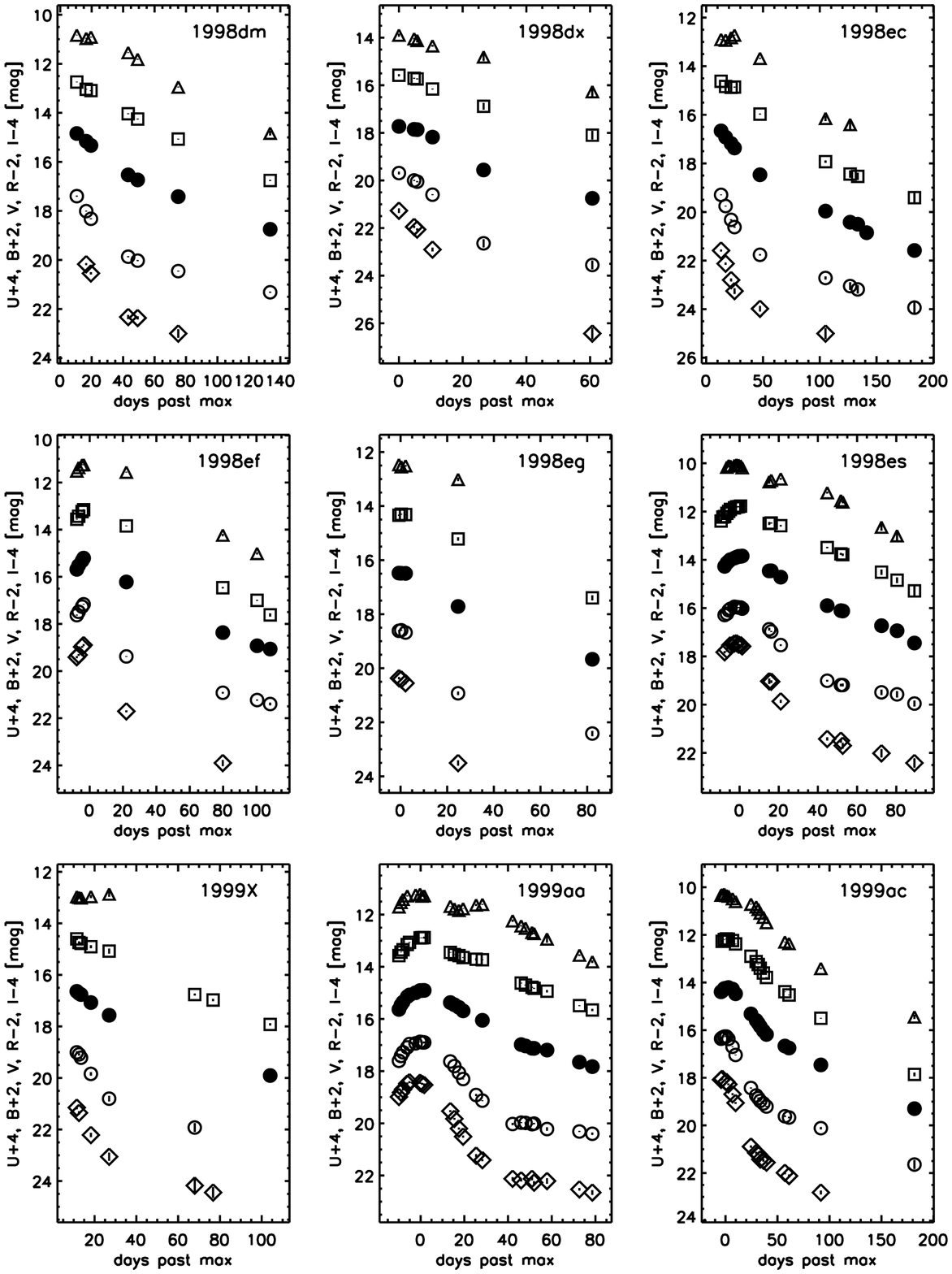}
\end{center}
\caption[Light curves continued]{Continued}
\end{figure}

\addtocounter{figure}{-1}
\begin{figure}
\begin{center}
\includegraphics[height=7.5in]{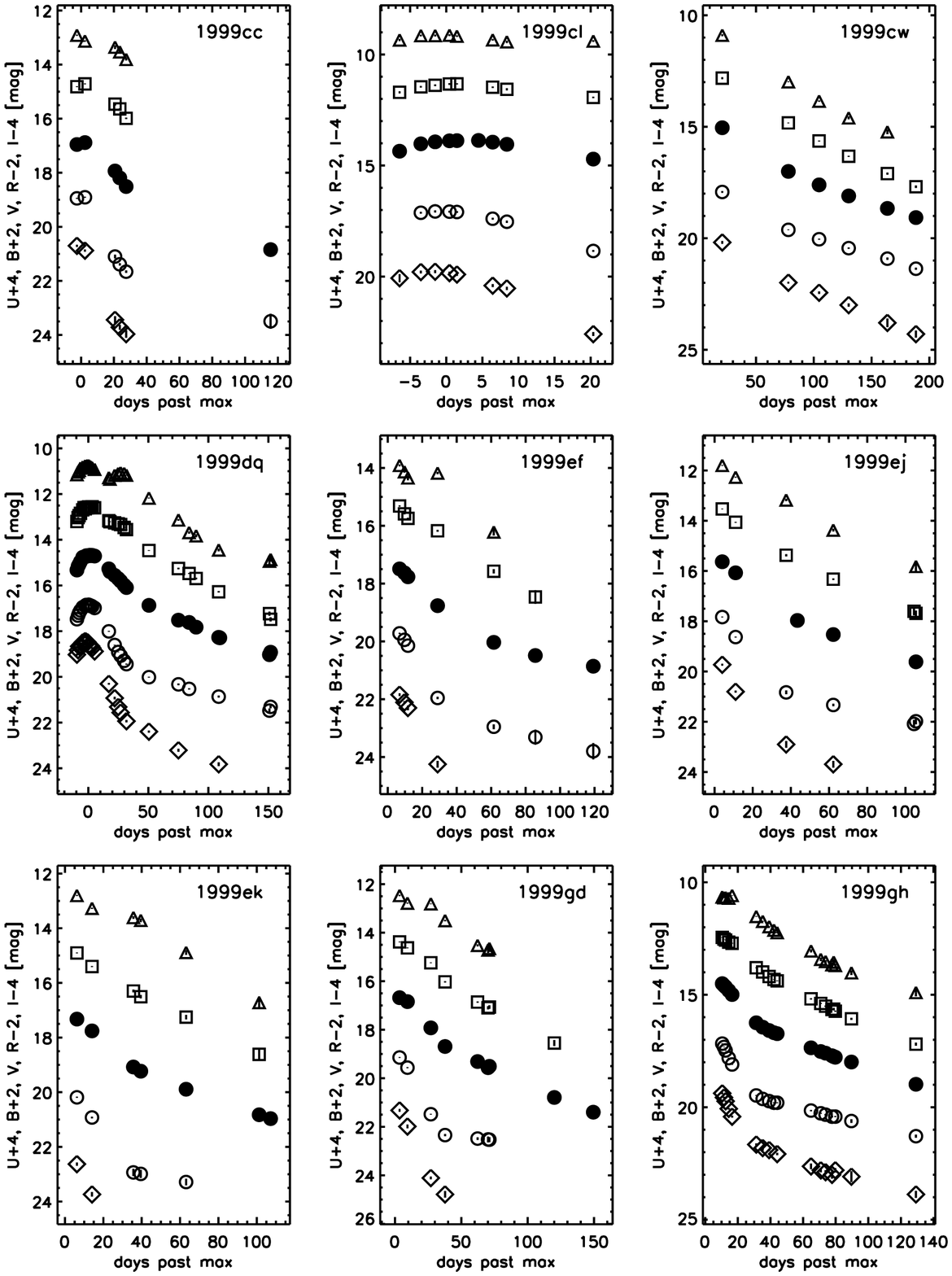}
\end{center}
\caption[Light curves continued]{Continued}
\end{figure}

\addtocounter{figure}{-1}
\begin{figure}
\begin{center}
\includegraphics[height=7.5in]{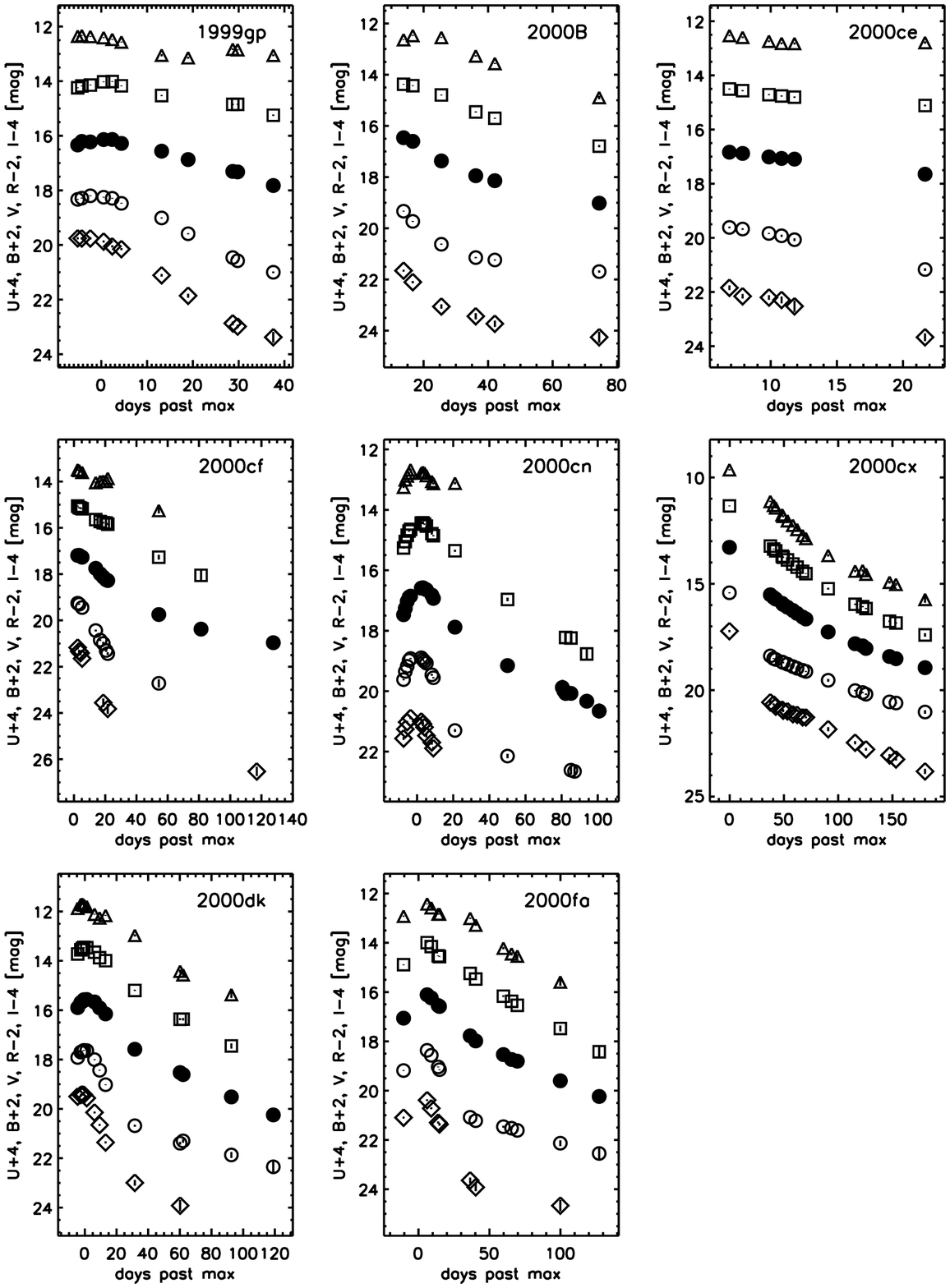}
\end{center}
\caption[Light curves continued]{Continued}
\end{figure}

\clearpage

\begin{figure}
\begin{center}
\includegraphics[width=6.0in]{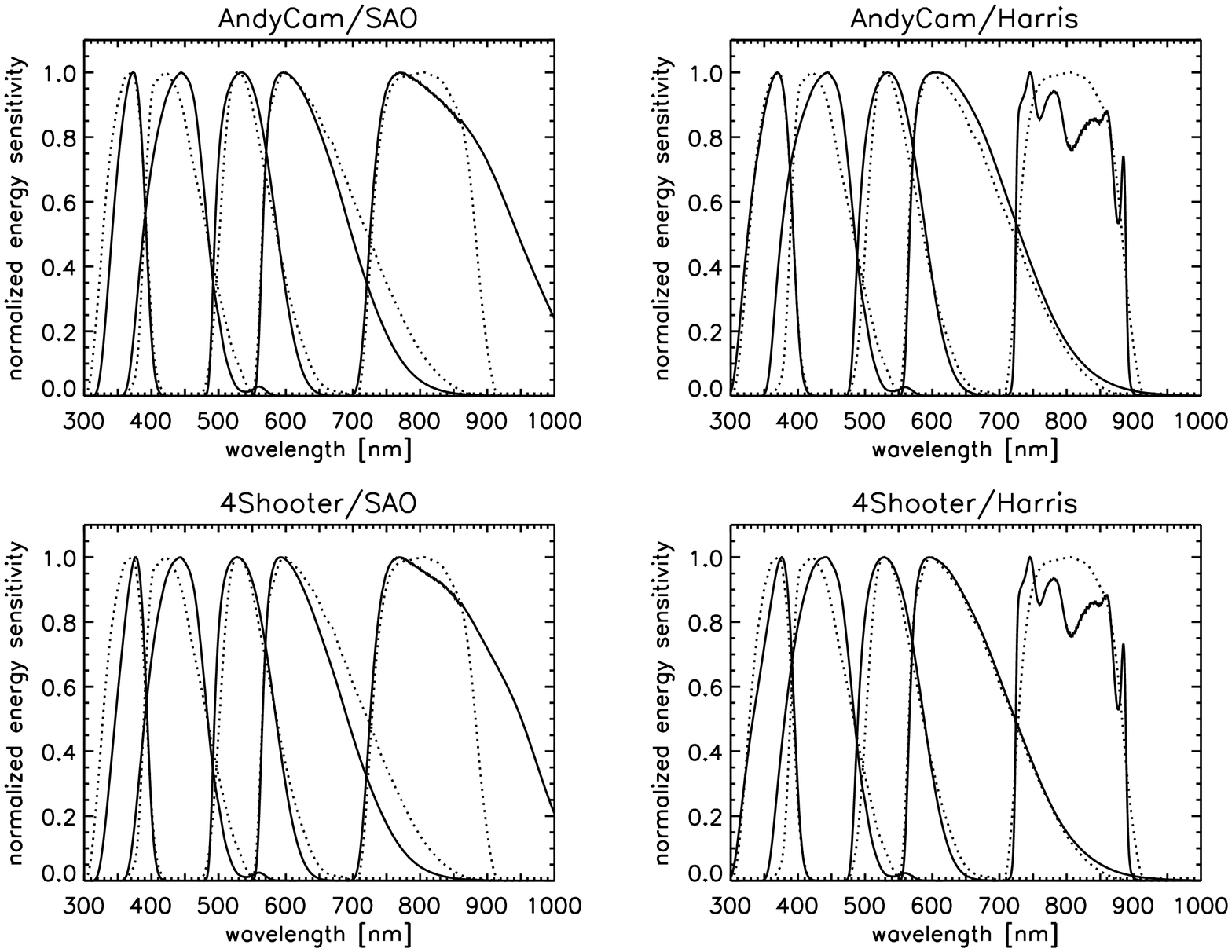}
\end{center}
\caption[Natural system passbands]{\singlespace Synthesized natural
system \ubvri passbands ({\emph{solid curves}}) with the standard $U\!X$
and \bvri passbands ({\emph{dotted curves}}) of Bessell~(1990) shown
for each detector/filterset combination. \label{ch3-fig-allpb}}
\end{figure}

\clearpage

\begin{figure}
\begin{center}
\includegraphics[width=6.0in]{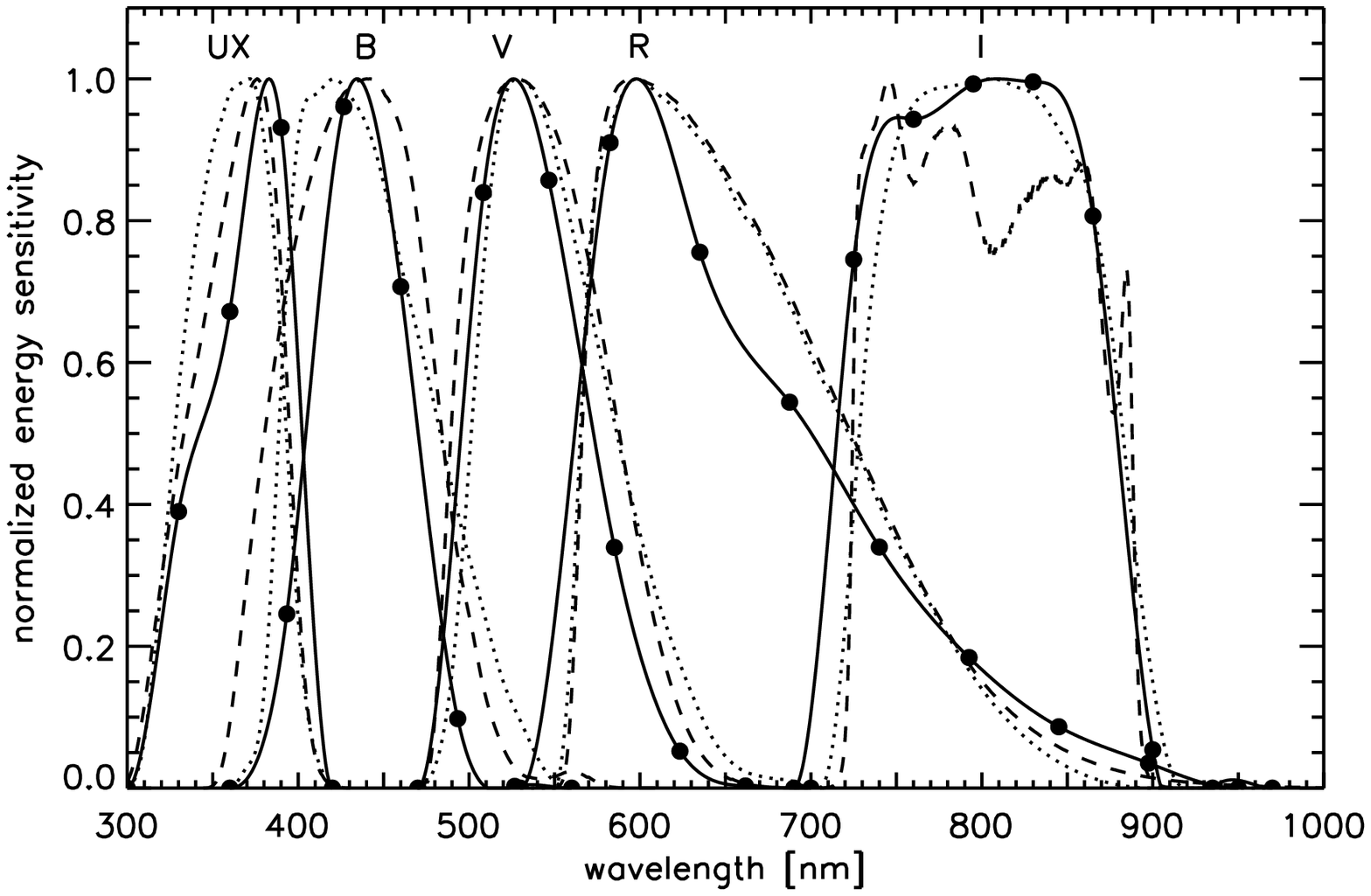}
\end{center}
\caption[Modeled natural system passbands]{\singlespace Model
4Shooter/Harris \ubvri passbands derived from observations of
spectrophotometric standard stars ({\emph{solid curves}}), calculated
passbands from optics + filter transmission + detector response
({\emph{dashed curves}}), and the standard $U\!X$ and \bvri passbands of
Bessell~(1990; {\emph{dotted curves}}). The solid points show the
location of the model spline points; see text for
details. \label{ch3-fig-specphot}}
\end{figure}

\clearpage

\begin{figure}
\begin{center}
\includegraphics[width=6.0in]{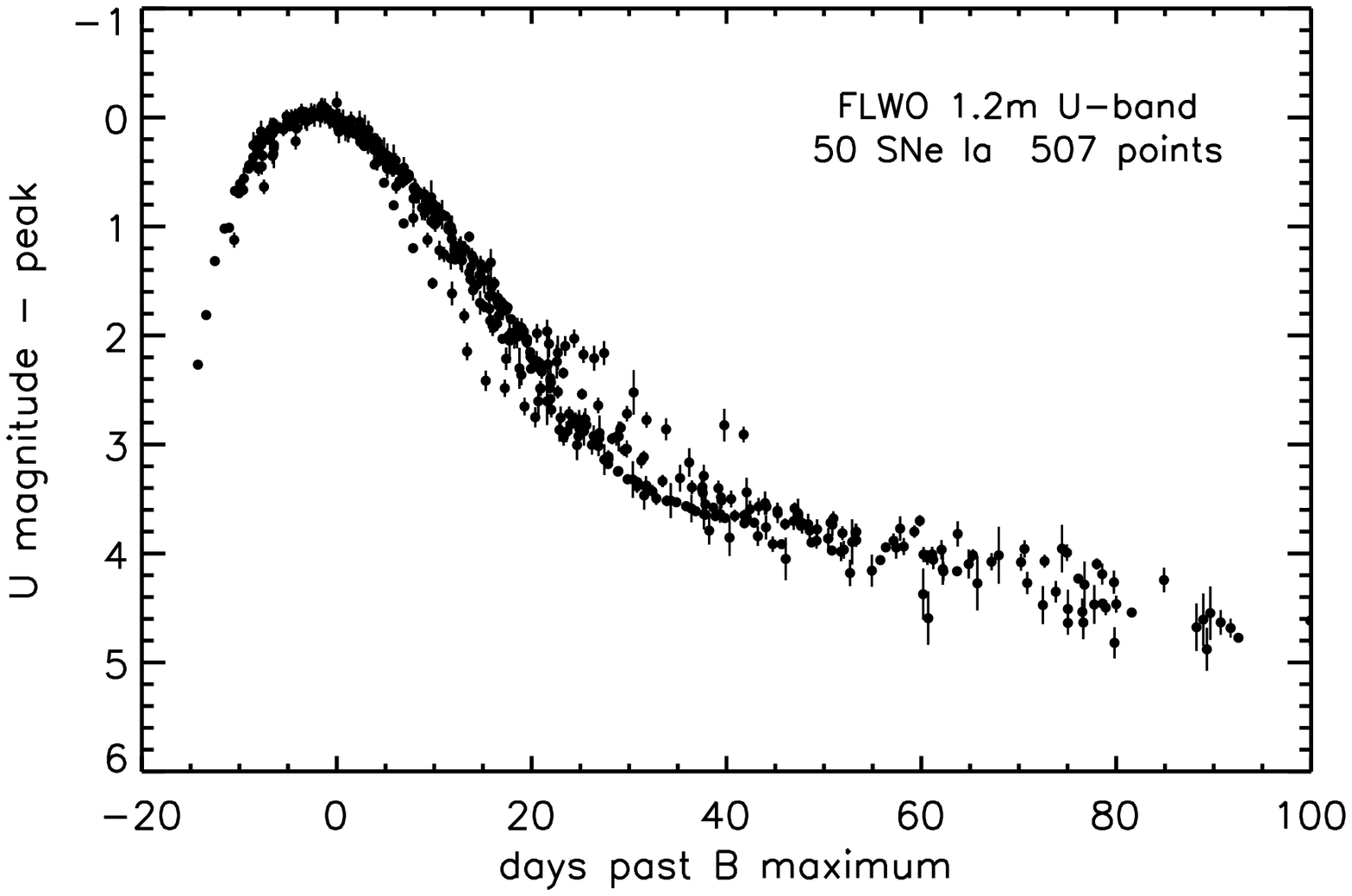}
\end{center}
\caption[U-band composite light curve]{\singlespace Composite $U$-band
light curve of 50 SN Ia observed with the FLWO 1.2m telescope. The
data were K-corrected and time-dilated to the SN rest frame. There are
507 individual points in the time interval displayed, from $-$20 to
$+$100 days after maximum light in $B$. \label{ch3-fig-ucomp}}
\end{figure}

\clearpage

\begin{figure}
\begin{center}
\includegraphics[width=6.0in]{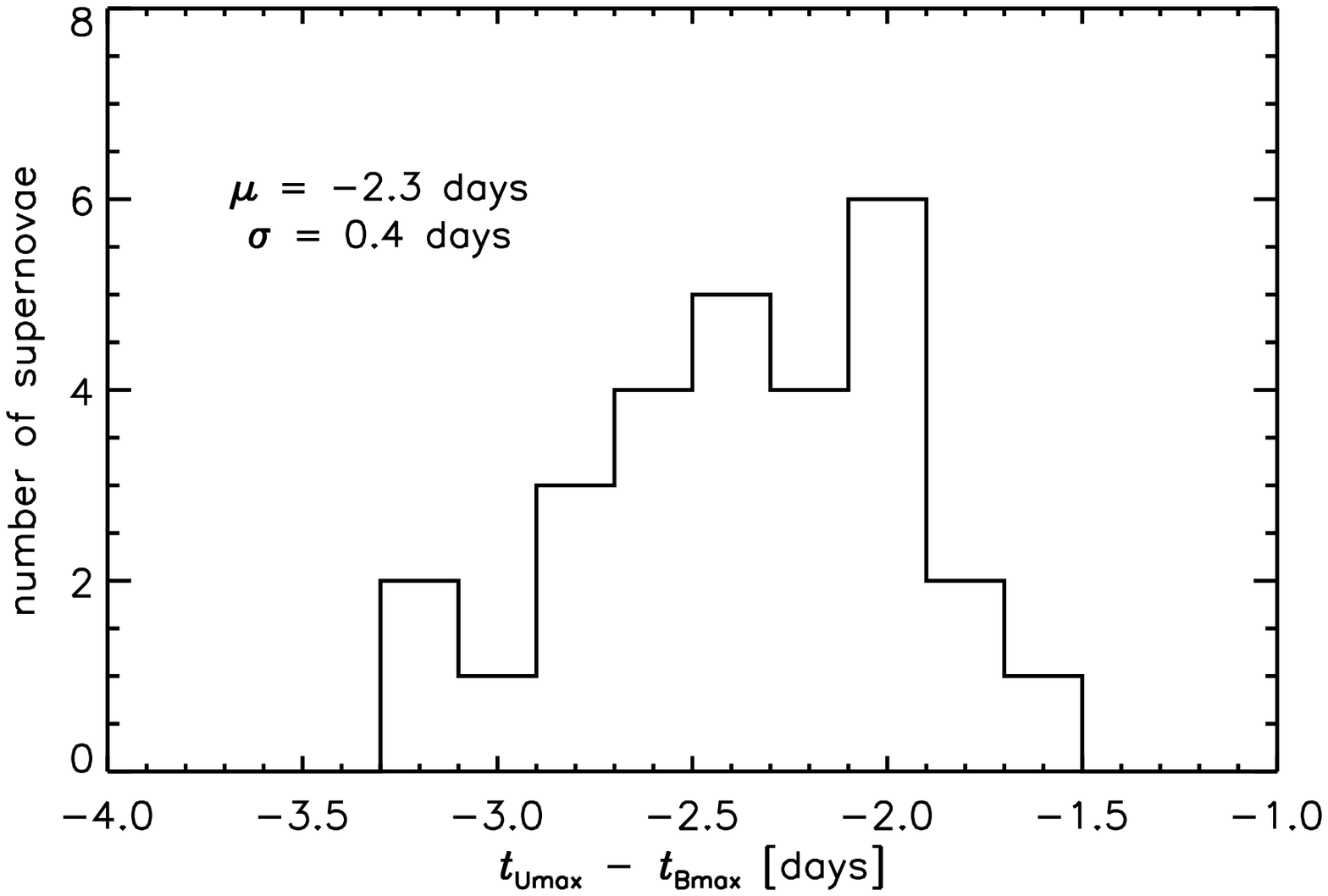}
\end{center}
\caption[Time of U maximum]{\singlespace Distribution of the time of
maximum light in $U$ relative to the time of maximum light in $B$,
measured in the SN rest frame. \label{ch3-fig-tu}}
\end{figure}

\clearpage
\begin{figure}
\begin{center}
\includegraphics[height=7.5in]{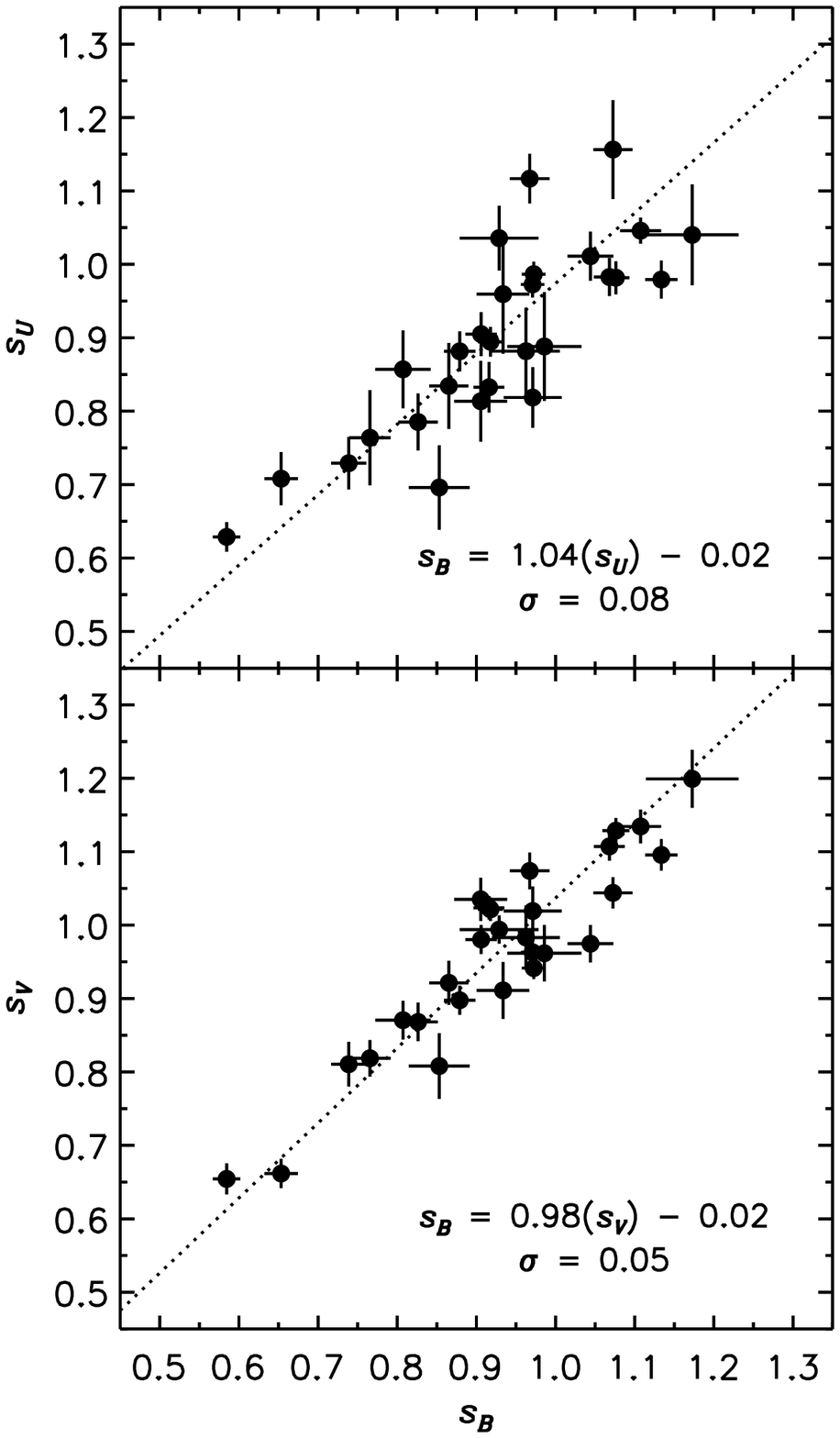}
\end{center}
\caption[\ubv Stretch]{\singlespace Relation between timescale stretch
factors in \ubv based on the stretch-corrected SN 1998aq
templates (see text for details). The best linear fits are: $s_B =
(1.04 \pm 0.06)s_U - (0.02 \pm 0.05)$ and $s_B = (0.98 \pm 0.05)s_V -
(0.02 \pm 0.05)$. \label{ch3-fig-subv}}
\end{figure}

\clearpage
\begin{figure}
\begin{center}
\includegraphics[height=7.5in]{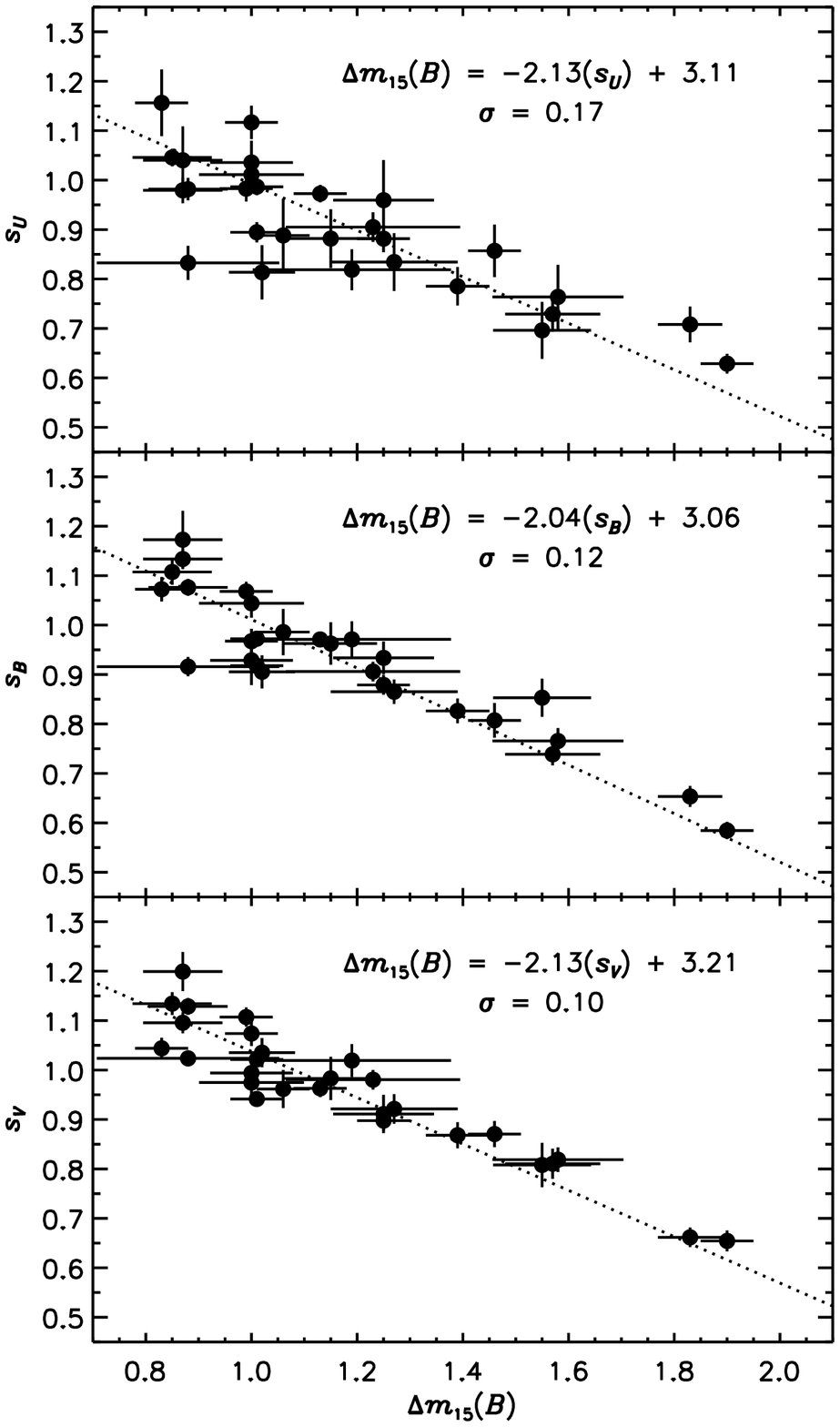}
\end{center}
\caption[\ubv Stretch-\dmf]{\singlespace Relation between timescale
stretch factors and \dmf. The best linear fits are \dmf\ $= (-2.13 \pm
0.14)s_U + (3.11 \pm 0.13)$, \dmf\ $= (-2.04 \pm 0.11)s_B + (3.06 \pm
0.10)$, and \dmf $= (-2.13 \pm 0.12)s_V + (3.21 \pm
0.11)$. \label{ch3-fig-sdm15}}
\end{figure}

\clearpage
\begin{figure}
\begin{center}
\includegraphics[height=7.2in]{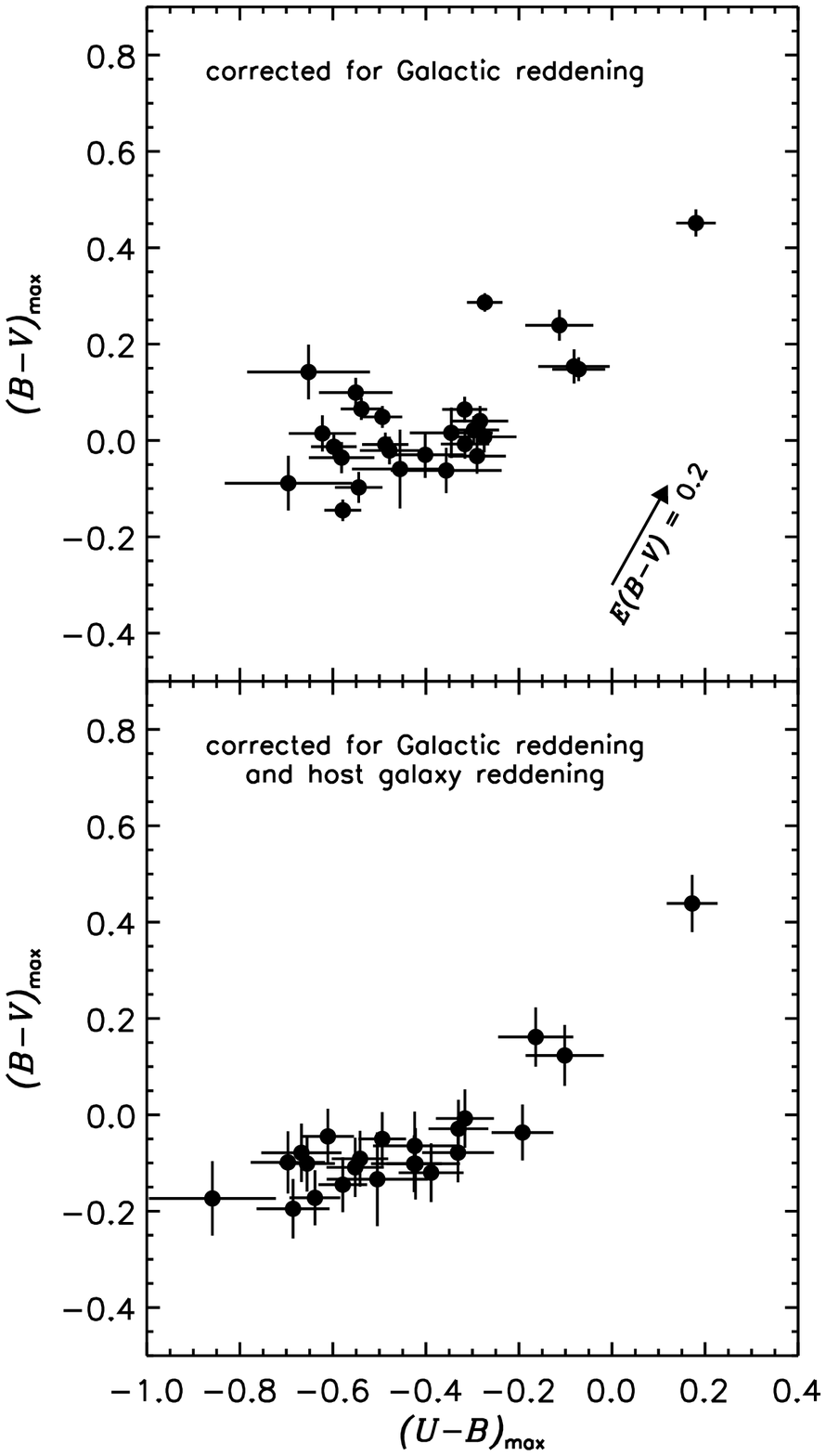}
\end{center}
\caption[Colors Plot]{\singlespace SN Ia colors at the epoch of
maximum light in $B$. The top panel shows maximum light colors
corrected for Galactic extinction, while the bottom panel includes a
correction for extinction in the host galaxy. The arrow indicates a
reddening vector corresponding to $E(\bv)_{\rm true} = 0.2$
mag. \label{ch3-fig-ubbv}}
\end{figure}

\clearpage
\begin{figure}
\begin{center}
\includegraphics[height=7.2in]{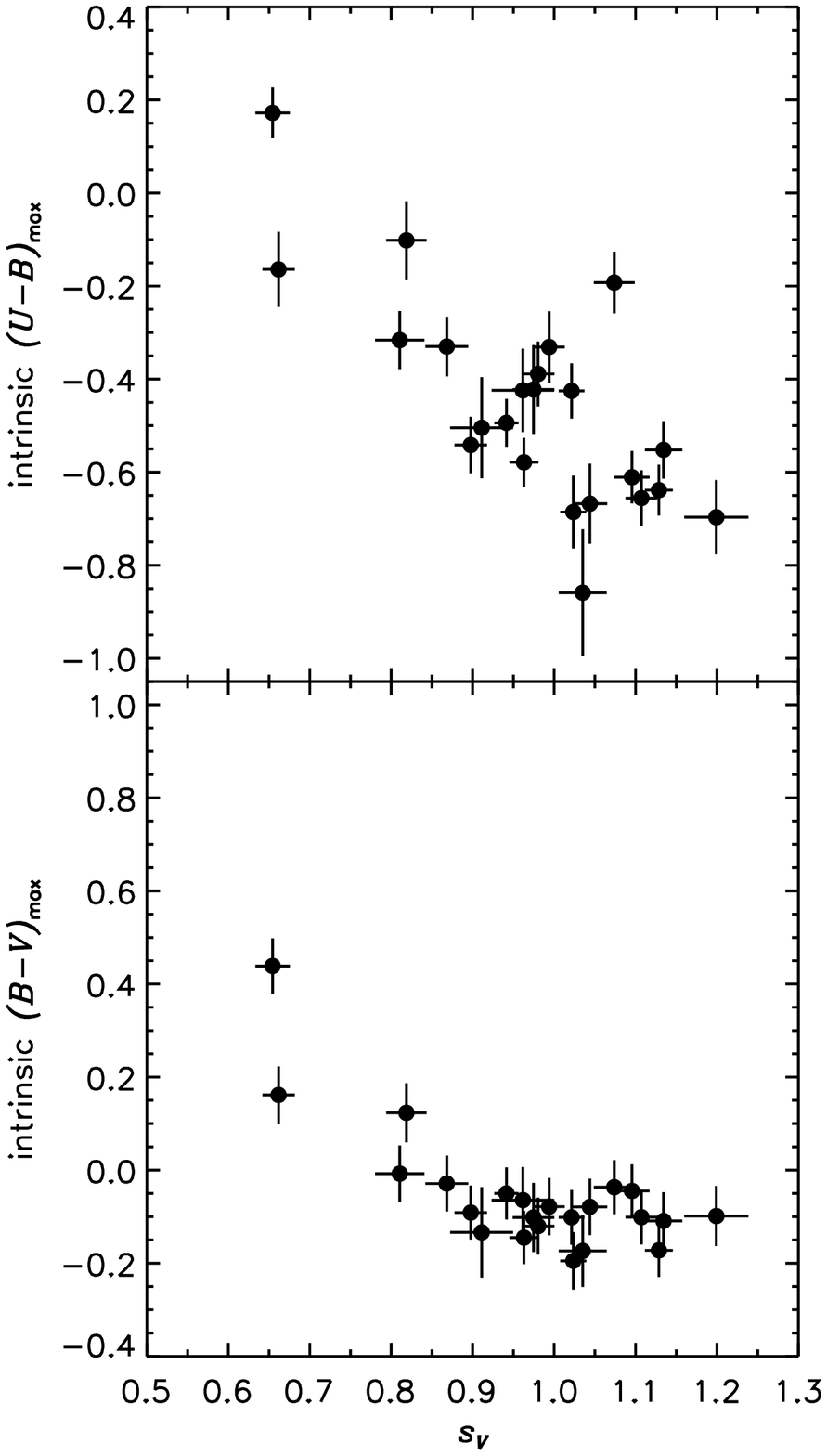}
\end{center}
\caption[Colors vs Stretch]{\singlespace Relation between SN Ia
\ubv colors at maximum light in $B$ and the $V$-band timescale stretch
factor. The colors have been corrected for Galactic and host-galaxy
reddening. \label{ch3-fig-ubbvsv}}
\end{figure}

\clearpage
\addtolength{\textheight}{0.2in}

\tabletypesize{\tiny}


\begin{deluxetable}{llllclc}
\tabletypesize{\scriptsize}
\singlespace
\tablewidth{0pt}
\tablecaption{SN Ia Discovery and Classification Data \label{ch3-tab-disc}}
\tablehead{\colhead{SN Ia} & \colhead{Galaxy} & \colhead{Discovery
Date} & \colhead{Discoverer} & \colhead{IAUC} &
\colhead{Spectroscopic ID} & \colhead{IAUC}}
\startdata
1997E  & NGC 2258      & 1997-01-14.5  & R. Kushida                    & 6538  & P. Garnavich \& R. Kirshner (CfA) & 6538 \\
1997Y  & NGC 4675      & 1997-02-02    & W. Li et al.~(BAO)            & 6556  & A. Filippenko et al., P. Garnavich et al.~(CfA) & 6557 \\
1997bp & NGC 4680      & 1997-04-06.5  & R. Evans                      & 6613  & M. Phillips et al.       & 6613 \\
1997bq & NGC 3147      & 1997-04-08.0  & S. Laurie (UK)                & 6616  & P. Challis (CfA)            & 6616 \\
1997br & ESO 576-40    & 1997-04-10.6  & Q. Qiao et al.~(BAO)          & 6623  & Q. Qiao et al.        & 6623 \\
1997cn & NGC 5490      & 1997-05-14.6  & W. Li et al.~(BAO)            & 6661  & M. Turatto et al.           & 6667 \\
1997cw & NGC 105       & 1997-07-10.8  & Q. Qiao et al.~(BAO)          & 6699  & Q. Qiao et al.        & 6699 \\
1997dg & anonymous     & 1997-09-27.7  & Q. Qiao et al.~(BAO)          & 6749  & S. Jha et al. (CfA)         & 6749 \\
\enddata
\end{deluxetable}


\begin{deluxetable}{ccccccccc}
\tablecolumns{9}
\singlespace
\tablewidth{0pt}
\tablecaption{Comparison Star Photometry \label{ch3-tab-compstar}}
\tablehead{\colhead{Star} & \colhead{$\alpha$ (J2000)} &
\colhead{$\delta$ (J2000)} & \colhead{$V$} & \colhead{\ub} & \colhead{\bv}
& \colhead{\vr} & \colhead{\vi} & \colhead{$N$}}
\startdata

\cutinhead{SN 1997E} \\*
SN       & 06:47:38.16 & $+$74:29:51.0 \\*
 1  & 06:47:27.41 & $+$74:30:02.6 & 14.402(014) & {\phs}0.086(041) & {\phs}0.675(011) & {\phs}0.427(007) & {\phs}0.846(010) & 3 \\*
 2  & 06:47:52.40 & $+$74:31:52.6 & 15.139(015) & {\phs}0.123(040) & {\phs}0.715(011) & {\phs}0.436(007) & {\phs}0.863(010) & 3 \\*
 3  & 06:47:18.41 & $+$74:31:40.7 & 15.895(017) & {\phs}0.018(040) & {\phs}0.584(013) & {\phs}0.365(008) & {\phs}0.747(009) & 3 \\*
 4  & 06:47:23.92 & $+$74:31:01.6 & 15.410(022) & {\phs}0.580(045) & {\phs}0.946(019) & {\phs}0.539(007) & {\phs}1.021(014) & 3 \\*
 5  & 06:47:00.39 & $+$74:28:13.5 & 15.358(014) & {\phs}0.594(041) & {\phs}0.907(012) & {\phs}0.511(009) & {\phs}0.968(009) & 3 \\*
 6  & 06:47:03.36 & $+$74:29:06.4 & 15.152(013) & {\phs}0.143(040) & {\phs}0.733(012) & {\phs}0.439(008) & {\phs}0.877(010) & 3 \\
\enddata
\end{deluxetable}


\clearpage

\begin{deluxetable}{lrrc}
\tabletypesize{\scriptsize}
\singlespace
\tablewidth{0pt}
\tablecaption{Photometric Color Terms \label{ch3-tab-colterm}}
\tablehead{\colhead{Detector/Filterset} & \colhead{Color Term} &
 \colhead{Value} & \colhead{Nights}}
\startdata
AndyCam/SAO    &      (\vV)/(\bv) &    $+0.0340 \pm 0.0042$ & 7 \\
AndyCam/SAO    &      (\iub)/(\ub) &   $ 0.9312 \pm 0.0039$ & 6 \\
AndyCam/SAO    &      (\ibv)/(\bv) &   $ 0.9293 \pm 0.0029$ & 7 \\
AndyCam/SAO    &      (\ivr)/(\vr) &   $ 0.9824 \pm 0.0053$ & 7 \\
AndyCam/SAO    &      (\ivi)/(\vi) &   $ 1.0739 \pm 0.0040$ & 7 \\
 \\
AndyCam/Harris+$I_{\rm SAO}$\tablenotemark{a} & (\ivi)/(\vi) & $1.0639 \pm 0.0124$ & 2\tablenotemark{b} \\
 \\
\enddata
\tablecomments{ The lowercase/uppercase letters in the color terms refer to
instrumental/standard magnitudes. All color terms implicitly contain
an additive constant. For example, for the AndyCam/SAO
combination: (\vV)$ = +0.0340$(\bv)$\, + $ const, and (\iub)$ =
0.9312$(\ub)$\, + $ const.}
\tablenotetext{a}{This filterset consists of the Harris \ubvr\ filters
and the SAO $I$ filter.}
\tablenotetext{b}{These nights were not photometric; the color terms
were derived from the calibrated comparison stars. See text for details.}
\end{deluxetable}

\clearpage


\begin{deluxetable}{lcccccc}
\singlespace
\tablewidth{0pt}
\tablecaption{Photometry of SN 1997E\label{ch3-tab-sn97E}}
\tablehead{\colhead{HJD} & \colhead{$U$} & 
\colhead{$B$} & \colhead{$V$} & \colhead{$R$} & \colhead{$I$} & 
\colhead{Detector/Filterset}}
\startdata
2450464.92 & \nodata & 15.694(017) & 15.598(011) & 15.377(015) & \nodata & AndyCam/SAO \\
2450465.69 & 15.441(038) & 15.667(015) & 15.544(010) & 15.357(013) & 15.423(014) & AndyCam/SAO \\
2450466.78 & 15.414(036) & 15.656(013) & 15.502(008) & 15.323(011) & 15.468(011) & AndyCam/SAO \\
2450468.66 & 15.500(037) & 15.620(014) & 15.492(010) & 15.322(012) & 15.480(013) & AndyCam/SAO \\
2450472.66 & 15.789(037) & 15.779(013) & 15.491(008) & 15.323(013) & 15.638(012) & AndyCam/SAO \\
2450476.89 & 16.330(061) & 16.166(032) & 15.741(023) & 15.716(030) & 15.940(032) & AndyCam/SAO \\
2450479.87 & 16.802(037) & 16.547(013) & 15.933(008) & 15.906(011) & 16.006(011) & AndyCam/SAO \\
2450485.62 & 17.725(041) & 17.381(015) & 16.301(009) & 16.003(013) & 15.847(018) & AndyCam/SAO \\
2450489.77 & \nodata & 17.891(022) & 16.627(014) & 16.131(019) & 15.823(019) & AndyCam/SAO \\
2450512.65 & 19.247(060) & 18.929(025) & 17.829(016) & 17.448(026) & 17.296(023) & AndyCam/SAO \\
2450521.62 & 19.383(186) & 19.147(028) & 18.101(016) & 17.811(023) & 17.736(024) & AndyCam/SAO \\
2450523.66 & 19.668(114) & 19.147(051) & 18.113(034) & 17.933(048) & 17.831(074) & AndyCam/SAO \\
2450545.65 & 20.159(120) & 19.495(070) & 18.783(046) & 18.622(066) & 18.717(129) & AndyCam/SAO \\
\enddata
\end{deluxetable}

\begin{deluxetable}{lcccccc}
\singlespace
\tablewidth{0pt}
\tablecaption{Photometry of SN 1997Y\label{ch3-tab-sn97Y}}
\tablehead{\colhead{HJD} & \colhead{$U$} & 
\colhead{$B$} & \colhead{$V$} & \colhead{$R$} & \colhead{$I$} & 
\colhead{Detector/Filterset}}
\startdata
2450488.96 & 15.073(037) & 15.379(015) & 15.368(010) & 15.278(015) & 15.592(014) & AndyCam/SAO \\
2450489.91 & 15.157(038) & 15.421(016) & 15.384(011) & 15.284(016) & 15.600(015) & AndyCam/SAO \\
2450515.94 & \nodata & 18.000(025) & 16.852(013) & 16.332(019) & 15.928(018) & AndyCam/SAO \\
2450518.84 & 18.429(056) & 18.225(022) & 17.012(010) & 16.563(015) & 16.168(014) & AndyCam/SAO \\
2450520.86 & 18.549(056) & 18.256(024) & 17.141(012) & 16.681(023) & 16.339(017) & AndyCam/SAO \\
2450536.84 & 18.827(069) & 18.707(030) & 17.690(011) & 17.396(018) & 17.174(020) & AndyCam/SAO \\
2450545.87 & 18.989(099) & 18.821(046) & 17.944(024) & 17.672(037) & 17.530(032) & AndyCam/SAO \\
\enddata
\end{deluxetable}

\begin{deluxetable}{lcccccc}
\singlespace
\tablewidth{0pt}
\tablecaption{Photometry of SN 1997bp\label{ch3-tab-sn97bp}}
\tablehead{\colhead{HJD} & \colhead{$U$} & 
\colhead{$B$} & \colhead{$V$} & \colhead{$R$} & \colhead{$I$} & 
\colhead{Detector/Filterset}}
\startdata
2450546.79 & 14.077(036) & 14.122(012) & 14.018(007) & 13.921(010) & 14.144(011) & AndyCam/SAO \\
2450547.80 & 14.079(038) & 14.097(014) & 13.980(009) & 13.894(012) & 14.133(014) & AndyCam/SAO \\
2450548.79 & 14.110(036) & 14.085(012) & 13.934(007) & 13.838(009) & 14.105(011) & AndyCam/SAO \\
2450549.80 & 14.175(037) & 14.109(014) & 13.902(009) & 13.812(015) & 14.119(013) & AndyCam/SAO \\
2450550.76 & 14.229(035) & 14.124(013) & 13.879(008) & 13.792(013) & 14.150(018) & AndyCam/SAO \\
2450552.82 & 14.340(037) & 14.179(015) & 13.875(009) & 13.762(012) & 14.169(013) & AndyCam/SAO \\
2450567.77 & 15.914(039) & 15.680(017) & 14.594(010) & 14.494(015) & 14.564(024) & AndyCam/SAO \\
2450569.71 & 16.114(037) & 15.866(016) & 14.648(011) & 14.483(014) & 14.523(014) & AndyCam/SAO \\
2450573.79 & 16.494(037) & 16.200(012) & 14.822(007) & 14.461(009) & 14.412(010) & AndyCam/SAO \\
2450575.73 & 16.689(039) & 16.311(015) & 14.890(008) & 14.475(013) & 14.366(012) & AndyCam/SAO \\
2450579.74 & 16.983(046) & 16.655(016) & 15.135(010) & 14.655(013) & 14.417(015) & AndyCam/SAO \\
2450597.73 & 17.733(041) & 17.118(016) & 15.962(009) & 15.540(013) & 15.482(012) & AndyCam/SAO \\
2450610.69 & 17.859(040) & 17.226(020) & 16.284(013) & 16.025(017) & 16.058(018) & AndyCam/SAO \\
\enddata
\end{deluxetable}

\begin{deluxetable}{lcccccc}
\singlespace
\tablewidth{0pt}
\tablecaption{Photometry of SN 1997bq\label{ch3-tab-sn97bq}}
\tablehead{\colhead{HJD} & \colhead{$U$} & 
\colhead{$B$} & \colhead{$V$} & \colhead{$R$} & \colhead{$I$} & 
\colhead{Detector/Filterset}}
\startdata
2450547.62 & \nodata & 15.676(020) & 15.641(012) & 15.493(016) & 15.522(017) & AndyCam/SAO \\
2450548.81 & 14.971(042) & 15.350(023) & 15.401(017) & 15.234(022) & 15.270(021) & AndyCam/SAO \\
2450549.82 & 14.735(043) & 15.135(023) & 15.168(017) & 15.054(023) & 15.080(023) & AndyCam/SAO \\
2450550.70 & 14.551(070) & 14.994(042) & 15.013(020) & 14.902(028) & 14.914(027) & AndyCam/SAO \\
2450551.83 & 14.451(044) & 14.824(021) & 14.865(009) & 14.740(012) & 14.807(014) & AndyCam/SAO \\
2450568.66 & 15.323(046) & 15.255(018) & 14.675(011) & 14.711(015) & 15.032(017) & AndyCam/SAO \\
2450570.76 & 15.650(039) & 15.501(012) & 14.819(007) & 14.827(010) & 15.052(011) & AndyCam/SAO \\
2450573.70 & 16.078(041) & 15.860(016) & 15.009(011) & 14.918(014) & 15.000(016) & AndyCam/SAO \\
2450575.68 & 16.380(038) & 16.114(014) & 15.102(008) & 14.951(012) & 14.954(013) & AndyCam/SAO \\
2450578.66 & \nodata & 16.429(020) & 15.259(011) & 15.027(019) & 14.908(014) & AndyCam/SAO \\
2450582.78 & 17.162(089) & 16.824(022) & 15.473(013) & 15.068(017) & 14.853(018) & AndyCam/SAO \\
2450600.72 & 17.968(045) & 17.575(017) & 16.422(011) & 16.030(015) & 15.834(015) & AndyCam/SAO \\
2450609.66 & 18.047(054) & 17.679(021) & 16.636(011) & 16.355(016) & 16.237(016) & AndyCam/SAO \\
\enddata
\end{deluxetable}

\begin{deluxetable}{lcccccc}
\singlespace
\tablewidth{0pt}
\tablecaption{Photometry of SN 1997br\label{ch3-tab-sn97br}}
\tablehead{\colhead{HJD} & \colhead{$U$} & 
\colhead{$B$} & \colhead{$V$} & \colhead{$R$} & \colhead{$I$} & 
\colhead{Detector/Filterset}}
\startdata
2450551.84 & 13.713(045) & 14.397(016) & 14.154(008) & 13.983(011) & 13.922(011) & AndyCam/SAO \\
2450552.83 & 13.695(053) & 14.317(028) & 14.078(016) & 13.847(023) & 13.811(021) & AndyCam/SAO \\
2450567.79 & 14.278(043) & 14.371(015) & 13.800(009) & 13.637(013) & 13.759(012) & AndyCam/SAO \\
2450569.79 & 14.537(042) & 14.574(014) & 13.855(008) & 13.717(012) & 13.825(012) & AndyCam/SAO \\
2450571.81 & 14.826(037) & 14.811(016) & 13.936(010) & 13.849(013) & 13.833(015) & AndyCam/SAO \\
2450573.75 & 15.205(039) & 15.060(015) & 14.080(009) & 13.890(012) & 13.838(014) & AndyCam/SAO \\
2450575.76 & 15.532(041) & 15.315(014) & 14.208(008) & 13.985(012) & 13.828(014) & AndyCam/SAO \\
2450578.73 & 15.988(040) & 15.672(019) & 14.416(013) & 14.048(017) & 13.789(017) & AndyCam/SAO \\
2450582.80 & 16.381(056) & 16.002(021) & 14.609(012) & 14.162(017) & 13.875(017) & AndyCam/SAO \\
2450597.75 & 17.156(046) & 16.810(017) & 15.331(009) & 14.825(012) & 14.350(014) & AndyCam/SAO \\
2450607.75 & 17.334(046) & 17.009(017) & 15.701(009) & 15.262(013) & 14.915(013) & AndyCam/SAO \\
\enddata
\end{deluxetable}
 
\begin{deluxetable}{lcccccc}
\singlespace
\tablewidth{0pt}
\tablecaption{Photometry of SN 1997cn\label{ch3-tab-sn97cn}}
\tablehead{\colhead{HJD} & \colhead{$U$} & 
\colhead{$B$} & \colhead{$V$} & \colhead{$R$} & \colhead{$I$} & 
\colhead{Detector/Filterset}}
\startdata
2450597.76 & 18.897(068) & 18.523(027) & 17.216(014) & 16.708(019) & 16.465(019) & AndyCam/SAO \\
2450598.83 & 18.915(068) & 18.653(029) & 17.291(012) & 16.761(016) & 16.428(016) & AndyCam/SAO \\
2450599.71 & \nodata & 18.780(048) & \nodata & \nodata & \nodata & AndyCam/SAO \\
2450600.75 & 19.119(060) & 18.858(026) & 17.497(014) & 16.948(019) & 16.563(021) & AndyCam/SAO \\
2450602.73 & 19.471(099) & 19.081(038) & 17.694(016) & 17.160(022) & 16.724(022) & AndyCam/SAO \\
2450604.82 & 19.440(076) & 19.290(037) & 17.911(020) & 17.402(028) & 16.941(031) & AndyCam/SAO \\
2450605.88 & 19.422(099) & \nodata & \nodata & 17.487(041) & 17.078(039) & AndyCam/SAO \\
2450607.77 & 19.559(080) & \nodata & 18.062(014) & 17.623(018) & 17.191(021) & AndyCam/SAO \\
2450608.70 & 19.493(072) & 19.421(045) & \nodata & 17.710(039) & 17.265(039) & AndyCam/SAO \\
2450609.67 & 19.643(065) & 19.454(030) & 18.189(020) & 17.746(026) & 17.286(027) & AndyCam/SAO \\
2450610.79 & 19.679(106) & 19.595(053) & 18.238(023) & 17.810(030) & 17.391(031) & AndyCam/SAO \\
2450611.82 & 19.632(101) & \nodata & 18.301(019) & 17.837(025) & 17.447(029) & AndyCam/SAO \\
2450627.67 & \nodata & 20.076(051) & 18.911(020) & 18.616(031) & 18.345(029) & AndyCam/SAO \\
2450640.75 & \nodata & \nodata & 19.238(034) & 19.135(052) & 18.932(063) & AndyCam/SAO \\
\enddata
\end{deluxetable}
 
\begin{deluxetable}{lcccccc}
\singlespace
\tablewidth{0pt}
\tablecaption{Photometry of SN 1997cw\label{ch3-tab-sn97cw}}
\tablehead{\colhead{HJD} & \colhead{$U$} & 
\colhead{$B$} & \colhead{$V$} & \colhead{$R$} & \colhead{$I$} & 
\colhead{Detector/Filterset}}
\startdata
2450642.94 & 17.041(039) & 16.892(016) & 16.146(008) & 15.948(011) & 16.109(014) & AndyCam/SAO \\
2450643.88 & 17.227(038) & 17.024(015) & 16.184(009) & 16.002(012) & 16.155(013) & AndyCam/SAO \\
2450644.85 & 17.319(056) & 17.096(027) & 16.222(015) & 16.079(020) & 16.189(024) & AndyCam/SAO \\
2450645.98 & 17.469(048) & 17.213(020) & 16.341(010) & 16.132(014) & 16.197(020) & AndyCam/SAO \\
2450646.98 & 17.594(059) & 17.282(019) & 16.359(009) & 16.199(012) & 16.182(013) & AndyCam/SAO \\
2450647.98 & 17.804(076) & 17.424(045) & 16.429(008) & 16.215(014) & 16.167(013) & AndyCam/SAO \\
2450652.92 & 18.453(119) & 17.971(046) & 16.643(018) & 16.268(024) & 16.113(027) & AndyCam/SAO \\
2450653.94 & 18.484(079) & 18.145(028) & 16.707(018) & 16.290(025) & 16.038(025) & AndyCam/SAO \\
2450665.99 & 19.368(116) & 18.808(033) & 17.227(013) & 16.619(019) & 16.196(018) & AndyCam/SAO \\
2450673.99 & 19.612(187) & 19.229(050) & 17.612(021) & 17.044(027) & 16.627(027) & AndyCam/SAO \\
2450680.99 & \nodata & 19.258(097) & 17.920(058) & 17.427(075) & 16.956(073) & AndyCam/SAO \\
2450682.97 & \nodata & 19.293(094) & 17.983(057) & 17.469(072) & 17.194(070) & AndyCam/SAO \\
2450685.97 & \nodata & 19.379(051) & 18.075(030) & 17.545(042) & 17.299(040) & AndyCam/SAO \\
2450688.00 & \nodata & 19.329(051) & 18.138(037) & 17.655(047) & 17.444(047) & AndyCam/SAO \\
2450693.79 & \nodata & 19.499(036) & 18.260(021) & 17.828(028) & 17.769(040) & AndyCam/SAO \\
2450783.74 & \nodata & 21.126(092) & 20.068(051) & 20.421(093) & \nodata & AndyCam/SAO \\
2450810.62 & \nodata & \nodata & 20.472(086) & 20.690(160) & \nodata & AndyCam/SAO \\
\enddata
\end{deluxetable}

\begin{deluxetable}{lcccccc}
\singlespace
\tablewidth{0pt}
\tablecaption{Photometry of SN 1997dg\label{ch3-tab-sn97dg}}
\tablehead{\colhead{HJD} & \colhead{$U$} & 
\colhead{$B$} & \colhead{$V$} & \colhead{$R$} & \colhead{$I$} & 
\colhead{Detector/Filterset}}
\startdata
2450720.90 & 16.806(047) & 17.186(027) & 17.095(018) & 16.921(024) & 17.141(026) & AndyCam/SAO \\
2450722.74 & 16.919(044) & 17.178(021) & 17.098(014) & 16.927(019) & 17.231(023) & AndyCam/SAO \\
2450726.83 & 17.405(044) & 17.364(022) & 17.230(013) & 17.078(019) & 17.392(022) & AndyCam/SAO \\
2450730.65 & 17.958(049) & 17.745(022) & 17.387(012) & 17.344(017) & 17.714(024) & AndyCam/SAO \\
2450747.71 & \nodata & 19.612(132) & 18.416(029) & 17.910(038) & 17.587(039) & AndyCam/SAO \\
2450761.79 & 19.311(275) & 20.406(191) & 19.205(047) & 18.668(066) & 18.406(063) & AndyCam/SAO \\
2450783.72 & 20.664(144) & \nodata & 19.786(034) & 19.420(046) & 19.424(111) & AndyCam/SAO \\
2450810.60 & \nodata & 21.392(192) & 20.383(116) & 20.227(183) & \nodata & AndyCam/SAO \\
2450842.61 & \nodata & 22.062(176) & 21.224(127) & 21.404(222) & \nodata & 4Sh-chip1/SAO \\
\enddata
\end{deluxetable}

\begin{deluxetable}{lcccccc}
\singlespace
\tablewidth{0pt}
\tablecaption{Photometry of SN 1997do\label{ch3-tab-sn97do}}
\tablehead{\colhead{HJD} & \colhead{$U$} & 
\colhead{$B$} & \colhead{$V$} & \colhead{$R$} & \colhead{$I$} & 
\colhead{Detector/Filterset}}
\startdata
2450760.01 & \nodata & 14.849(029) & 14.880(022) & 14.716(028) & 14.873(027) & AndyCam/SAO \\
2450762.84 & 14.455(038) & 14.626(014) & 14.635(008) & 14.454(011) & 14.691(013) & AndyCam/SAO \\
2450778.02 & \nodata & \nodata & 14.807(009) & 14.773(012) & \nodata & AndyCam/SAO \\
2450780.90 & 15.959(037) & 15.607(013) & 15.003(007) & 15.021(009) & 15.288(010) & AndyCam/SAO \\
2450783.81 & 16.190(038) & 15.960(013) & 15.131(008) & 15.096(011) & 15.203(011) & AndyCam/SAO \\
2450810.78 & 17.991(065) & 17.722(039) & 16.480(008) & 16.063(012) & 15.833(013) & AndyCam/SAO \\
2450842.74 & 18.458(059) & 18.116(024) & 17.329(009) & 17.093(014) & 17.223(019) & 4Sh-chip1/SAO \\
\enddata
\end{deluxetable}

\begin{deluxetable}{lcccccc}
\singlespace
\tablewidth{0pt}
\tablecaption{Photometry of SN 1997dt\label{ch3-tab-sn97dt}}
\tablehead{\colhead{HJD} & \colhead{$U$} & 
\colhead{$B$} & \colhead{$V$} & \colhead{$R$} & \colhead{$I$} & 
\colhead{Detector/Filterset}}
\startdata
2450777.60 & \nodata & \nodata & 15.919(007) & 15.402(010) & \nodata & AndyCam/SAO \\
2450781.68 & 15.863(038) & 15.801(014) & 15.369(009) & 14.928(011) & 14.636(013) & AndyCam/SAO \\
2450810.58 & 18.539(051) & 17.847(015) & 16.242(008) & 15.570(011) & 14.986(012) & AndyCam/SAO \\
2450842.59 & \nodata & 18.915(044) & 17.510(019) & 16.945(027) & 16.474(035) & 4Sh-chip1/SAO \\
\enddata
\end{deluxetable}

\begin{deluxetable}{lcccccc}
\singlespace
\tablewidth{0pt}
\tablecaption{Photometry of SN 1998D\label{ch3-tab-sn98D}}
\tablehead{\colhead{HJD} & \colhead{$U$} & 
\colhead{$B$} & \colhead{$V$} & \colhead{$R$} & \colhead{$I$} & 
\colhead{Detector/Filterset}}
\startdata
2450874.04 & 18.902(057) & 18.588(028) & 17.433(013) & 16.971(020) & 16.571(024) & 4Sh-chip1/SAO \\
2450893.93 & 19.445(079) & 19.094(032) & 18.156(022) & 17.801(048) & 17.558(031) & AndyCam/SAO \\
2450925.89 & \nodata & 19.703(037) & 19.081(021) & 19.096(036) & 18.797(045) & 4Sh-chip1/SAO \\
2450980.80 & \nodata & 20.604(082) & 20.494(059) & 20.692(086) & 20.134(198) & 4Sh-chip1/SAO \\
2450987.78 & \nodata & 20.809(083) & 20.590(060) & \nodata & \nodata & 4Sh-chip1/SAO \\
\enddata
\end{deluxetable}

\begin{deluxetable}{lcccccc}
\singlespace
\tablewidth{0pt}
\tablecaption{Photometry of SN 1998V\label{ch3-tab-sn98V}}
\tablehead{\colhead{HJD} & \colhead{$U$} & 
\colhead{$B$} & \colhead{$V$} & \colhead{$R$} & \colhead{$I$} & 
\colhead{Detector/Filterset}}
\startdata
2450894.00 & 15.738(039) & 15.917(016) & 15.698(010) & 15.486(017) & 15.757(015) & AndyCam/SAO \\
2450894.98 & 15.918(037) & 15.972(014) & 15.761(009) & 15.521(011) & \nodata & AndyCam/SAO \\
2450896.01 & 15.960(038) & 15.996(014) & 15.736(007) & 15.558(010) & 15.807(011) & AndyCam/SAO \\
2450900.93 & 16.362(036) & 16.337(011) & 15.913(007) & 15.812(008) & 16.036(012) & AndyCam/SAO \\
2450904.00 & 16.906(036) & 16.601(011) & 16.159(007) & 16.090(010) & 16.248(013) & AndyCam/SAO \\
2450906.99 & 17.296(041) & 16.975(018) & 16.285(012) & 16.140(016) & 16.213(018) & AndyCam/SAO \\
2450921.99 & 18.995(076) & 18.526(032) & 17.147(011) & 16.526(017) & 16.100(025) & 4Sh-chip1/SAO \\
2450925.95 & 19.186(056) & 18.761(028) & 17.408(010) & 16.779(016) & 16.364(022) & 4Sh-chip1/SAO \\
2450957.98 & 19.952(199) & 19.345(060) & 18.395(022) & 18.072(038) & 17.895(063) & 4Sh-chip1/SAO \\
2450980.86 & 20.362(160) & 19.683(035) & 18.960(021) & 18.710(030) & 18.735(043) & 4Sh-chip1/SAO \\
2450987.83 & \nodata & 19.823(035) & 19.137(018) & 18.987(036) & 19.057(049) & 4Sh-chip1/SAO \\
2451013.79 & \nodata & 20.172(049) & 19.702(032) & 19.700(065) & 19.815(111) & AndyCam/SAO \\
2451077.66 & \nodata & 21.350(266) & 20.626(199) & 21.282(280) & \nodata & 4Sh-chip1/SAO \\
\enddata
\end{deluxetable}
\clearpage

\begin{deluxetable}{lcccccc}
\singlespace
\tablewidth{0pt}
\tablecaption{Photometry of SN 1998ab\label{ch3-tab-sn98ab}}
\tablehead{\colhead{HJD} & \colhead{$U$} & 
\colhead{$B$} & \colhead{$V$} & \colhead{$R$} & \colhead{$I$} & 
\colhead{Detector/Filterset}}
\startdata
2450906.82 & 15.974(038) & 16.602(015) & 16.570(008) & 16.478(013) & 16.565(015) & AndyCam/SAO \\
2450920.66 & 16.216(044) & 16.384(019) & 16.216(008) & 16.082(011) & 16.473(017) & 4Sh-chip1/SAO \\
2450922.65 & 16.438(040) & 16.533(015) & 16.277(009) & 16.153(012) & 16.520(014) & 4Sh-chip1/SAO \\
2450925.84 & 16.850(040) & 16.850(015) & 16.388(008) & 16.280(012) & 16.606(014) & 4Sh-chip1/SAO \\
2450930.76 & 17.673(045) & 17.431(018) & 16.746(009) & 16.503(013) & 16.605(015) & 4Sh-chip1/SAO \\
2450931.81 & 17.865(046) & 17.583(020) & 16.781(011) & 16.571(015) & 16.583(017) & 4Sh-chip1/SAO \\
2450936.68 & 18.425(065) & 18.132(027) & 17.106(011) & 16.695(017) & 16.617(021) & 4Sh-chip1/SAO \\
2450938.71 & 18.840(104) & 18.300(032) & 17.211(015) & 16.772(021) & 16.668(023) & 4Sh-chip1/SAO \\
2450952.69 & \nodata & 19.096(037) & 17.914(015) & 17.311(024) & 17.030(028) & 4Sh-chip1/SAO \\
2450955.65 & 19.568(075) & \nodata & 18.035(012) & 17.452(021) & 17.213(031) & 4Sh-chip1/SAO \\
2450961.69 & 19.522(071) & 19.365(028) & 18.255(013) & 17.686(019) & 17.520(024) & AndyCam/SAO \\
2450964.65 & \nodata & 19.505(086) & 18.349(029) & 17.824(043) & 17.683(042) & 4Sh-chip1/SAO \\
2450980.69 & 19.721(113) & 19.649(063) & 18.702(026) & 18.369(042) & 18.330(058) & 4Sh-chip1/SAO \\
2450987.73 & 19.882(075) & 19.661(052) & 18.863(016) & 18.542(028) & 18.406(067) & 4Sh-chip1/SAO \\
\enddata
\end{deluxetable}

\begin{deluxetable}{lcccccc}
\singlespace
\tablewidth{0pt}
\tablecaption{Photometry of SN 1998bp\label{ch3-tab-sn98bp}}
\tablehead{\colhead{HJD} & \colhead{$U$} & 
\colhead{$B$} & \colhead{$V$} & \colhead{$R$} & \colhead{$I$} & 
\colhead{Detector/Filterset}}
\startdata
2450934.88 & 15.655(037) & 15.641(014) & 15.439(008) & 15.162(011) & 15.230(014) & 4Sh-chip1/SAO \\
2450937.90 & 15.801(040) & 15.672(017) & 15.313(010) & 15.070(015) & 15.134(016) & 4Sh-chip1/SAO \\
2450938.95 & 15.923(037) & 15.739(014) & 15.298(008) & 15.054(012) & 15.135(013) & 4Sh-chip1/SAO \\
2450939.93 & \nodata & 15.849(029) & 15.307(019) & 15.052(026) & 15.160(025) & 4Sh-chip1/SAO \\
2450949.93 & 17.921(053) & 17.416(026) & 16.077(011) & 15.607(015) & 15.295(019) & 4Sh-chip1/SAO \\
2450951.87 & 18.184(069) & 17.636(038) & 16.266(013) & 15.751(021) & 15.403(025) & 4Sh-chip1/SAO \\
2450953.85 & 18.250(049) & 17.883(029) & 16.416(012) & 15.870(019) & 15.482(023) & 4Sh-chip1/SAO \\
2450955.89 & 18.417(053) & 18.028(032) & 16.653(013) & 16.058(019) & 15.622(022) & 4Sh-chip1/SAO \\
2450959.94 & 18.700(046) & 18.239(023) & 16.941(016) & 16.441(021) & 16.022(022) & AndyCam/SAO \\
2450962.88 & 18.770(062) & 18.422(042) & 17.125(019) & 16.680(026) & 16.237(035) & 4Sh-chip1/SAO \\
2450980.84 & 19.263(088) & 18.939(064) & 17.787(039) & 17.481(051) & 17.203(053) & 4Sh-chip1/SAO \\
2450983.81 & 19.434(053) & 18.958(031) & 17.891(013) & 17.610(022) & 17.307(029) & 4Sh-chip1/SAO \\
2450987.82 & 19.476(083) & 19.025(038) & 18.014(017) & 17.770(027) & 17.529(038) & 4Sh-chip1/SAO \\
2450996.78 & \nodata & 19.204(049) & 18.301(025) & 18.121(036) & 17.836(059) & 4Sh-chip1/SAO \\
2451013.74 & 20.311(106) & 19.456(037) & 18.721(025) & 18.835(053) & 18.538(106) & AndyCam/SAO \\
2451077.63 & \nodata & 20.705(082) & \nodata & \nodata & 20.114(175) & 4Sh-chip1/SAO \\
\enddata
\end{deluxetable}

\begin{deluxetable}{lcccccc}
\singlespace
\tablewidth{0pt}
\tablecaption{Photometry of SN 1998co\label{ch3-tab-sn98co}}
\tablehead{\colhead{HJD} & \colhead{$U$} & 
\colhead{$B$} & \colhead{$V$} & \colhead{$R$} & \colhead{$I$} & 
\colhead{Detector/Filterset}}
\startdata
2450990.90 & 15.934(042) & 15.930(016) & 15.739(010) & 15.608(017) & 15.897(016) & 4Sh-chip1/SAO \\
2450993.96 & 16.319(045) & 16.228(016) & 15.867(009) & 15.796(012) & 15.961(020) & 4Sh-chip1/SAO \\
2451077.71 & 20.381(238) & 19.951(141) & 19.484(084) & 19.522(135) & 18.896(173) & 4Sh-chip1/SAO \\
2451110.66 & \nodata & 20.441(156) & 20.279(100) & \nodata & \nodata & 4sh-chip3/SAO \\
2451136.58 & \nodata & 20.971(192) & 20.920(134) & \nodata & 19.618(175) & AndyCam/SAO \\
\enddata
\end{deluxetable}

\begin{deluxetable}{lcccccc}
\singlespace
\tablewidth{0pt}
\tablecaption{Photometry of SN 1998de\label{ch3-tab-sn98de}}
\tablehead{\colhead{HJD} & \colhead{$U$} & 
\colhead{$B$} & \colhead{$V$} & \colhead{$R$} & \colhead{$I$} & 
\colhead{Detector/Filterset}}
\startdata
2451021.95 & 18.587(049) & 17.979(019) & 17.336(010) & 17.085(015) & 17.175(016) & 4Sh-chip1/SAO \\
2451022.96 & 18.389(064) & 17.781(022) & 17.176(011) & 16.952(024) & 16.937(021) & 4Sh-chip1/SAO \\
2451024.98 & 18.291(056) & 17.572(029) & 16.974(017) & 16.740(023) & 16.701(026) & 4Sh-chip1/SAO \\
2451047.85 & \nodata & 19.972(072) & 18.554(042) & 17.866(056) & 17.419(057) & 4Sh-chip1/SAO \\
2451055.90 & \nodata & 20.193(040) & 18.879(017) & 18.321(024) & 17.925(028) & 4Sh-chip1/SAO \\
2451074.74 & \nodata & 20.621(055) & 19.489(027) & 19.176(037) & 18.598(085) & 4Sh-chip1/SAO \\
2451077.84 & \nodata & 20.663(075) & 19.619(046) & 19.315(060) & 18.839(071) & 4Sh-chip1/SAO \\
2451110.84 & \nodata & 21.203(112) & 20.540(067) & 20.582(101) & 19.893(188) & 4sh-chip3/SAO \\
2451136.81 & \nodata & 21.760(133) & 21.139(085) & 21.654(198) & 20.523(175) & AndyCam/SAO \\
\enddata
\end{deluxetable}

\begin{deluxetable}{lcccccc}
\singlespace
\tablewidth{0pt}
\tablecaption{Photometry of SN 1998dh\label{ch3-tab-sn98dh}}
\tablehead{\colhead{HJD} & \colhead{$U$} & 
\colhead{$B$} & \colhead{$V$} & \colhead{$R$} & \colhead{$I$} & 
\colhead{Detector/Filterset}}
\startdata
2451021.94 & 14.427(039) & 14.745(015) & 14.718(007) & 14.475(010) & 14.453(012) & 4Sh-chip1/SAO \\
2451026.89 & 13.929(038) & 14.193(013) & 14.138(008) & 13.987(011) & 14.105(011) & 4Sh-chip1/SAO \\
2451045.84 & 15.743(052) & 15.534(036) & 14.665(024) & \nodata & 14.709(047) & 4Sh-chip1/SAO \\
2451047.84 & 15.994(045) & 15.753(021) & 14.782(009) & 14.618(014) & 14.580(015) & 4Sh-chip1/SAO \\
2451055.89 & 16.804(042) & 16.574(019) & 15.235(007) & 14.790(011) & 14.472(015) & 4Sh-chip1/SAO \\
2451069.87 & 17.455(051) & 17.293(023) & 16.045(010) & 15.634(014) & 15.343(016) & 4Sh-chip1/SAO \\
2451077.76 & 17.652(042) & 17.407(022) & 16.289(010) & 15.922(014) & 15.739(016) & 4Sh-chip1/SAO \\
2451110.75 & 18.467(059) & 17.855(038) & 17.128(025) & 17.049(036) & 17.058(040) & 4sh-chip3/SAO \\
2451136.68 & 19.097(045) & 18.229(017) & 17.730(011) & 17.847(015) & 17.971(032) & AndyCam/SAO \\
\enddata
\end{deluxetable}

\begin{deluxetable}{lcccccc}
\singlespace
\tablewidth{0pt}
\tablecaption{Photometry of SN 1998dk\label{ch3-tab-sn98dk}}
\tablehead{\colhead{HJD} & \colhead{$U$} & 
\colhead{$B$} & \colhead{$V$} & \colhead{$R$} & \colhead{$I$} & 
\colhead{Detector/Filterset}}
\startdata
2451075.72 & 16.700(055) & 16.494(023) & 15.566(012) & 15.434(020) & 15.456(019) & 4Sh-chip1/SAO \\
2451077.82 & 16.961(042) & 16.750(020) & 15.667(008) & 15.435(013) & 15.402(014) & 4Sh-chip1/SAO \\
2451079.76 & 17.234(048) & 16.941(026) & 15.743(013) & 15.469(017) & 15.369(021) & 4Sh-chip1/SAO \\
2451102.62 & 18.321(081) & 18.192(028) & 16.951(013) & 16.498(020) & 16.200(023) & 4sh-chip3/SAO \\
2451110.79 & 18.500(065) & 18.311(022) & 17.196(013) & 16.760(019) & 16.588(021) & 4sh-chip3/SAO \\
2451136.73 & 19.231(058) & 18.661(018) & 17.899(011) & 17.613(018) & 17.762(018) & AndyCam/SAO \\
\enddata
\end{deluxetable}

\begin{deluxetable}{lcccccc}
\singlespace
\tablewidth{0pt}
\tablecaption{Photometry of SN 1998dm\label{ch3-tab-sn98dm}}
\tablehead{\colhead{HJD} & \colhead{$U$} & 
\colhead{$B$} & \colhead{$V$} & \colhead{$R$} & \colhead{$I$} & 
\colhead{Detector/Filterset}}
\startdata
2451071.82 & \nodata & 15.398(023) & 14.843(006) & 14.743(010) & 14.832(011) & 4Sh-chip1/SAO \\
2451077.86 & 16.171(040) & 16.008(017) & 15.158(008) & 15.040(013) & 14.962(014) & 4Sh-chip1/SAO \\
2451080.88 & 16.545(046) & 16.321(021) & 15.328(009) & 15.094(015) & 14.909(014) & 4Sh-chip1/SAO \\
2451104.74 & 18.325(083) & 17.866(022) & 16.529(008) & 16.037(016) & 15.544(012) & 4sh-chip3/SAO \\
2451110.80 & 18.363(062) & 18.027(021) & 16.736(011) & 16.251(019) & 15.816(017) & 4sh-chip3/SAO \\
2451136.77 & 19.000(173) & 18.454(017) & 17.417(011) & 17.071(016) & 16.947(016) & AndyCam/SAO \\
2451195.64 & \nodata & 19.317(023) & 18.749(014) & 18.759(021) & 18.831(094) & 4sh-chip3/Harris+$I_{\rm SAO}$ \\
\enddata
\end{deluxetable}

\begin{deluxetable}{lcccccc}
\singlespace
\tablewidth{0pt}
\tablecaption{Photometry of SN 1998dx\label{ch3-tab-sn98dx}}
\tablehead{\colhead{HJD} & \colhead{$U$} & 
\colhead{$B$} & \colhead{$V$} & \colhead{$R$} & \colhead{$I$} & 
\colhead{Detector/Filterset}}
\startdata
2451072.62 & 17.269(095) & 17.689(057) & 17.729(030) & 17.578(045) & 17.895(062) & 4Sh-chip1/SAO \\
2451077.64 & 17.951(046) & 17.999(023) & 17.852(014) & 17.708(020) & 18.046(027) & 4Sh-chip1/SAO \\
2451078.64 & 18.059(046) & 18.050(024) & 17.865(015) & 17.735(022) & 18.118(085) & 4Sh-chip1/SAO \\
2451083.73 & 18.904(078) & 18.598(021) & 18.179(012) & 18.158(019) & 18.343(036) & AndyCam/SAO \\
2451100.62 & \nodata & 20.642(175) & 19.559(093) & 18.882(142) & 18.808(174) & 4sh-chip3/SAO \\
2451136.60 & 22.432(242) & 21.552(184) & 20.751(123) & 20.097(181) & 20.270(212) & AndyCam/SAO \\
\enddata
\end{deluxetable}

\begin{deluxetable}{lcccccc}
\singlespace
\tablewidth{0pt}
\tablecaption{Photometry of SN 1998ec\label{ch3-tab-sn98ec}}
\tablehead{\colhead{HJD} & \colhead{$U$} & 
\colhead{$B$} & \colhead{$V$} & \colhead{$R$} & \colhead{$I$} & 
\colhead{Detector/Filterset}}
\startdata
2451101.96 & 17.587(044) & 17.296(014) & 16.663(008) & 16.620(011) & 16.901(014) & 4sh-chip3/SAO \\
2451106.02 & 18.128(049) & 17.757(017) & 16.920(009) & 16.836(013) & 16.928(017) & 4sh-chip3/SAO \\
2451110.97 & 18.800(059) & 18.324(024) & 17.182(013) & 16.873(019) & 16.823(018) & 4sh-chip3/SAO \\
2451114.01 & 19.255(116) & 18.622(026) & 17.366(013) & 16.859(019) & 16.724(020) & 4sh-chip3/SAO \\
2451136.88 & 19.983(113) & 19.762(032) & 18.469(012) & 17.972(019) & 17.678(023) & AndyCam/SAO \\
2451195.74 & 21.003(226) & 20.717(064) & 19.964(036) & 19.929(048) & 20.149(108) & 4sh-chip3/Harris+$I_{\rm SAO}$ \\
2451217.84 & \nodata & 21.050(160) & 20.421(121) & 20.439(151) & 20.401(210) & AndyCam/Harris+$I_{\rm SAO}$ \\
2451224.68 & \nodata & 21.187(171) & 20.504(121) & 20.535(152) & \nodata & AndyCam/Harris+$I_{\rm SAO}$ \\
2451232.67 & \nodata & \nodata & 20.853(298) & \nodata & \nodata & 4sh-chip3/Harris+$I_{\rm SAO}$ \\
2451275.62 & \nodata & 21.938(296) & 21.584(153) & 21.411(261) & \nodata & 4sh-chip3/Harris+$I_{\rm SAO}$ \\
\enddata
\end{deluxetable}

\begin{deluxetable}{lcccccc}
\singlespace
\tablewidth{0pt}
\tablecaption{Photometry of SN 1998ef\label{ch3-tab-sn98ef}}
\tablehead{\colhead{HJD} & \colhead{$U$} & 
\colhead{$B$} & \colhead{$V$} & \colhead{$R$} & \colhead{$I$} & 
\colhead{Detector/Filterset}}
\startdata
2451106.62 & 15.412(043) & 15.628(018) & 15.684(011) & 15.557(016) & 15.513(017) & 4sh-chip3/SAO \\
2451107.75 & 15.317(040) & 15.487(013) & 15.521(007) & 15.431(010) & 15.367(012) & AndyCam/SAO \\
2451109.86 & 14.972(038) & 15.266(016) & 15.320(010) & 15.233(013) & 15.251(014) & AndyCam/SAO \\
2451110.78 & 14.898(039) & 15.167(016) & 15.209(010) & 15.157(013) & 15.242(013) & 4sh-chip3/SAO \\
2451136.86 & 17.707(041) & 17.380(014) & 16.220(008) & 15.845(013) & 15.564(014) & AndyCam/SAO \\
2451195.70 & 19.901(130) & 18.921(038) & 18.369(023) & 18.465(031) & 18.233(047) & 4sh-chip3/Harris+$I_{\rm SAO}$ \\
2451216.62 & \nodata & 19.230(030) & 18.928(017) & 19.000(024) & 19.011(046) & AndyCam/Harris+$I_{\rm SAO}$ \\
2451224.59 & \nodata & 19.399(056) & 19.066(034) & 19.618(067) & \nodata & AndyCam/Harris+$I_{\rm SAO}$ \\
\enddata
\end{deluxetable}

\begin{deluxetable}{lcccccc}
\singlespace
\tablewidth{0pt}
\tablecaption{Photometry of SN 1998eg\label{ch3-tab-sn98eg}}
\tablehead{\colhead{HJD} & \colhead{$U$} & 
\colhead{$B$} & \colhead{$V$} & \colhead{$R$} & \colhead{$I$} & 
\colhead{Detector/Filterset}}
\startdata
2451110.67 & 16.367(042) & 16.615(022) & 16.480(014) & 16.340(021) & 16.467(020) & 4sh-chip3/SAO \\
2451111.66 & 16.409(041) & 16.594(019) & 16.486(011) & 16.303(017) & 16.545(018) & 4sh-chip3/SAO \\
2451113.58 & 16.552(043) & 16.680(022) & 16.490(012) & 16.314(018) & 16.510(028) & 4sh-chip3/SAO \\
2451136.64 & 19.506(099) & 18.926(064) & 17.714(048) & 17.217(062) & 17.014(062) & AndyCam/SAO \\
2451195.61 & \nodata & 20.415(096) & 19.669(062) & 19.397(093) & \nodata & 4sh-chip3/Harris+$I_{\rm SAO}$ \\
\enddata
\end{deluxetable}

\begin{deluxetable}{lcccccc}
\singlespace
\tablewidth{0pt}
\tablecaption{Photometry of SN 1998es\label{ch3-tab-sn98es}}
\tablehead{\colhead{HJD} & \colhead{$U$} & 
\colhead{$B$} & \colhead{$V$} & \colhead{$R$} & \colhead{$I$} & 
\colhead{Detector/Filterset}}
\startdata
2451133.84 & \nodata & \nodata & \nodata & 14.391(039) & \nodata & AndyCam/SAO \\
2451134.80 & \nodata & \nodata & \nodata & 14.228(039) & \nodata & AndyCam/SAO \\
2451135.67 & 13.827(045) & 14.288(021) & 14.270(012) & 14.205(018) & \nodata & AndyCam/SAO \\
2451136.84 & 13.725(084) & 14.231(032) & 14.131(019) & 14.073(031) & 14.159(027) & AndyCam/SAO \\
2451137.77 & \nodata & 14.090(027) & 14.061(014) & 13.994(021) & 14.095(019) & 4sh-chip3/SAO \\
2451138.64 & 13.537(047) & 14.032(022) & 13.995(013) & 13.913(023) & 14.110(021) & 4sh-chip3/SAO \\
2451140.78 & 13.461(045) & 13.945(023) & 13.927(016) & 13.850(024) & 14.079(027) & 4sh-chip3/SAO \\
2451141.82 & 13.450(065) & 13.958(026) & 13.889(019) & 13.807(025) & 14.056(027) & 4sh-chip3/SAO \\
\enddata
\end{deluxetable}

\begin{deluxetable}{lcccccc}
\singlespace
\tablewidth{0pt}
\tablecaption{Photometry of SN 1999X\label{ch3-tab-sn99X}}
\tablehead{\colhead{HJD} & \colhead{$U$} & 
\colhead{$B$} & \colhead{$V$} & \colhead{$R$} & \colhead{$I$} & 
\colhead{Detector/Filterset}}
\startdata
2451217.81 & 17.150(065) & 17.007(041) & 16.640(031) & 16.602(046) & 16.969(054) & AndyCam/Harris+$I_{\rm SAO}$ \\
2451218.89 & 17.356(063) & 17.086(046) & 16.717(034) & 16.744(042) & 16.991(043) & AndyCam/Harris+$I_{\rm SAO}$ \\
2451219.86 & \nodata & 17.215(043) & 16.765(032) & 16.779(038) & 17.016(038) & AndyCam/Harris+$I_{\rm SAO}$ \\
2451224.71 & 18.214(082) & 17.842(067) & 17.074(047) & 16.907(058) & 16.963(059) & AndyCam/Harris+$I_{\rm SAO}$ \\
2451233.68 & 19.056(158) & 18.803(105) & 17.568(080) & 17.078(097) & 16.867(103) & 4sh-chip3/Harris+$I_{\rm SAO}$ \\
2451275.67 & 20.175(259) & 19.924(172) & \nodata & 18.764(038) & \nodata & 4sh-chip3/Harris+$I_{\rm SAO}$ \\
2451284.69 & 20.444(209) & \nodata & \nodata & 18.976(046) & \nodata & 4sh-chip3/Harris+$I_{\rm SAO}$ \\
2451312.68 & \nodata & \nodata & 19.906(031) & 19.922(045) & \nodata & 4sh-chip3/Harris \\
\enddata
\end{deluxetable}
\clearpage

\begin{deluxetable}{lcccccc}
\singlespace
\tablewidth{0pt}
\tablecaption{Photometry of SN 1999aa\label{ch3-tab-sn99aa}}
\tablehead{\colhead{HJD} & \colhead{$U$} & 
\colhead{$B$} & \colhead{$V$} & \colhead{$R$} & \colhead{$I$} & 
\colhead{Detector/Filterset}}
\startdata
2451222.90 & 14.986(036) & 15.603(013) & 15.633(008) & 15.573(011) & 15.704(013) & AndyCam/Harris+$I_{\rm SAO}$ \\
2451223.84 & 14.818(037) & 15.417(012) & 15.466(007) & 15.439(009) & 15.534(011) & AndyCam/Harris+$I_{\rm SAO}$ \\
2451224.79 & 14.693(043) & 15.276(022) & 15.351(009) & 15.346(012) & 15.415(020) & AndyCam/Harris+$I_{\rm SAO}$ \\
2451226.76 & 14.477(041) & 15.088(021) & 15.147(016) & 15.145(019) & 15.282(020) & AndyCam/Harris+$I_{\rm SAO}$ \\
2451227.80 & 14.422(036) & 14.960(013) & 15.079(009) & 15.064(012) & \nodata & AndyCam/Harris+$I_{\rm SAO}$ \\
2451230.64 & \nodata & 14.933(014) & 14.994(007) & \nodata & 15.238(010) & 4sh-chip3/Harris+$I_{\rm SAO}$ \\
2451232.69 & 14.442(038) & 14.876(015) & 14.911(007) & 14.890(008) & 15.214(010) & 4sh-chip3/Harris+$I_{\rm SAO}$ \\
2451233.62 & 14.486(036) & 14.895(012) & 14.909(008) & 14.888(011) & 15.290(023) & 4sh-chip3/Harris+$I_{\rm SAO}$ \\
\enddata
\end{deluxetable}

\begin{deluxetable}{lcccccc}
\singlespace
\tablewidth{0pt}
\tablecaption{Photometry of SN 1999ac\label{ch3-tab-sn99ac}}
\tablehead{\colhead{HJD} & \colhead{$U$} & 
\colhead{$B$} & \colhead{$V$} & \colhead{$R$} & \colhead{$I$} & 
\colhead{Detector/Filterset}}
\startdata
2451246.89 & 14.081(035) & 14.361(012) & 14.392(007) & \nodata & 14.335(011) & 4sh-chip3/Harris+$I_{\rm SAO}$ \\
2451247.95 & 14.039(046) & 14.316(028) & 14.356(022) & 14.271(027) & 14.303(027) & 4sh-chip3/Harris+$I_{\rm SAO}$ \\
2451249.99 & \nodata & 14.274(012) & 14.253(007) & 14.200(010) & 14.290(011) & AndyCam/Harris+$I_{\rm SAO}$ \\
2451252.00 & 14.168(037) & 14.270(012) & 14.237(008) & 14.167(010) & 14.344(012) & AndyCam/Harris+$I_{\rm SAO}$ \\
2451253.99 & 14.276(038) & 14.372(015) & 14.211(008) & 14.178(012) & 14.373(013) & AndyCam/Harris+$I_{\rm SAO}$ \\
2451258.00 & 14.690(045) & 14.698(015) & 14.278(010) & 14.237(012) & 14.482(014) & 4sh-chip3/Harris+$I_{\rm SAO}$ \\
2451260.96 & 15.057(038) & 15.043(013) & 14.473(007) & 14.383(010) & 14.600(011) & 4sh-chip3/Harris+$I_{\rm SAO}$ \\
2451275.90 & 16.888(047) & 16.430(022) & 15.319(009) & 14.901(024) & 14.712(017) & 4sh-chip3/Harris+$I_{\rm SAO}$ \\
\enddata
\end{deluxetable}

\begin{deluxetable}{lcccccc}
\singlespace
\tablewidth{0pt}
\tablecaption{Photometry of SN 1999cc\label{ch3-tab-sn99cc}}
\tablehead{\colhead{HJD} & \colhead{$U$} & 
\colhead{$B$} & \colhead{$V$} & \colhead{$R$} & \colhead{$I$} & 
\colhead{Detector/Filterset}}
\startdata
2451312.84 & 16.702(039) & 16.947(017) & 16.955(011) & 16.818(015) & 16.914(022) & 4sh-chip3/Harris \\
2451317.90 & 16.878(040) & 16.907(019) & 16.884(013) & 16.716(017) & 17.128(025) & 4sh-chip3/Harris \\
2451336.80 & 19.439(136) & 19.099(068) & 17.935(042) & 17.462(051) & 17.355(067) & 4sh-chip3/Harris \\
2451339.79 & 19.718(091) & 19.386(042) & 18.190(021) & 17.640(026) & 17.530(037) & 4sh-chip3/Harris \\
2451343.79 & 19.966(139) & 19.659(046) & 18.510(023) & 17.987(028) & 17.809(039) & 4sh-chip3/Harris \\
2451434.65 & \nodata & 21.494(276) & 20.843(181) & \nodata & \nodata & 4sh-chip3/Harris \\
\enddata
\end{deluxetable}

\begin{deluxetable}{lcccccc}
\singlespace
\tablewidth{0pt}
\tablecaption{Photometry of SN 1999cl\label{ch3-tab-sn99cl}}
\tablehead{\colhead{HJD} & \colhead{$U$} & 
\colhead{$B$} & \colhead{$V$} & \colhead{$R$} & \colhead{$I$} & 
\colhead{Detector/Filterset}}
\startdata
2451335.69 & 16.071(184) & \nodata & 14.354(035) & 13.702(046) & 13.346(049) & 4sh-chip3/Harris \\
2451338.65 & 15.793(061) & 15.126(043) & 14.022(028) & 13.455(034) & 13.147(035) & 4sh-chip3/Harris \\
2451340.65 & 15.777(064) & 15.066(046) & 13.933(030) & 13.387(036) & 13.147(042) & 4sh-chip3/Harris \\
2451342.65 & 15.838(060) & 15.076(041) & 13.887(027) & 13.326(038) & 13.131(041) & 4sh-chip3/Harris \\
2451343.69 & 15.904(060) & 15.091(037) & 13.878(022) & 13.323(028) & 13.175(031) & 4sh-chip3/Harris \\
2451346.70 & \nodata & \nodata & 13.869(031) & \nodata & \nodata & 4sh-chip3/Harris \\
2451348.68 & 16.405(066) & 15.395(043) & 13.946(027) & 13.471(032) & 13.340(035) & 4sh-chip3/Harris \\
2451350.66 & 16.531(072) & 15.530(054) & 14.044(036) & 13.567(051) & 13.429(054) & 4sh-chip3/Harris \\
2451362.72 & 18.588(087) & 16.846(040) & 14.709(026) & 13.934(032) & 13.394(036) & 4sh-chip3/Harris \\
\enddata
\end{deluxetable}

\begin{deluxetable}{lcccccc}
\singlespace
\tablewidth{0pt}
\tablecaption{Photometry of SN 1999cw\label{ch3-tab-sn99cw}}
\tablehead{\colhead{HJD} & \colhead{$U$} & 
\colhead{$B$} & \colhead{$V$} & \colhead{$R$} & \colhead{$I$} & 
\colhead{Detector/Filterset}}
\startdata
2451376.96 & 16.185(064) & 15.925(043) & 15.044(022) & 14.821(041) & 14.895(044) & 4sh-chip3/Harris \\
2451434.86 & 17.991(042) & 17.624(016) & 17.003(007) & 16.828(015) & 16.982(019) & 4sh-chip3/Harris \\
2451461.79 & 18.441(057) & 18.044(023) & 17.606(016) & 17.639(020) & 17.849(028) & 4sh-chip3/Harris \\
2451487.71 & 18.998(074) & 18.445(030) & 18.103(018) & 18.325(024) & 18.602(065) & 4sh-chip3/Harris \\
2451521.60 & 19.794(190) & 18.916(050) & 18.663(033) & 19.102(046) & 19.229(103) & 4sh-chip3/Harris \\
2451546.62 & 20.297(161) & 19.367(045) & 19.073(030) & 19.689(056) & \nodata & 4sh-chip3/Harris \\
\enddata
\end{deluxetable}

\begin{deluxetable}{lcccccc}
\singlespace
\tablewidth{0pt}
\tablecaption{Photometry of SN 1999dq\label{ch3-tab-sn99dq}}
\tablehead{\colhead{HJD} & \colhead{$U$} & 
\colhead{$B$} & \colhead{$V$} & \colhead{$R$} & \colhead{$I$} & 
\colhead{Detector/Filterset}}
\startdata
2451426.99 & 15.027(043) & 15.477(022) & 15.325(014) & 15.202(017) & 15.147(022) & 4sh-chip3/Harris \\
2451427.97 & 14.831(037) & 15.322(015) & 15.162(010) & 15.027(013) & 15.029(015) & 4sh-chip3/Harris \\
2451428.88 & 14.663(054) & 15.199(021) & 15.074(014) & 14.943(018) & 14.960(020) & 4sh-chip3/Harris \\
2451429.99 & 14.613(038) & 15.092(018) & 14.955(010) & 14.843(014) & 14.861(017) & 4sh-chip3/Harris \\
2451431.93 & 14.491(046) & 14.953(023) & 14.766(016) & 14.689(020) & 14.808(026) & 4sh-chip3/Harris \\
2451432.97 & 14.472(064) & \nodata & \nodata & 14.594(038) & \nodata & 4sh-chip3/Harris \\
2451434.00 & 14.431(040) & 14.895(020) & 14.740(013) & 14.644(016) & 14.760(021) & 4sh-chip3/Harris \\
2451434.95 & 14.466(039) & 14.853(021) & 14.697(015) & 14.587(021) & 14.788(021) & 4sh-chip3/Harris \\
\enddata
\end{deluxetable}

\begin{deluxetable}{lcccccc}
\singlespace
\tablewidth{0pt}
\tablecaption{Photometry of SN 1999ef\label{ch3-tab-sn99ef}}
\tablehead{\colhead{HJD} & \colhead{$U$} & 
\colhead{$B$} & \colhead{$V$} & \colhead{$R$} & \colhead{$I$} & 
\colhead{Detector/Filterset}}
\startdata
2451464.81 & 17.842(047) & 17.720(022) & 17.491(015) & 17.316(019) & 17.909(031) & 4sh-chip3/Harris \\
2451467.73 & 18.109(041) & 17.945(018) & 17.627(011) & 17.586(016) & 18.131(029) & 4sh-chip3/Harris \\
2451469.85 & 18.301(081) & 18.146(020) & 17.765(012) & 17.742(020) & 18.343(080) & 4sh-chip3/Harris \\
2451487.74 & 20.248(116) & 19.955(039) & 18.763(024) & 18.173(031) & 18.174(041) & 4sh-chip3/Harris \\
2451521.62 & \nodata & 20.953(081) & 20.032(046) & 19.573(074) & 20.215(120) & 4sh-chip3/Harris \\
2451546.68 & \nodata & 21.313(258) & 20.487(177) & 20.464(234) & \nodata & 4sh-chip3/Harris \\
2451581.62 & \nodata & 21.798(280) & 20.858(177) & \nodata & \nodata & 4sh-chip3/Harris \\
\enddata
\end{deluxetable}

\begin{deluxetable}{lcccccc}
\singlespace
\tablewidth{0pt}
\tablecaption{Photometry of SN 1999ej\label{ch3-tab-sn99ej}}
\tablehead{\colhead{HJD} & \colhead{$U$} & 
\colhead{$B$} & \colhead{$V$} & \colhead{$R$} & \colhead{$I$} & 
\colhead{Detector/Filterset}}
\startdata
2451487.75 & 15.728(038) & 15.833(016) & 15.632(010) & 15.531(013) & 15.795(014) & 4sh-chip3/Harris \\
2451494.83 & 16.801(041) & 16.629(018) & 16.072(011) & 16.060(014) & 16.267(016) & 4sh-chip3/Harris \\
2451521.70 & 18.902(135) & 18.831(075) & \nodata & 17.373(044) & 17.174(045) & 4sh-chip3/Harris \\
2451527.64 & \nodata & \nodata & 17.968(256) & \nodata & \nodata & 4sh-chip3/Harris \\
2451546.71 & 19.690(133) & 19.332(048) & 18.529(019) & 18.326(029) & 18.372(052) & 4sh-chip3/Harris \\
2451589.59 & \nodata & 20.087(164) & \nodata & 19.604(161) & \nodata & 4sh-chip3/Harris \\
2451590.61 & \nodata & 19.981(118) & 19.615(074) & 19.684(110) & 19.818(197) & 4sh-chip3/Harris \\
\enddata
\end{deluxetable}

\begin{deluxetable}{lcccccc}
\singlespace
\tablewidth{0pt}
\tablecaption{Photometry of SN 1999ek\label{ch3-tab-sn99ek}}
\tablehead{\colhead{HJD} & \colhead{$U$} & 
\colhead{$B$} & \colhead{$V$} & \colhead{$R$} & \colhead{$I$} & 
\colhead{Detector/Filterset}}
\startdata
2451487.91 & 18.629(058) & 18.186(021) & 17.324(011) & 16.905(015) & 16.803(017) & 4sh-chip3/Harris \\
2451495.91 & 19.737(083) & 18.924(030) & 17.757(012) & 17.401(016) & 17.272(022) & 4sh-chip3/Harris \\
2451517.87 & \nodata & 20.934(130) & 19.077(038) & 18.296(049) & 17.609(049) & 4sh-chip3/Harris \\
2451521.81 & \nodata & 20.988(150) & 19.231(044) & 18.505(060) & 17.712(064) & 4sh-chip3/Harris \\
2451545.84 & \nodata & 21.286(121) & 19.890(067) & 19.251(083) & 18.888(085) & 4sh-chip3/Harris \\
2451584.62 & \nodata & \nodata & 20.825(161) & 20.616(209) & 20.719(224) & 4sh-chip3/Harris \\
2451590.70 & \nodata & \nodata & 20.966(201) & \nodata & \nodata & 4sh-chip3/Harris \\
\enddata
\end{deluxetable}

\begin{deluxetable}{lcccccc}
\singlespace
\tablewidth{0pt}
\tablecaption{Photometry of SN 1999gd\label{ch3-tab-sn99gd}}
\tablehead{\colhead{HJD} & \colhead{$U$} & 
\colhead{$B$} & \colhead{$V$} & \colhead{$R$} & \colhead{$I$} & 
\colhead{Detector/Filterset}}
\startdata
2451521.93 & 17.325(059) & 17.144(015) & 16.680(010) & 16.381(014) & 16.470(027) & 4sh-chip3/Harris \\
2451527.95 & 17.987(078) & 17.563(020) & 16.842(011) & 16.627(014) & 16.783(019) & 4sh-chip3/Harris \\
2451545.91 & 20.109(055) & 19.482(016) & 17.923(007) & 17.242(011) & 16.812(012) & 4sh-chip3/Harris \\
2451556.84 & 20.785(135) & 20.343(045) & 18.691(021) & 18.028(026) & 17.502(028) & 4sh-chip3/Harris \\
2451581.82 & \nodata & 20.486(052) & 19.312(021) & 18.860(027) & 18.525(038) & 4sh-chip3/Harris \\
2451589.86 & \nodata & 20.521(118) & 19.563(046) & 19.096(065) & 18.716(120) & 4sh-chip3/Harris \\
2451590.89 & \nodata & 20.531(101) & 19.508(045) & 19.068(071) & 18.651(140) & 4sh-chip3/Harris \\
2451640.77 & \nodata & \nodata & 20.791(091) & 20.549(123) & \nodata & 4sh-chip3/Harris \\
2451670.68 & \nodata & \nodata & 21.398(257) & \nodata & \nodata & 4sh-chip3/Harris \\
\enddata
\end{deluxetable}
\clearpage

\begin{deluxetable}{lcccccc}
\singlespace
\tablewidth{0pt}
\tablecaption{Photometry of SN 1999gh\label{ch3-tab-sn99gh}}
\tablehead{\colhead{HJD} & \colhead{$U$} & 
\colhead{$B$} & \colhead{$V$} & \colhead{$R$} & \colhead{$I$} & 
\colhead{Detector/Filterset}}
\startdata
2451520.95 & 15.381(040) & 15.172(011) & 14.507(007) & 14.450(008) & 14.651(013) & 4sh-chip3/Harris \\
2451522.01 & 15.568(040) & 15.321(011) & 14.587(007) & 14.524(011) & 14.687(015) & 4sh-chip3/Harris \\
2451523.00 & 15.733(037) & 15.477(014) & 14.657(010) & 14.570(012) & 14.679(015) & 4sh-chip3/Harris \\
2451524.97 & 16.067(038) & 15.814(015) & 14.813(010) & 14.661(012) & 14.707(015) & 4sh-chip3/Harris \\
2451527.04 & 16.420(043) & 16.099(016) & 14.991(009) & 14.712(012) & 14.589(014) & 4sh-chip3/Harris \\
2451541.95 & 17.656(051) & 17.477(017) & 16.247(009) & 15.799(012) & 15.520(016) & 4sh-chip3/Harris \\
2451545.95 & 17.810(113) & 17.631(015) & 16.442(010) & 15.986(014) & 15.737(019) & 4sh-chip3/Harris \\
2451549.92 & 17.902(052) & 17.728(019) & 16.589(008) & 16.190(012) & 15.973(020) & 4sh-chip3/Harris \\
\enddata
\end{deluxetable}

\begin{deluxetable}{lcccccc}
\singlespace
\tablewidth{0pt}
\tablecaption{Photometry of SN 1999gp\label{ch3-tab-sn99gp}}
\tablehead{\colhead{HJD} & \colhead{$U$} & 
\colhead{$B$} & \colhead{$V$} & \colhead{$R$} & \colhead{$I$} & 
\colhead{Detector/Filterset}}
\startdata
2451545.81 & 15.756(037) & 16.330(012) & 16.338(008) & 16.243(011) & 16.352(012) & 4sh-chip3/Harris \\
2451546.74 & 15.758(041) & 16.285(016) & 16.206(008) & 16.177(012) & 16.330(012) & 4sh-chip3/Harris \\
2451548.65 & 15.760(036) & 16.199(013) & 16.226(009) & 16.145(012) & 16.366(015) & 4sh-chip3/Harris \\
2451551.63 & 15.874(037) & 16.251(013) & 16.140(007) & 16.023(011) & 16.406(016) & 4sh-chip3/Harris \\
2451553.57 & 16.050(038) & 16.291(013) & 16.136(008) & 16.015(011) & 16.472(014) & 4sh-chip3/Harris \\
2451555.62 & 16.150(037) & 16.475(014) & 16.280(009) & 16.171(012) & 16.568(014) & 4sh-chip3/Harris \\
2451564.64 & 17.109(048) & 17.011(019) & 16.564(012) & 16.527(017) & 17.048(041) & 4sh-chip3/Harris \\
2451570.58 & 17.857(072) & 17.589(021) & 16.872(011) & \nodata & 17.142(033) & 4sh-chip3/Harris \\
2451580.58 & 18.870(066) & 18.463(031) & 17.306(017) & 16.851(026) & 16.831(029) & 4sh-chip3/Harris \\
2451581.67 & 18.985(063) & 18.578(017) & 17.324(010) & 16.849(017) & 16.850(017) & 4sh-chip3/Harris \\
2451589.67 & 19.378(193) & 18.999(028) & 17.822(016) & 17.249(024) & 17.053(032) & 4sh-chip3/Harris \\
\enddata
\end{deluxetable}%

\clearpage

\end{document}